\documentclass[apjfonts]{emulateapj}

\newcommand{\mum}{~\mu\mbox{m}}
\newcommand{\Ang}{~\mbox{\AA}}

\newcommand{\nJy}{~\mbox{nJy}}

\newcommand{\invHz}{~\mbox{Hz}^{-1}}
\newcommand{\erg}{~\mbox{erg}}
\newcommand{\s}{~\mbox{s}}

\newcommand{\Myr}{~\mbox{Myr}}

\newcommand{\Gpc}{~\mbox{Gpc}}
\newcommand{\cMpch}{~h^{-1}~\mbox{Mpc}~\mbox{comoving}}

\newcommand{\pc}{~\mbox{pc}}
\newcommand{\ckpch}{~h^{-1}~\mbox{kpc}~\mbox{comoving}}

\newcommand{\invpcsq}{~\mbox{pc}^{-2}}

\newcommand{\cmci}{~\mbox{cm}^{-3}}

\newcommand{\kpc}{~\mbox{kpc}}
\newcommand{\K}{~\mbox{K}}

\newcommand{\invMpc}{~\mbox{Mpc}^{-1}}

\newcommand{\invyr}{~\mbox{yr}^{-1}}
\newcommand{\invs}{~\mbox{s}^{-1}}

\newcommand{\Msun}{~\mbox{M}_{\odot}}
\newcommand{\Zsun}{~Z_{\odot}}

\newcommand{\kms}{~\mbox{km s}^{-1}}

\newcommand{\cmbfast}{{\sc cmbfast}}

\slugcomment{Submitted to ApJ}

\begin{document}

\shorttitle{THE FIRST DISK GALAXIES}
\shortauthors{PAWLIK ET AL.}

\title{The First Galaxies: Assembly of Disks and Prospects for Direct Detection}

\author{Andreas H. Pawlik, Milo\v s Milosavljevi\'c, and Volker Bromm}

\affil{Department of Astronomy and Texas Cosmology Center, The University of Texas at Austin, TX 78712}

\begin{abstract}
The {\it James Webb Space Telescope} ({\it JWST}) will enable
observations of galaxies at redshifts $z \gtrsim 10$ and hence allow
to test our current understanding of structure formation at very early
times. Previous work has shown that the very first galaxies inside
halos with virial temperatures $T_{\rm vir} \lesssim 10^4 \K$ and
masses $M_{\rm vir} \lesssim 10^8 \Msun$ at $z \gtrsim 10$ are
probably too faint, by at least one order of magnitude, to be detected
even in deep exposures with {\it JWST}. The light collected with {\it
JWST} may therefore be dominated by radiation from galaxies inside ten
times more massive halos. We use cosmological zoomed smoothed particle
hydrodynamics simulations to investigate the assembly of such galaxies
and assess their observability with {\it JWST}. We compare two
simulations that are identical except for the inclusion of
non-equilibrium H/D chemistry and radiative cooling by molecular
hydrogen. In both simulations a large fraction of the halo gas settles
in two nested, extended gas disks which surround a compact massive gas
core. The presence of molecular hydrogen allows the disk gas to reach
low temperatures and to develop marked spiral structure but does not
qualitatively change its stability against fragmentation. We post-process the
simulated galaxies by combining idealized models for star formation
with stellar population synthesis models to estimate the luminosities
in nebular recombination lines as well as in the ultraviolet
continuum. We demonstrate that {\it JWST} will be able to constrain
the nature of the stellar populations in galaxies such as simulated
here based on the detection of the He1640 recombination
line. Extrapolation of our results to halos with masses both lower and
higher than those simulated shows that {\it JWST} may find up to a
thousand star-bursting galaxies in future deep exposures of the $z
\gtrsim 10$ universe.
\end{abstract}


\section{Introduction}
The hierarchical assembly of dark matter halos and the cooling and
condensation of the cosmic gas to form stars and galaxies inside them
(\citealp{Rees:1977}; \citealp{Silk:1977}; \citealp{White:1978};
\citealp{Blumenthal:1984}) are major pillars of the current 
cold dark matter (CDM) paradigm of structure formation in the universe 
with cosmological constant $\Lambda$.  Both (semi-)analytical
arguments (e.g., \citealp{Tegmark:1997}; \citealp{Reed:2005};
\citealp{Naoz:2006}) and simulations (e.g., \citealp{Bromm:2002};
\citealp{Abel:2002}; \citealp{Yoshida:2006}) suggest that the first
stars have formed at redshifts as high as $z \sim
30$, when the universe was just about a percent of its present age
(for reviews see, e.g., \citealp{Barkana:2001}; \citealp{Bromm:2004}; \citealp{Bromm:2009}).
Their light ended the Dark Ages that followed the release of the
cosmic microwave background (CMB) radiation at $z \approx 1100$ and
fundamentally transformed the universe during a landmark period called
the epoch of reionization (for reviews see, e.g., \citealp{Loeb:2001}; 
\citealp{Ciardi:2005}; \citealp{Barkana:2007}; \citealp{Trac:2009}; \citealp{Stiavelli:2009}; \citealp{Loeb:2010}).
\par  
Future observations with telescopes such as, for
example, {\it Planck}\footnote{sci.esa.int/planck/}, the {\it Low
Frequency Array}\footnote{http://www.lofar.org}, the {\it Murchison
Widefield Array}\footnote{http://www.haystack.mit.edu/ast/arrays/mwa/}, 
the {\it Atacama Large Millimeter
Array}\footnote{http://www.almaobservatory.org/}, and the {\it James
Webb Space Telescope} ({\it JWST})\footnote{http://www.jwst.nasa.gov/} 
will test our current theoretical understanding of the formation of
stars and galaxies at these early times. A fascinating prospect is the
detection of line and continuum radiation from the first galaxies with
{\it JWST}. The ratio of the hydrogen and helium Balmer line
luminosities from recombining gas has been proposed as a telltale
signature that distinguishes between first-generation, metal-free 
(Population~III) and subsequent metal-enriched (Population~II) star
formation or between stellar sources and black
holes (e.g., \citealp{Tumlinson:2000}; \citealp{Bromm:2001};
\citealp{Oh:2001}; \citealp{Tumlinson:2001}; \citealp{Schaerer:2002}; 
\citealp{Schaerer:2003}; \citealp{Johnson:2009}). In addition, the detection of Ly$\alpha$, 
molecular, or metal line cooling radiation from high
redshifts would probe the gravitational assembly of the gas
in the first halos (e.g., \citealp{Haiman:2000};
\citealp{Dijkstra:2009b}, \citealp{Mizusawa:2005}; \citealp{Appleton:2009}), offering
direct insights in the structure and dynamics of the first galaxies
and the surrounding intergalactic medium (e.g., \citealp{Santos:2004};
\citealp{Dijkstra:2006}; \citealp{Verhamme:2006};
\citealp{Dijkstra:2007}; \citealp{Laursen:2010}).
\par
The combination of upcoming observations with {\it JWST} and other
future telescopes with detailed numerical supercomputer simulations of
the first galaxies and reionization will transform our knowledge of
structure formation in the universe. Most of the numerical effort in early galaxy
formation has concentrated on investigating the properties of the very
first building blocks of galaxy assembly, minihalos and dwarf galaxies
with virial temperatures $\lesssim 10^4 \K$, corresponding to halo
masses $\lesssim 10^8 \Msun$ at $z\gtrsim 10$ (e.g.,
\citealp{Abel:2002}; \citealp{Bromm:2002}; \citealp{Wise:2008};
\citealp{Turk:2009}; \citealp{Stacy:2010}; \citealp{Greif:2010}). A
key result emerging from the existing work is that the luminosities 
of these very low-mass objects are, unless magnified by
gravitational lensing, too low, by at least one order of magnitude, to
be detected in even very deep exposures with {\it JWST} (\citealp{Greif:2009};
\citealp{Johnson:2009}; see also, e.g., \citealp{Loeb:1997}; 
\citealp{Haiman:1998}; \citealp{Oh:1999}; \citealp{Oh:2001}; \citealp{Trenti:2009}). The light collected in future deep-field
observations with the {\it JWST} may thus be dominated by emission from
dwarf galaxies inside halos that are about ten times more massive, $M_{\rm vir}
\sim 10^9 \Msun$.
\par 
Our goal is to extend and complement existing numerical work on the
first galaxies by investigating the role played by
galaxies inside halos with masses $M_{\rm vir} \gtrsim 10^9 \Msun$,
at times before and during the epoch of reionization, when these galaxies
were assembling, possibly contributing a significant fraction (if not
most; e.g., \citealp{Loeb:2009}; \citealp{Wise:2009}; \citealp{Salvaterra:2010};
\citealp{Raicevic:2010}) of the ionizing emissivity in the universe,
to the present day, when these galaxies may be found in the Local
Group as fossil probes of the beginnings of galaxy formation (for
reviews see, e.g., \citealp{Mateo:1998}; \citealp{Tolstoy:2009}; \citealp{Ricotti:2010};
\citealp{Mayer:2010}). Indeed, the new field of `dwarf archaeology'
may hold the key to unravel the interplay of star and galaxy formation
at the end of the cosmic dark ages (\citealp{Frebel:2010}). Here, we
report our first steps towards achieving this goal by studying the
assembly of dwarf galaxies in halos reaching virial masses $M_{\rm
vir} \sim 10^9 \Msun$ at $z = 10$ using cosmological smoothed particle
hydrodynamics simulations.
\par
We utilize a zoomed simulation technique that allows us
to simulate the gravitational and hydrodynamical processes of dwarf
galaxy formation at high resolution while keeping information about
structure formation at large representative scales. Our simulations
include radiative cooling from atoms and molecules in gas of
primordial composition but ignore star formation and the associated
feedback.  We post-process our simulations with idealized models for star
formation and employ population synthesis models to estimate the
prospects for a direct detection of the first galaxies with the
upcoming {\it JWST}. Other aspects of the simulated galaxies, like
e.g., their role as reionization sources, will be investigated in
subsequent works, in which we will explicitly account for the effects of star
formation and associated feedback.
\par
\par
The structure of this paper is as follows. In
Section~\ref{Sec:Simulations} we describe the set-up of our
simulations. Then, in Section~\ref{Sec:Results}, we present our
results, and subsequently discuss the assembly of the simulated halo,
its structure and dynamics at redshift $z = 10$ and the properties of
the disks it hosts.  In Section~\ref{Sec:JWST} we combine our
simulations with assumptions about star formation to assess the
observability of the first galaxies in future observations with {\it
JWST}. In Section~\ref{Sec:Implications} and Section~\ref{Sec:Limitations} 
we discuss, respectively, implications and
limitations of our work. Finally, in Section~\ref{Sec:Summary}, we
summarize our work.
\par
Throughout this work we assume $\Lambda$CDM cosmological parameters 
$\Omega_{\rm{m}} = 0.258, \Omega_{\rm{b}} = 0.0441,
\Omega_\Lambda = 0.742, \sigma_8 = 0.796, n_{\rm{s}} = 0.963$, and $h =
0.719$, which are consistent with the 5-year
(\citealp{Komatsu:2009}) and 7-year (\citealp{Komatsu:2010})
analyses of the observations with the {\it Wilkinson Microwave Anisotropy Probe} satellite. 
Distances are expressed in physical (i.e., not
comoving) units, unless noted otherwise.
\section{Simulations}
\label{Sec:Simulations}
We use a modified version of the N-body/TreePM Smoothed Particle
Hydrodynamics (SPH) code {\sc gadget} (\citealp{Springel:2005};
\citealp{Springel:2001a}) to perform a suite of high-resolution
zoomed cosmological hydrodynamical simulations in a box of size
$L=3.125 \cMpch$. The box size was chosen after inspection of
published dark matter halo mass functions (e.g., \citealp{Reed:2007})
such that the box contains at least one halo of mass $\sim 10^9 \Msun$
at redshift $z = 10$. 
\par
We carry out SPH simulations of primordial metal-free gas including
non-equilibrium radiative cooling from both molecular and atomic
species and from atomic species only. In the following, these two
types of simulations will be distinguished by an additional {\it
NOMOL} at the end of the label of the atomic cooling simulation.  The
simulations are performed with the gravitational forces softened over
a sphere of Plummer-equivalent radius $\epsilon$. Our simulations 
\textit{Z4} and \textit{Z4NOMOL}, which 
are obtained by zooming into a parent cosmological 
simulation 4 times, use a force
softening radius $\epsilon = 0.1 \ckpch$ applied to all particles. 
\par
We
employ the entropy-conserving formulation of SPH
(\citealp{Springel:2002}) with $N_{\rm neigh} = 48$ neighbor particles
per SPH kernel. We limit the radius $h$ of the SPH kernel to above a
fraction $f_h$ of the softening length, $h \ge f_h \epsilon$, where
$f_h = 0.01$. The simulations are summarized in
Table~\ref{tbl:params}.
\begin{deluxetable}{rcccc}
\tablecolumns{5}
\tablewidth{3.5in}
\tablecaption{Simulation Parameters}
\tablehead{
  \colhead{Simulation} & 
  \colhead{$m_{\rm gas}$ \tablenotemark{a}} &
  \colhead{$m_{\rm DM}$ \tablenotemark{b}} &
  \colhead{Comoving $\epsilon$ \tablenotemark{c}} & 
  \colhead{Cooling} }
\startdata
\textit{Z4}            & $ 4.84 \times 10^2$  &$ 2.35 \times 10^3$ &  $ 0.1$      &  molecular \\
\textit{Z4NOMOL}                    & $4.84 \times 10^2$ & $ 2.35\times 10^3$  &  $0.1$      &   atomic 
\enddata
\tablenotetext{a}{Gas particle mass in the refinement region
  ($M_\odot$).}
\tablenotetext{b}{Dark matter particle mass in the refinement
  region ($M_\odot$).}
\tablenotetext{c}{Gravitational softening radius
  ($h^{-1}\,\textrm{kpc comoving}$).}
 \label{tbl:params}
\end{deluxetable}
\subsection{Initial Conditions}
\begin{figure}
\includegraphics[trim = 0 0 0 -10mm, width = 0.5\textwidth]{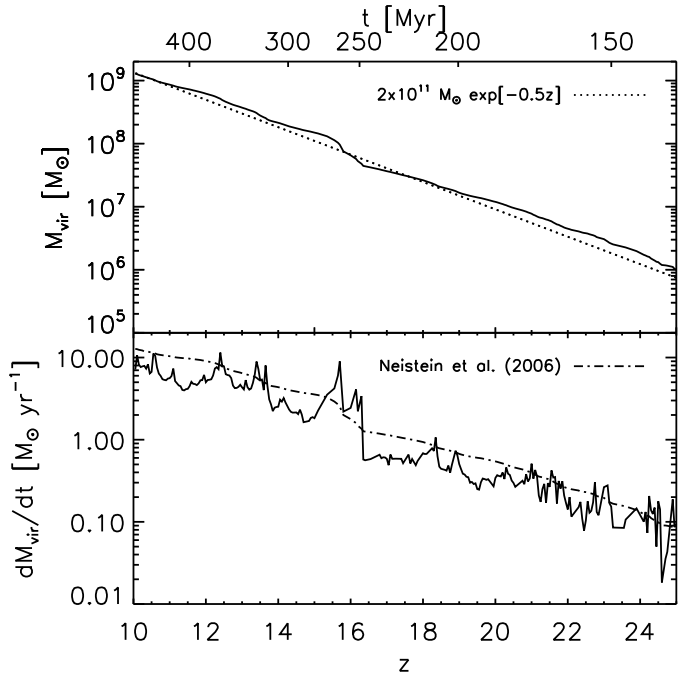}
\caption{Halo assembly history in the {\it Z4} simulation. {\it
Top panel}: Growth of the virial mass $M_{\rm vir}$ associated with the most massive progenitor FOF halo
(solid curve) is close to exponential (dotted line). {\it Bottom panel}: Accretion rate
corresponding to the halo growth shown in the top panel. The accretion rate is consistent with analytical estimates
of the mass growth rate of the main progenitor (\citealp{Neistein:2006}; dash-dotted curve).}
\label{Fig:mah}
\end{figure}
All simulations start at redshift $z = 127$. Initial particle
positions and velocities are obtained by applying the Zel'dovich
approximation (\citealp{Zeldovich:1970}) to particles arranged on a
Cartesian grid. We adopt a transfer function for matter perturbations
generated with \cmbfast\ (version 4.1; \citealp{Seljak:1996}).
\par
To achieve high resolution we make use of the zoomed simulation
technique (\citealp{Navarro:1994}; we use the same code as in
\citealp{Greif:2008}).  We first perform a simulation in which the
initial conditions are set up using $2 \times 64^3$ (dark matter and gas)
particles arranged on a uniform Cartesian grid. We then use the
friends-of-friends (FOF) halo finder, with linking parameter $b =
0.2$, built into the substructure finder {\sc subfind}
(\citealp{Springel:2001b}) to locate an FOF halo with mass $\gtrsim
10^9 \Msun$ at $z = 10$ in this simulation.  Using {\sc subfind}, we
determine the most bound particle of this FOF halo and let it mark the 
center of the region that we wish to resimulate. We compute the associated virial radius $r_{\rm vir}$ by
determining the radius of the sphere around the most bound particle within
which the average matter density equals $200$ times the critical
density at $z = 10$. The particles within a region of radius $r
\lesssim 3 r_{\rm vir}$ around the most bound particle are traced back to
their locations in the initial grid where they mark the region of
refinement.
\par
Subsequently, all parent particles in a cube enclosing the
refinement region are replaced by $8^{N_{\rm l}}$ daughter particles,
where $N_{\rm l}$ is the zoom level. Our simulations 
\textit{Z4} and \textit{Z4NOMOL} employ $N_{\rm l}= 4$
and hence in these simulations the daughter gas (dark matter)
particles have masses $m_{\rm g} = 484 \Msun$ ($m_{\rm DM} = 2350
\Msun$). To 
reduce numerical artifacts due to the large difference in the particle masses for particles 
inside and outside the refinement region (mass ratios $8^{N_{\rm l}}$), 
the refinement region is surrounded by $N_{\rm l} - 1$ concentric, nested layers 
in which the parent particles are successively replaced by $8^{N_{\rm l} - 1}$, $8^{N_{\rm l}-2}$,
$...$, $8$ daughter particles and, hence, the particle masses vary gradually, 
by discrete factors of 8, with increasing distance to the refinement region. 
The simulation is then re-run after applying the Zel'dovich approximation to evolve all 
particles to the starting redshift $z = 127$.
\subsection{Chemistry and Cooling}
All our simulations include radiative cooling in the optically thin
limit.  We assume that the gas is of primordial composition using a
hydrogen mass fraction $X = 0.752$ and a helium mass fraction $Y =
1-X$. We follow the non-equilibrium chemistry and cooling of $\rm
H_2$, $\rm D$, $\rm HD$, $\rm D^+$, $\rm H^+$, $\rm H$, $\rm D$, and
$\rm He$, and we include $\rm H^{-}$ and $\rm H_2^+$ assuming their
equilibrium abundances (\citealp{Johnson:2006}; \citealp{Greif:2010}).  
\par 
In simulation \textit{Z4}, gas cools 
by collisional ionization and excitation, the
emission of free-free and recombination radiation, Compton cooling
off the CMB, and emission of radiation by
molecular hydrogen and hydrogen deuteride (HD). If initial abundances are
expressed as number density with respect to hydrogen, where
$n_{\rm H} = X \rho_{\rm g}/m_{\rm H}$ with $\rho_{\rm g}$ being the gas
density at $z=127$, and $m_{\rm H}$ is the mass of the proton, we choose:
$\rm H_2 = 1.1\times 10^{-6}$, $\rm D = 2.6\times 10^{-5}$, $\rm HD =
10^{-9}$, $\rm D^+ = 1.2\times 10^{-8}$, $\rm H^+ = 3\times 10^{-4}$,
$\rm He^+ = 0$ and $\rm He^{++} = 0$, from which the initial
abundances of the remaining species ($\rm H$, $\rm D$, $\rm He$) as
well as the abundance of electrons are obtained through application
of conservation laws. Our choices for the initial abundances are
consistent with computations of cosmological
abundances in the early universe (e.g., \citealp{Lepp:1984};
\citealp{Galli:1998}). 
\par
Thanks to molecular cooling, gas in simulation
\textit{Z4} may reach temperatures as low as $\sim 10^2
\K$. Simulation \textit{Z4NOMOL} is 
identical to simulation \textit{Z4} except that the
formation of molecular hydrogen is suppressed, as would be the case 
in the presence of a strong photo-dissociating Lyman-Werner radiation
background (\citealp{Stecher:1967}; \citealp{Haiman:1997}). Gas in simulation \textit{Z4NOMOL} 
therefore cools only via atomic processes, which are
inefficient in reducing the thermal energy of primordial gas with
temperatures below $\sim 10^4 \K$.
\subsection{Jeans Floor}
\label{Sec:Simulations:Fragmention}
Simulations with mass resolutions insufficient to resolve the Jeans mass $M_{\rm
J} \equiv (4\pi/3) \rho_{\rm m} (\lambda_{\rm J}/2)^3$ by
at least $N_{\rm res} \equiv 2$ SPH kernel masses $M_{\rm K} \equiv N_{\rm neigh} m_{\rm
g}$ may suffer from artificial fragmentation (\citealp{Bate:1997}). 
Here, $\rho_{\rm m}$ is the total (dark matter and gas) mass density, 
$\lambda_{\rm J} \equiv c_{\rm s} \pi^{1/2}(G\rho_{\rm m})^{-1/2}$ 
the Jeans length, $c_{\rm s} = [\gamma k_{\rm B} T / (\mu m_{\rm H})]^{1/2} $ the
adiabatic speed of sound, $\gamma$ the ratio of specific heats, 
and $\mu$ the mean gas particle mass in units of the proton mass.
Our finite mass resolution implies a maximum density 
\begin{equation}
n_{\rm H, max} = 8 \times 10^5 \cmci f_{\rm g} \left(\frac{\gamma}{\mu}\right)^{3} \left(\frac{T}{ 10^4\K}\right)^3 
\label{Eq:Floor}
\end{equation}
up to which we satisfy the \cite{Bate:1997} criterion, where 
$f_{\rm g} \equiv \rho_{\rm g} / \rho_{\rm m}$. 
\par
To satisfy the \cite{Bate:1997} criterion independent of density we
make use of a density-dependent temperature floor
(\citealp{Robertson:2008}; see also, e.g., \citealp{Schaye:2008} for a
related approach).  At each time step and for all gas particles we
compute the Jeans mass $M_{\rm J}$ and compare it to the resolution
mass $N_{\rm res} M_{\rm K}$.  If the Jeans mass becomes smaller than
the resolution mass, then we increase the particle internal energy
and, hence, the particle temperature such that the Jeans mass becomes
equal to the resolution mass.
\section{Results}
\begin{figure*}
\includegraphics[trim = 20mm 0mm 25mm 0mm, width = 0.32\textwidth]{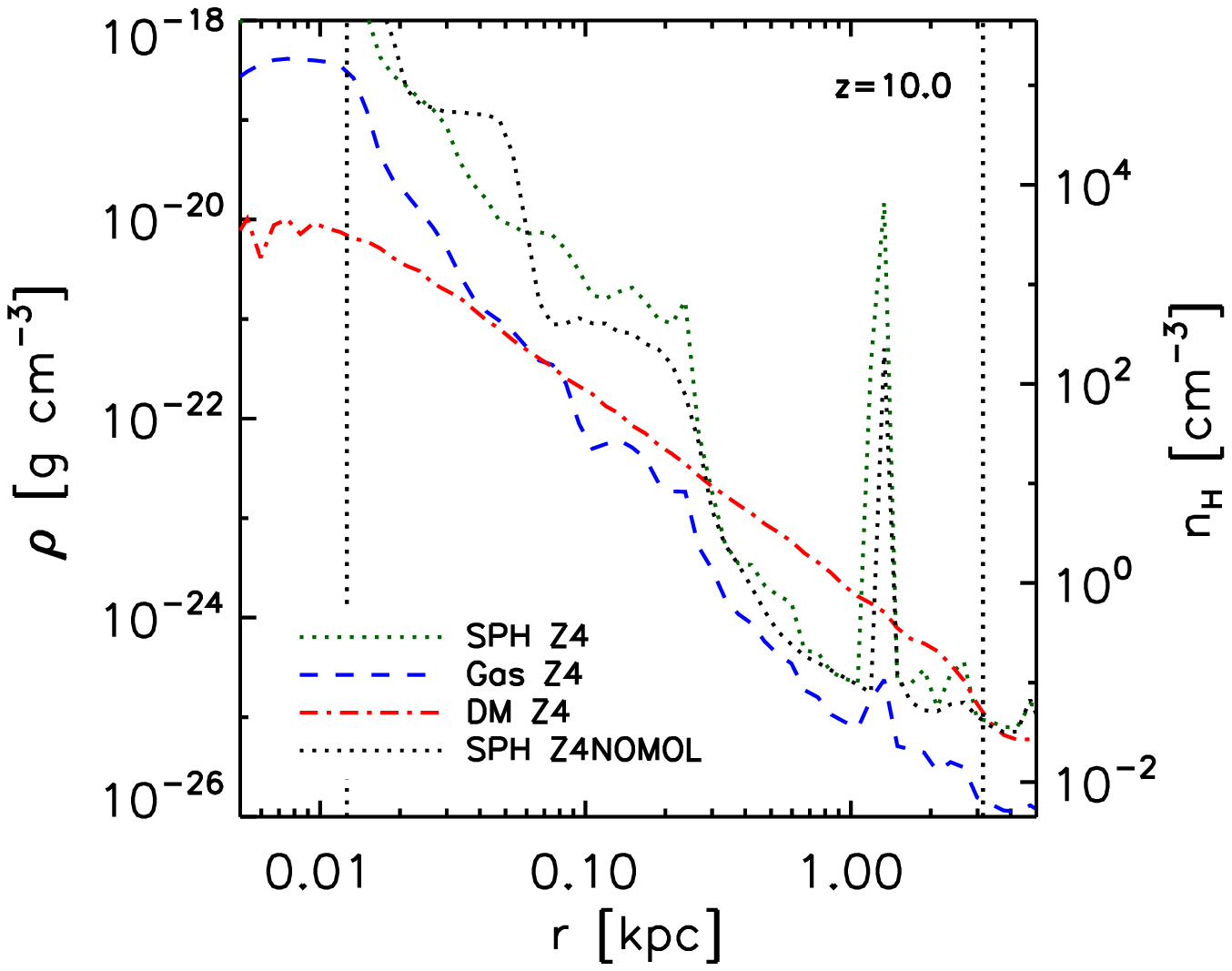}
\includegraphics[trim = 20mm 0mm 25mm 0mm, width = 0.32\textwidth]{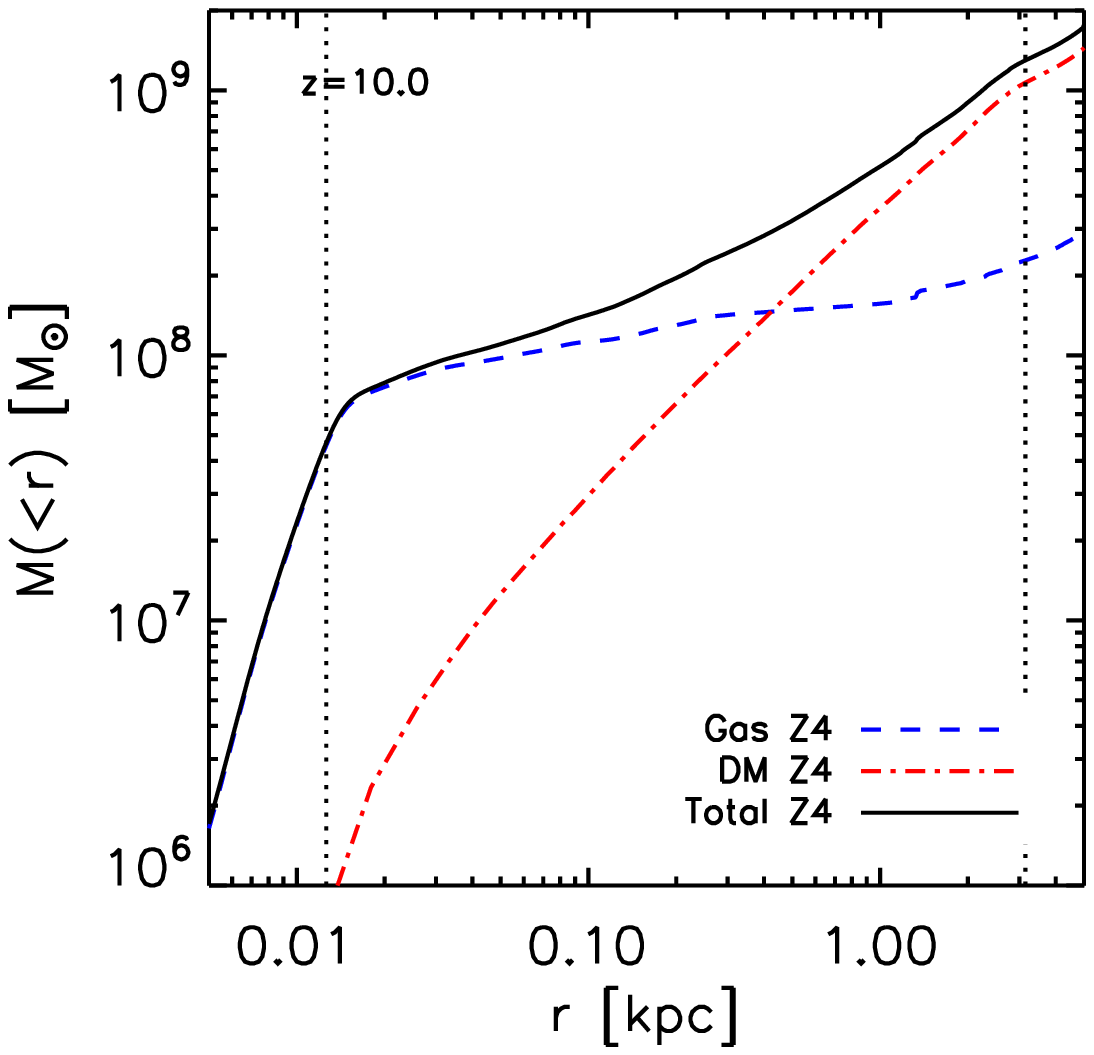}
\includegraphics[trim = 20mm 0mm 25mm 0mm, width = 0.32\textwidth]{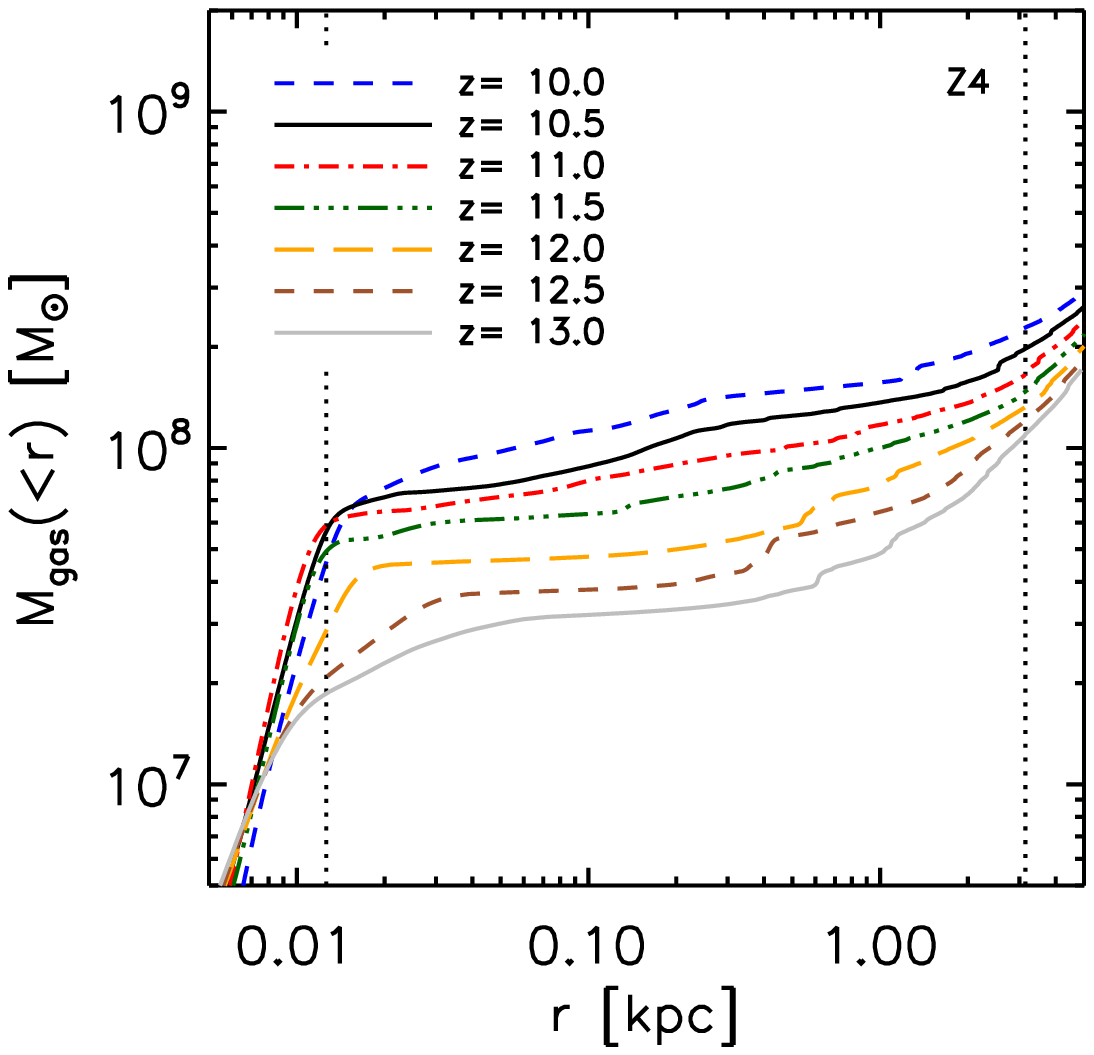}
\caption{Mass and density profiles centered on the most bound halo
particle in simulation \textit{Z4}. In all panels, the
vertical lines on the left mark the gravitational softening radius $\epsilon$ and the
vertical lines on the right mark the virial radius at $z = 10$, the final
simulation redshift. {\it Left panel}: Dark matter (red
dash-dotted curve) and gas density profiles at $z = 10$. The gas
densities were computed both by dividing the gas mass summed in shells
by the shell volumes (blue dashed curve; this was also done for the
dark matter density profiles) and by averaging the SPH particle
densities inside shells (green dotted curve). For comparison, the corresponding SPH
particle density profile from simulation {\it Z4NOMOL} is also shown
(black dotted curve). The dark matter profile is approximately singular-isothermal 
for all distances larger than the softening radius. The two
steepenings in the gas density profiles at $r \approx 0.07 \kpc$ and
$r \approx 0.3 \kpc$ are due to the presence of two nested
disks. {\it Middle panel}: Enclosed gas, dark matter and total masses
at $z = 10$. {\it Right panel}: Evolution of the enclosed
gas mass. There is a rapid inflow of mass into the central unresolved
region $r \lesssim \epsilon$ at $11.5 \lesssim z \lesssim 12.5$.
\label{Fig:Profile:Density}}
\end{figure*}

\label{Sec:Results}
In this section we describe the outcome of our simulations. We start 
in Section~\ref{Sec:Growth} by briefly presenting the growth
histories of the simulated halos.  We focus our subsequent discussion
on the halo properties at $z = 10$. In Section~\ref{Sec:Structure} we
discuss the structure of the halos, in Section~\ref{Sec:Dynamics}
we describe the dynamics of the gas inside them, and in
Section~\ref{Sec:Disks} we investigate the
emergence of nested gas disks at the halo centers. Throughout we will discuss differences and
similarities between simulation \textit{Z4} and simulation \textit{Z4NOMOL} 
in which molecular hydrogen formation is suppressed. 
\subsection{Growth}
\label{Sec:Growth}
We use FOF to locate the simulated halos 
at the final simulation redshift $z = 10$, and then compute
the halo properties as follows. Given a FOF halo, we use {\sc subfind} to identify its most bound
particle and let it mark the halo center. We then
compute the virial radius of the halo by determining the radius of the
spherical volume centered on the most bound particle within which the
average matter density is equal to $200$ times the critical density at
$z = 10$.  The total mass inside this volume defines the halo virial
mass.
\par
We find that $r_{\rm vir} \approx 3.1 \kpc$ and $M_{\rm vir} \approx
1.3\times 10^9\Msun$, independent of the inclusion of molecular
cooling. For the adopted cosmological parameters, this virial mass
corresponds to $\approx 3\sigma$ fluctuations in the linear theory
density field (e.g., \citealp{Barkana:2001}).  The circular velocity
$v_{\rm c} \equiv (G M_{\rm vir} / r_{\rm vir})^{1/2}$ and virial
temperature $T_{\rm vir} \equiv \mu m_{\rm H} v_{\rm vir}^2/(3k_{\rm
B})$ implied by the virial mass and the virial radius are $v_{\rm c}
\approx 40 \kms$ and $T_{\rm vir} \approx 42000 \K$, where we have
assumed $\mu = 0.6$ appropriate for ionized gas with primordial composition.
\par
After having located the halo at $z = 10$, we trace its history
to higher redshifts using the FOF halo finder together with {\sc
subfind}, as follows. Knowing the FOF halo, the descendant, at
redshift $z_i$ corresponding to simulation snapshot $i$, we locate the
FOF halo at redshift $z_{i-1} > z_i$ corresponding to snapshot $i-1$
that shares, among all FOF halos present at $z_{i-1}$, the most mass
with the descendant. We then use {\sc subfind} to identify the most bound particle within
this FOF halo and obtain the properties of the halo at $z_{i-1}$ 
by computing its virial radius and mass in a sphere of average matter
density $200$ times the critical density at $z_{i-1}$ centered on the
most bound particle. 
\par
Figure~\ref{Fig:mah} shows the evolution of the halo mass $M_{\rm
vir}$ (top panel) and its corresponding rate of growth $dM_{\rm
vir}/dt$ (bottom panel) in simulation \textit{Z4}. The
mass growth is described well by an exponential fit $M_{\rm vir}(z) =
M_0 \exp(-\alpha z)$ (\citealp{Wechsler:2002}) with $M_0 = 2\times
10^{11} \Msun$ and $\alpha = 0.5$. The derived growth rates are
consistent with analytical estimates of the
rate of growth of the main progenitor (\citealp{Lacey:1993}). The dot-dashed curve
shows the growth rate as given in equation (A15) of \cite{Neistein:2006} with $q = 2.3$.
\par

\subsection{Structure}
\label{Sec:Structure}

\begin{figure*}
\begin{center}
\begin{minipage}[c]{0.85\linewidth}
\includegraphics[width = 0.29\textwidth]{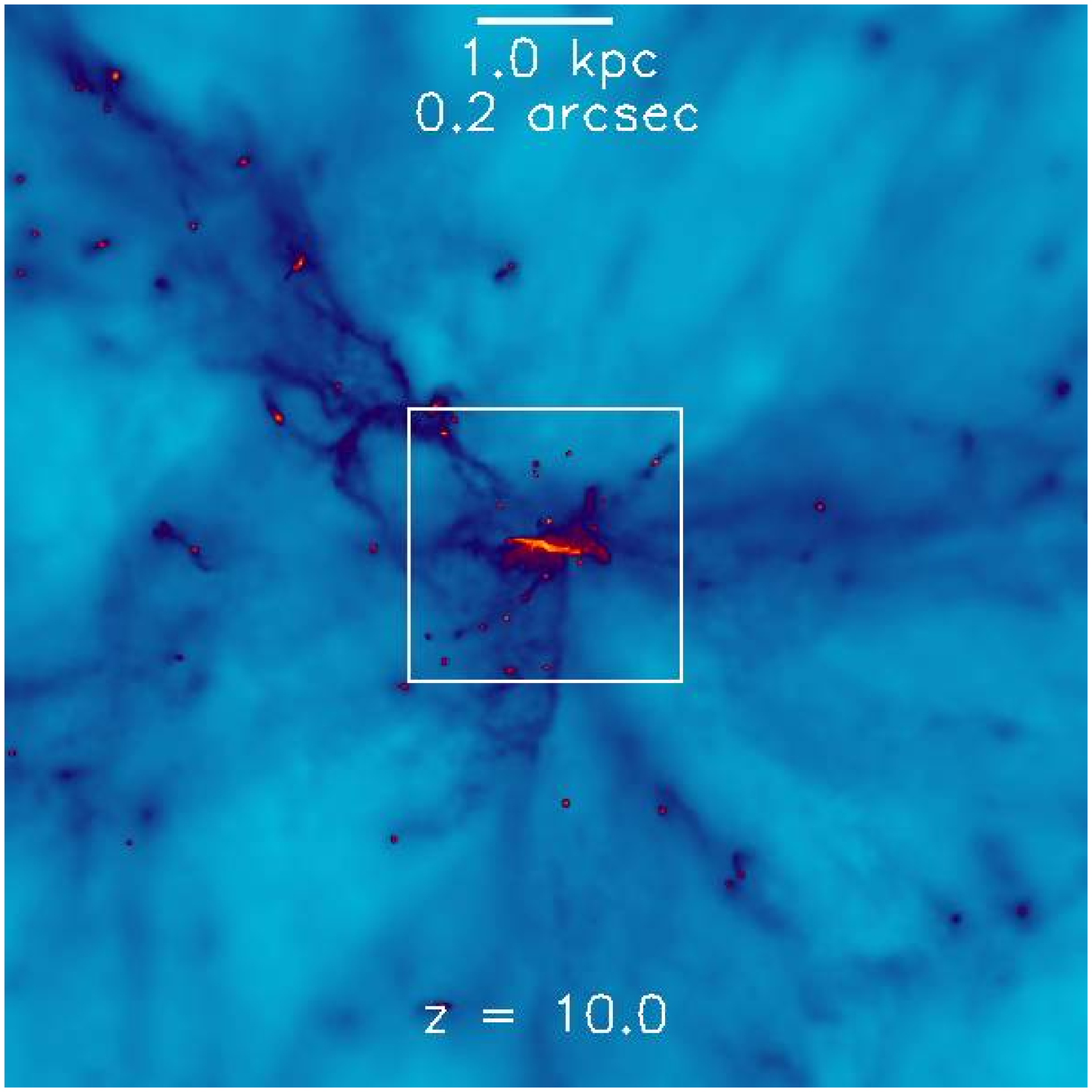}
\includegraphics[width = 0.29\textwidth]{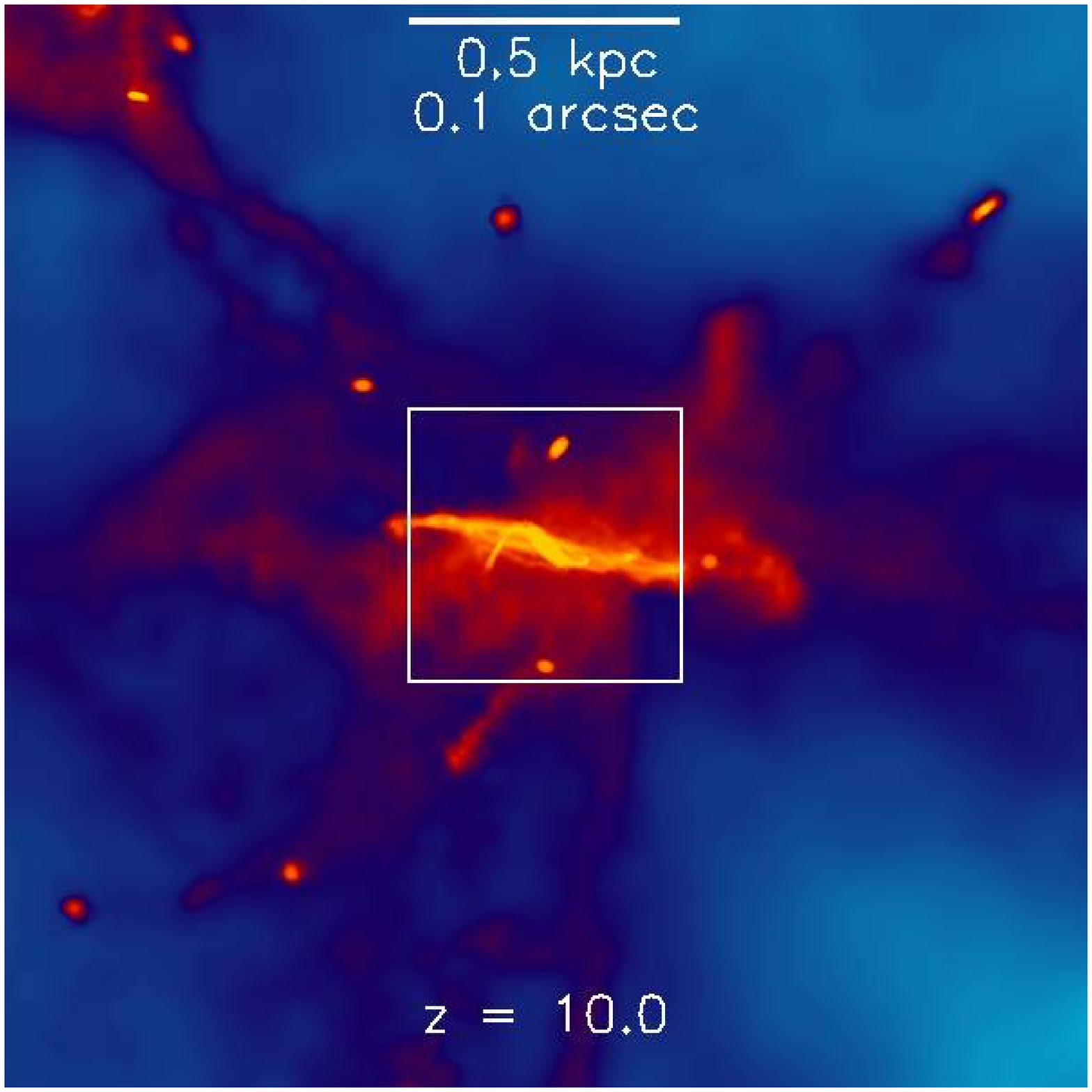}
\includegraphics[width = 0.29\textwidth]{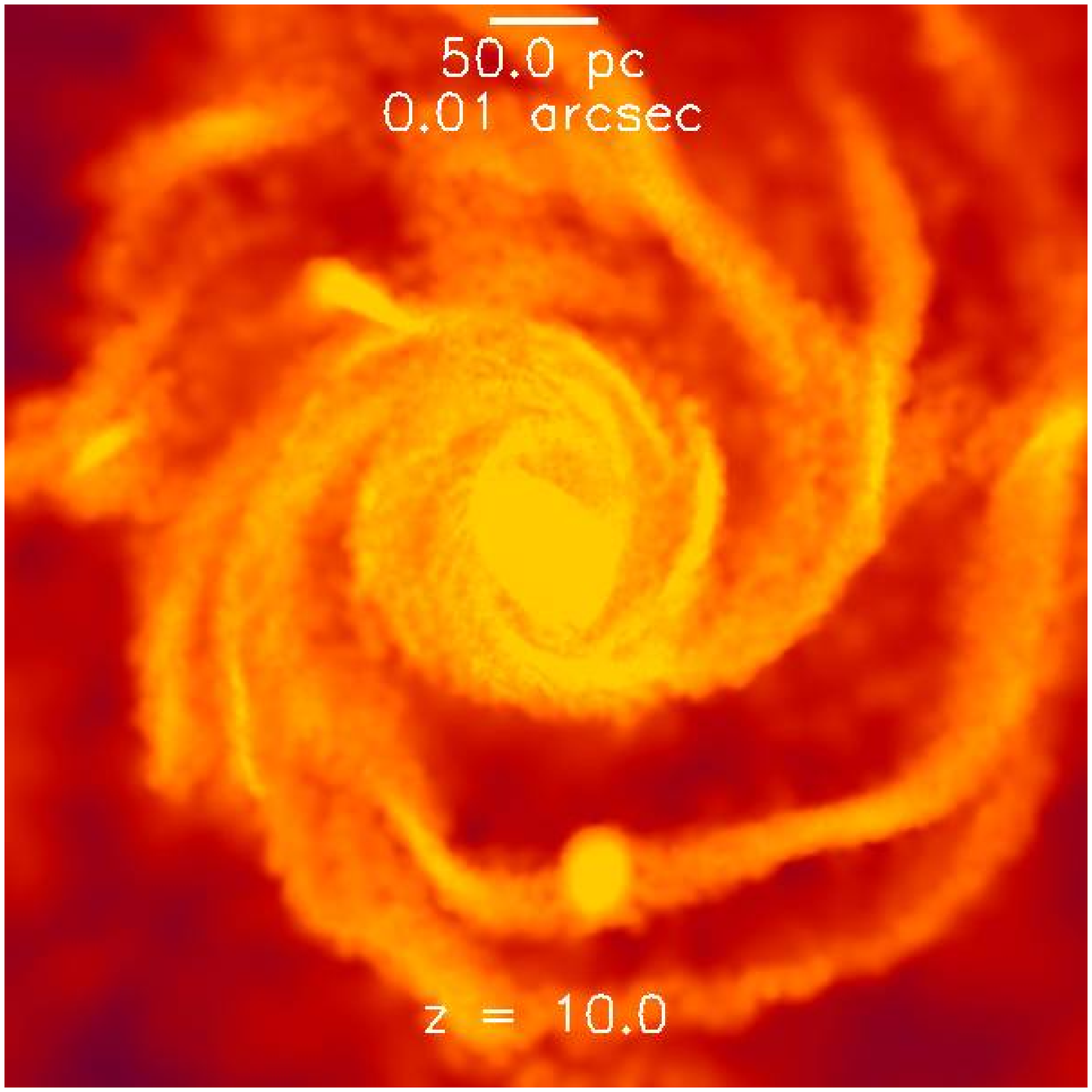}
\includegraphics[clip=true, trim = 320 0 50 0,height = 0.3\textwidth, width = 0.1\textwidth]{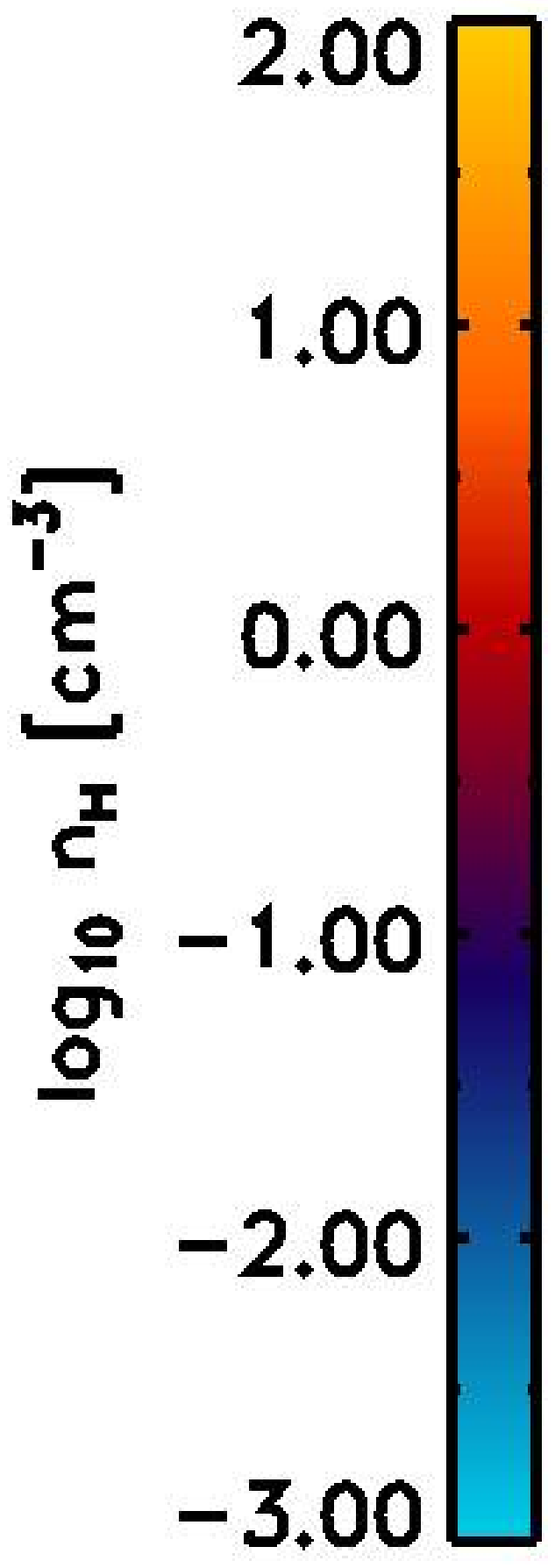}\\
\includegraphics[width = 0.29\textwidth]{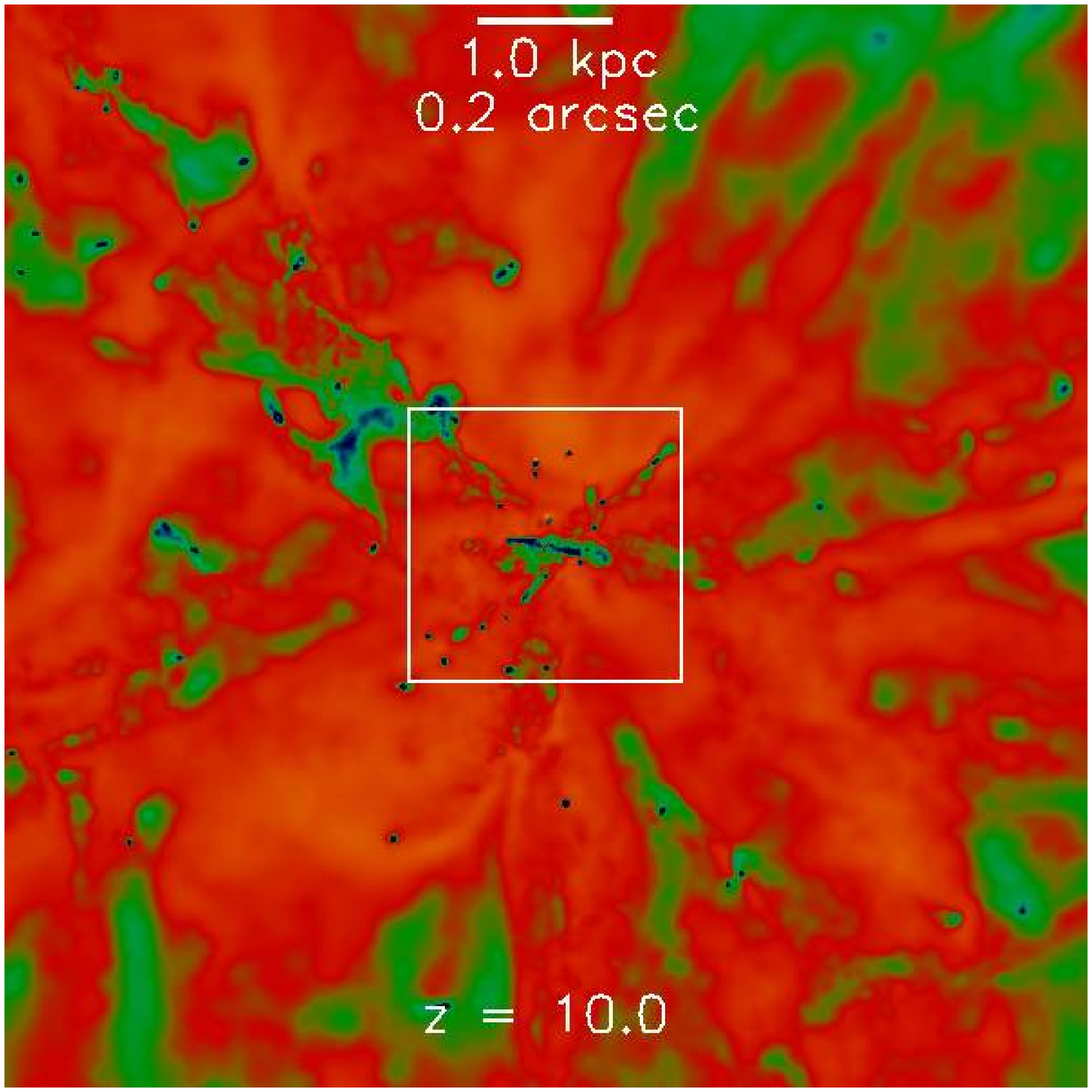}
\includegraphics[width = 0.29\textwidth]{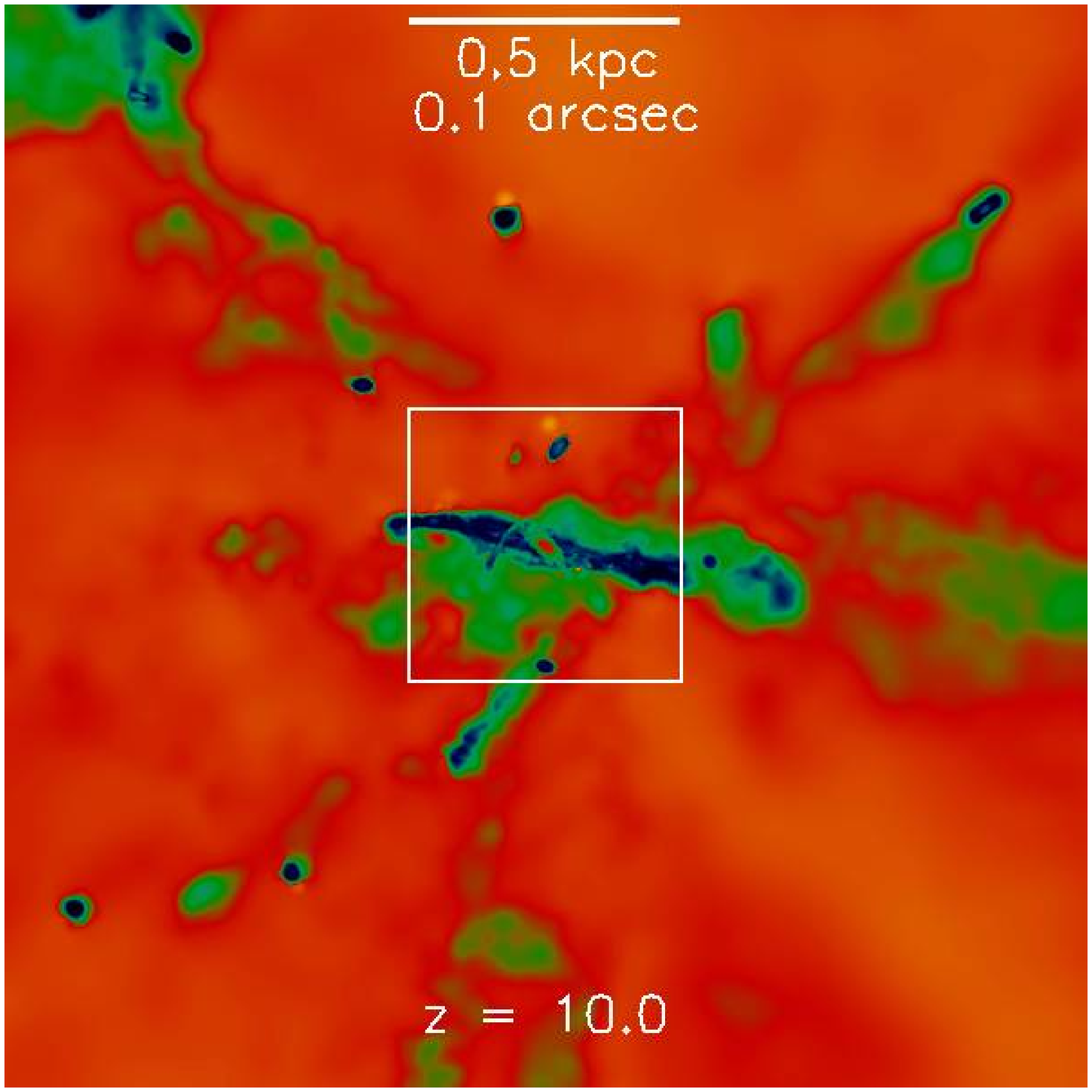}
\includegraphics[width = 0.29\textwidth]{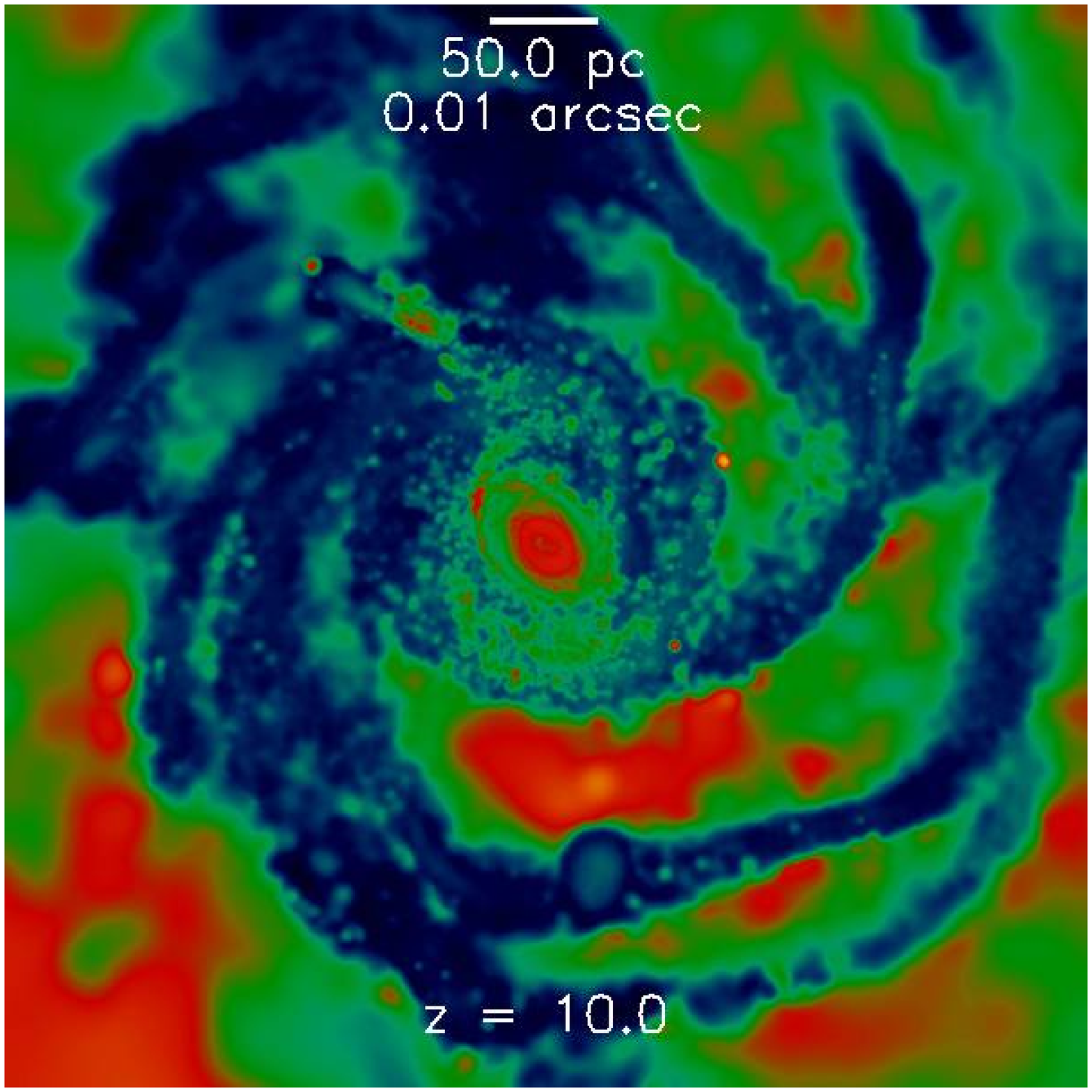}
\includegraphics[clip=true, trim = 320 0 50 0,height = 0.3\textwidth, width = 0.1\textwidth]{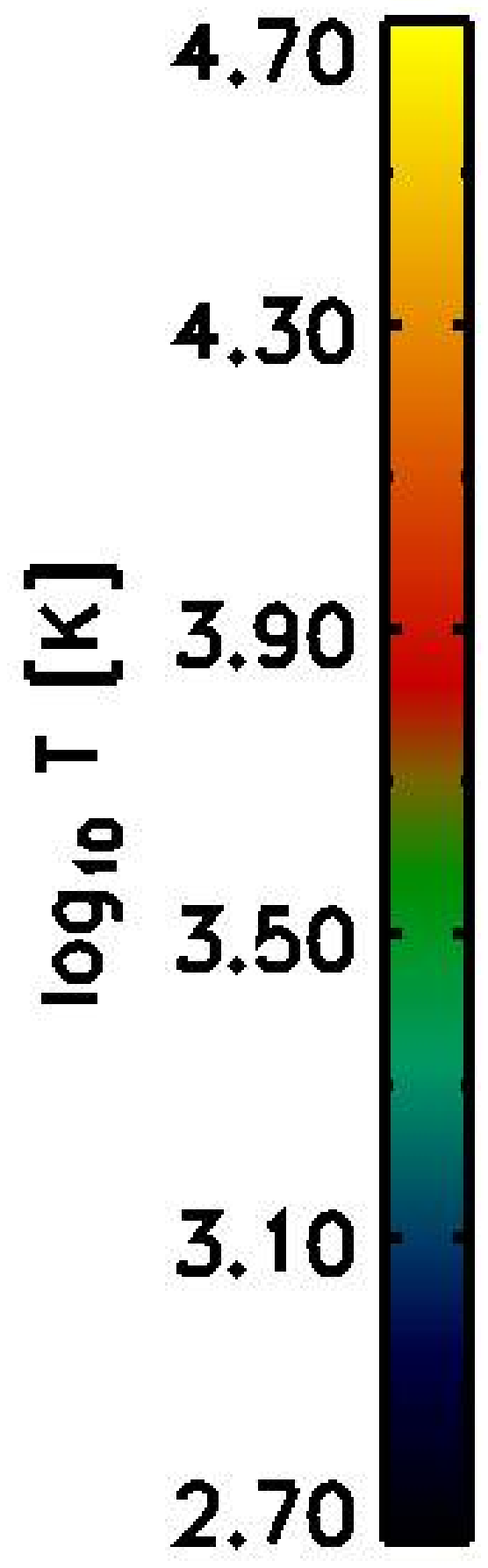}
\end{minipage}
\caption{Densities (top) and temperatures (bottom) at $z = 10$ in
simulation \textit{Z4} in cubical
slices centered on the most bound halo
particle. The left panels present edge-on views of the disks 
and encompass a volume slightly larger than the
virial region with radius $r_{\rm vir} \approx 3.1 \kpc$. The middle
panels are zooms into the cubical regions marked in the left
panels. The right panels are zooms into the cubical region marked
in the middle panels and are reoriented such that the outer disk is
seen face on. The temperature of the underresolved gas is artificially
elevated because of the use of a density-dependent temperature floor to prevent
artificial fragmentation. The spirals do not show signs of
fragmentation; the gas clump seen in the bottom of the face-on
view of the disk (right panels) is a gas-rich subhalo
in projection. \label{Fig:Images:Densities:Z4}}
\begin{minipage}[c]{0.85\linewidth}
\includegraphics[width = 0.29\textwidth]{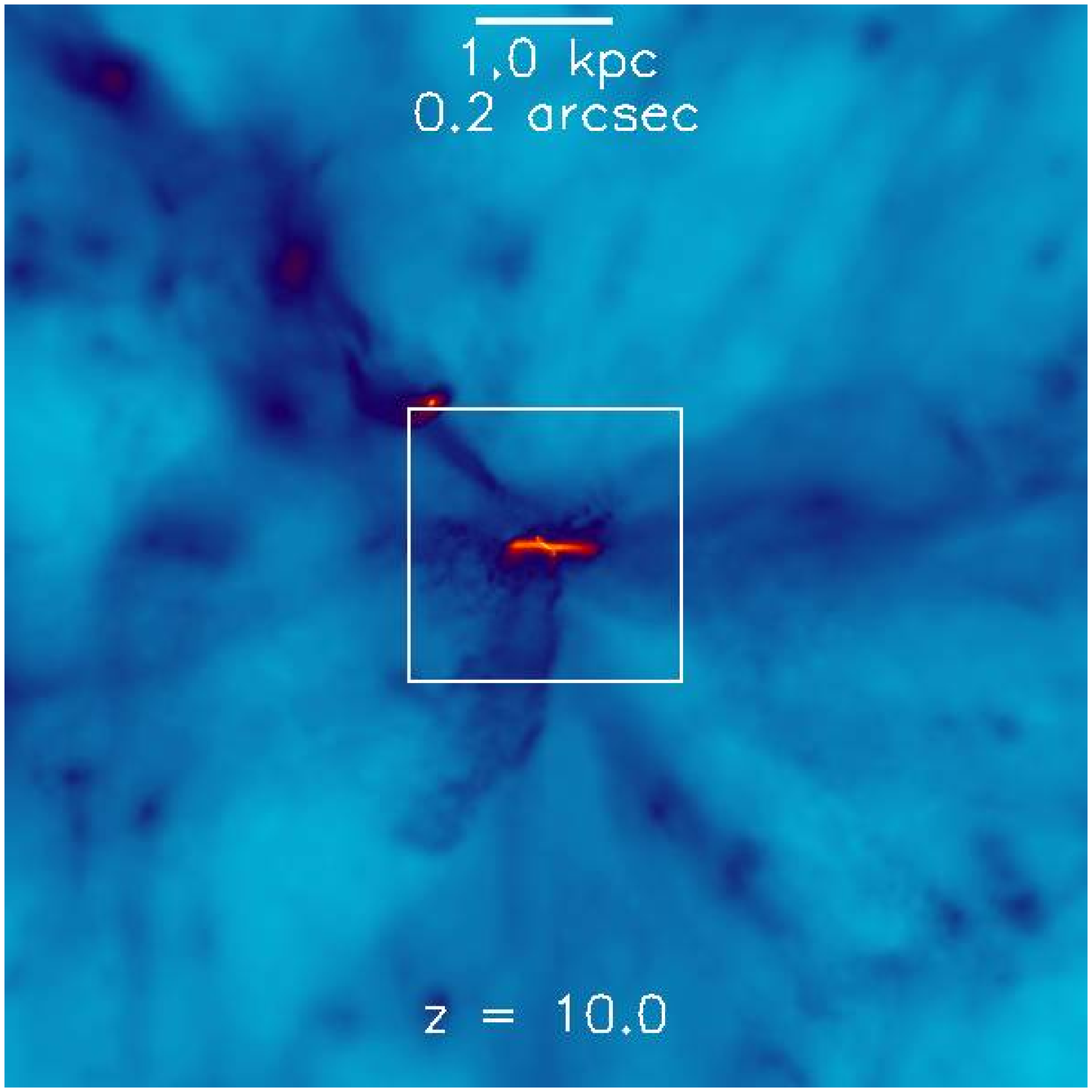}
\includegraphics[width = 0.29\textwidth]{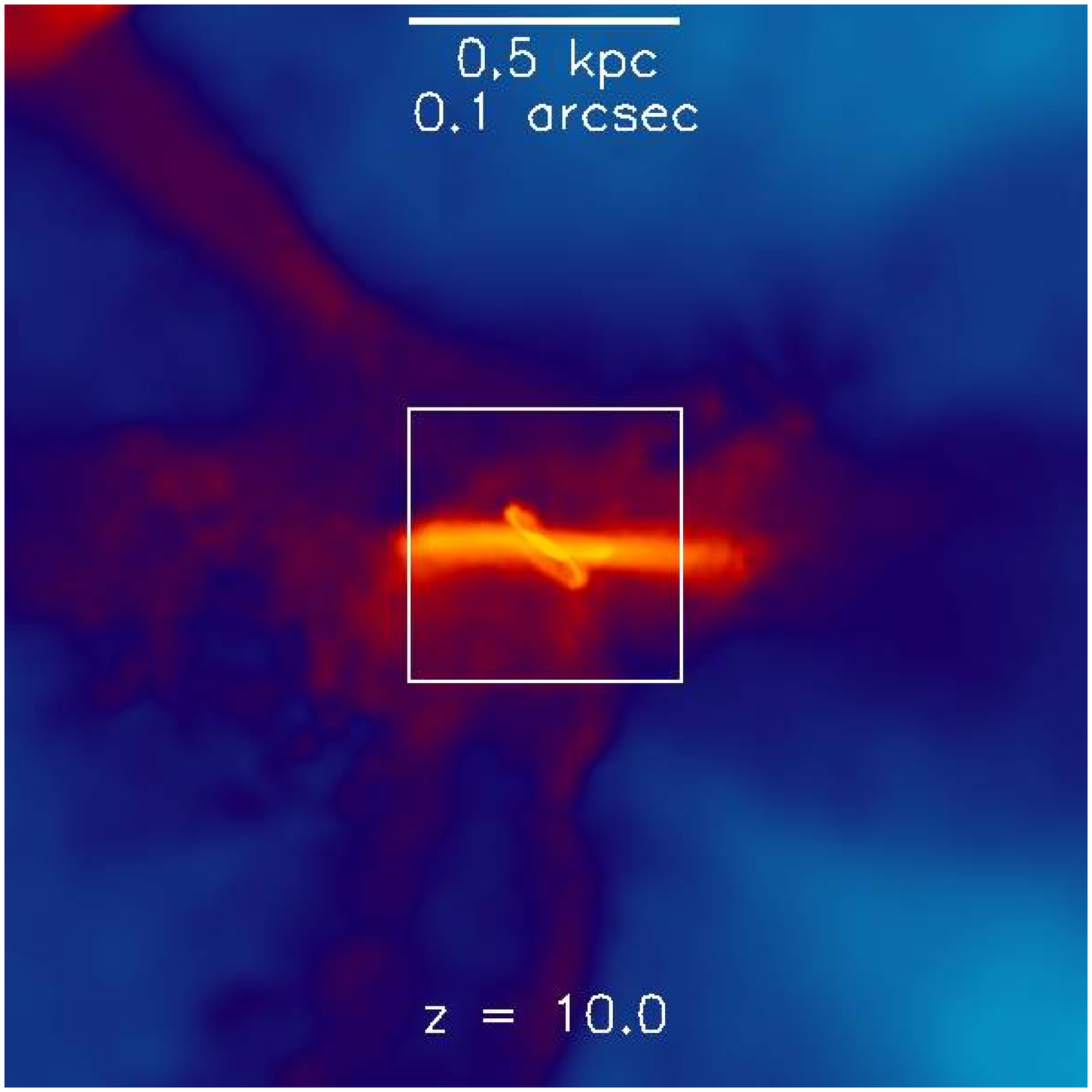}
\includegraphics[width = 0.29\textwidth]{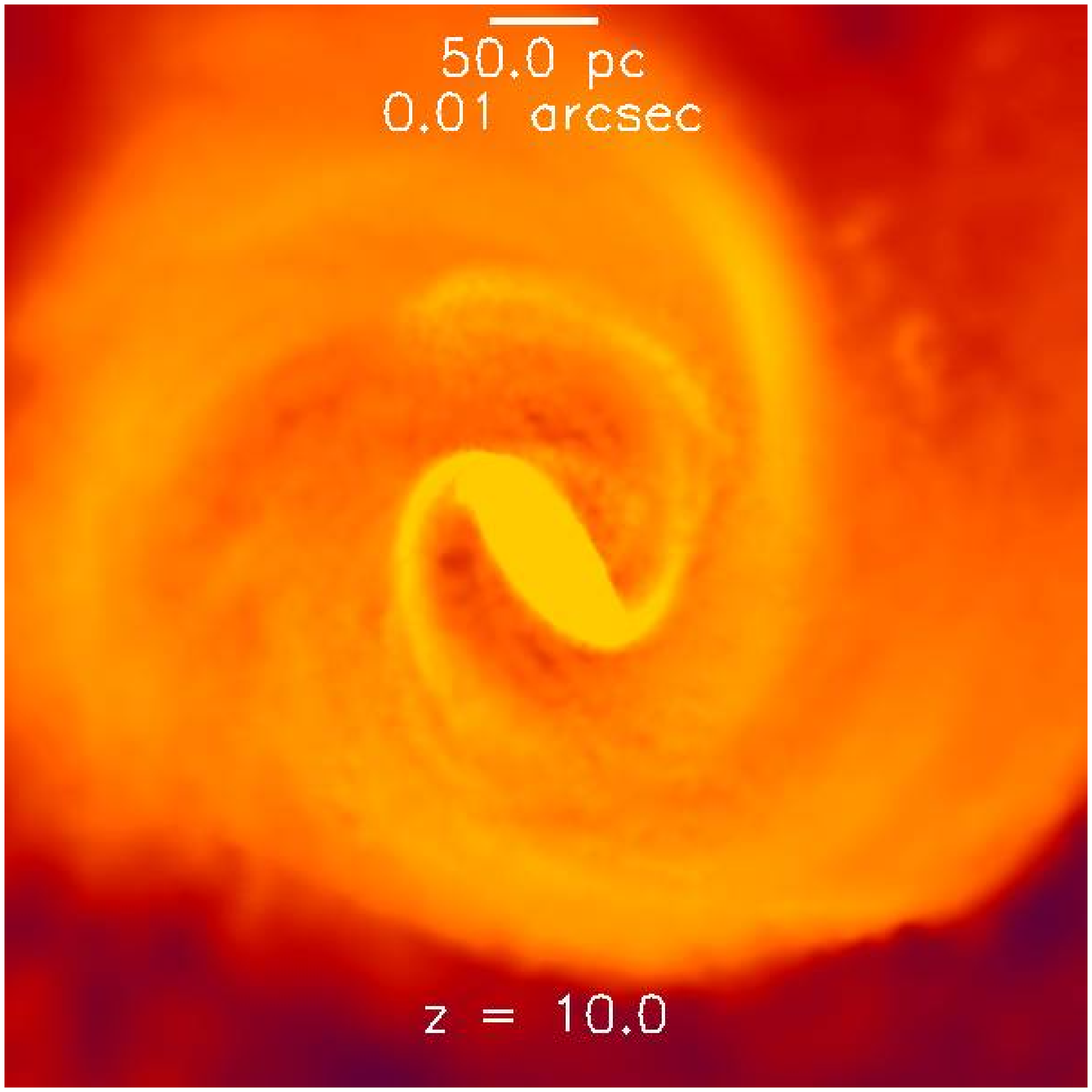}
\includegraphics[clip=true, trim = 320 0 50 0,height = 0.3\textwidth, width = 0.1\textwidth]{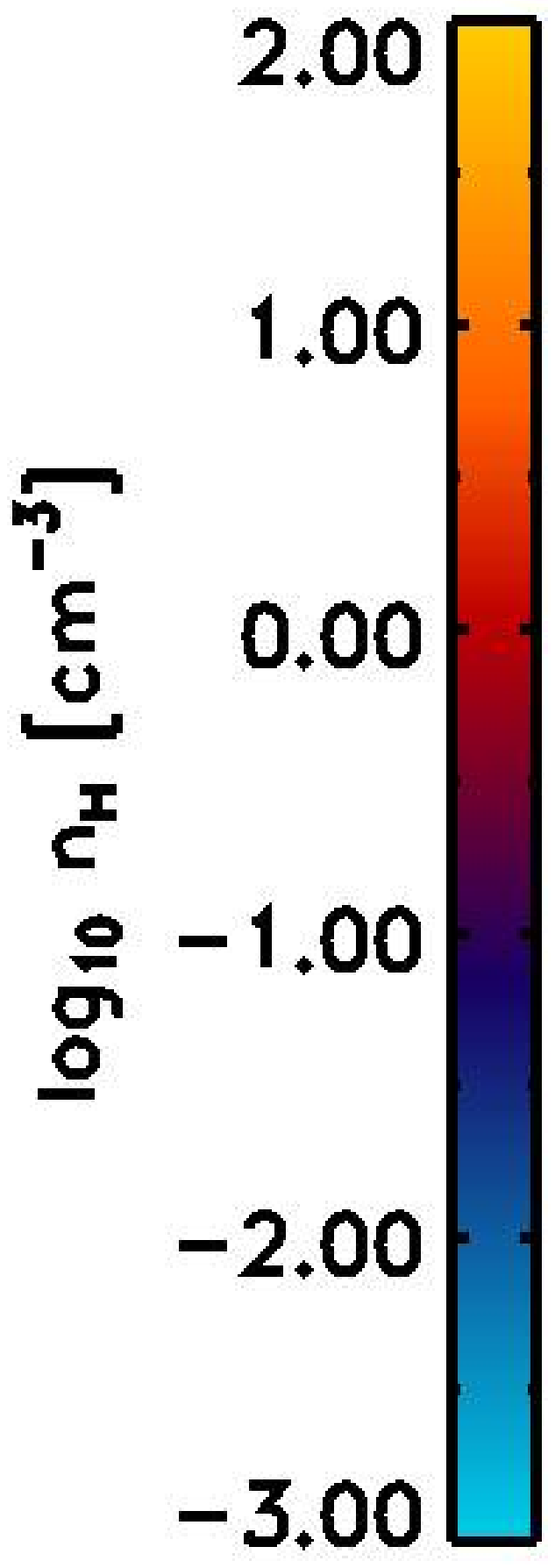}\\
\includegraphics[width = 0.29\textwidth]{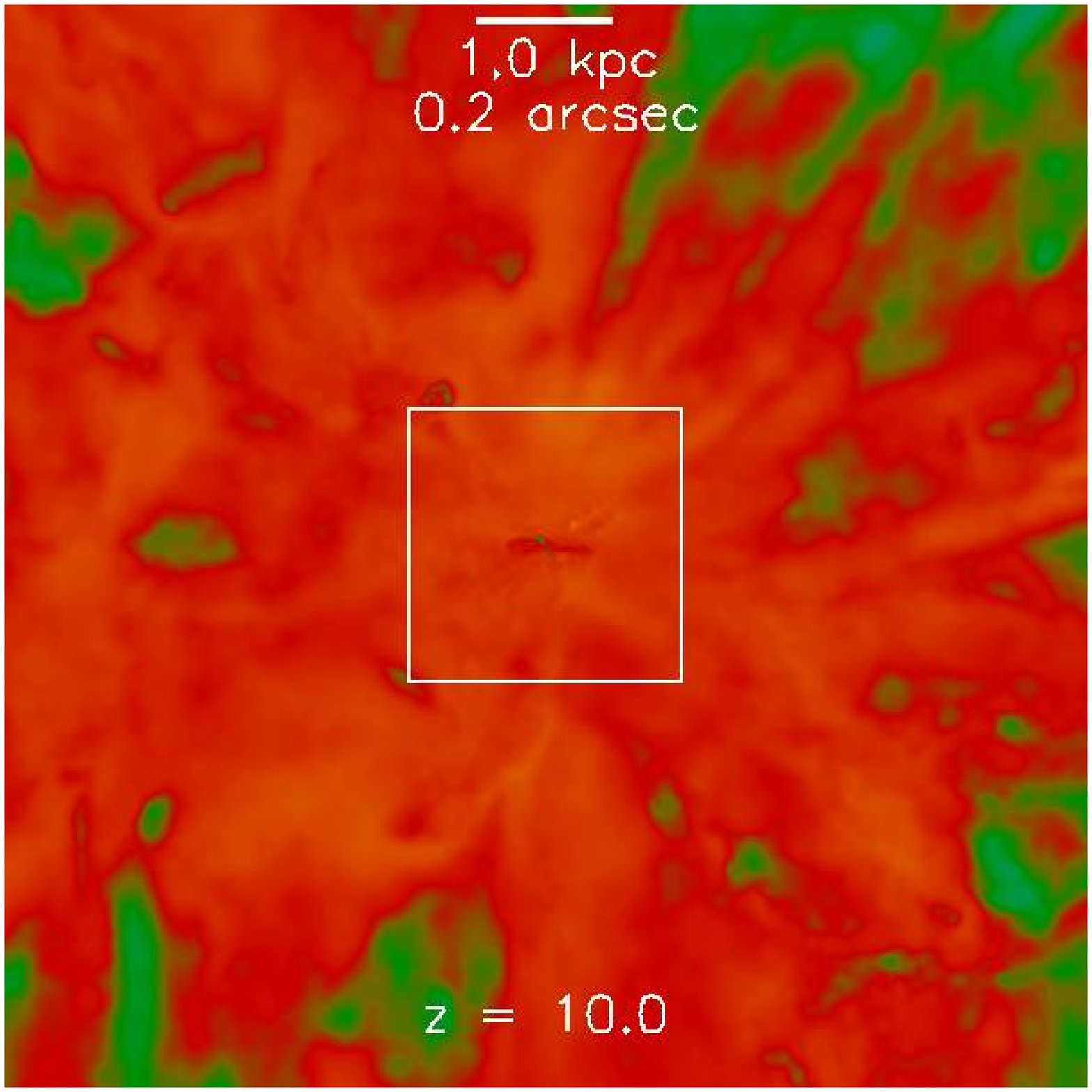}
\includegraphics[width = 0.29\textwidth]{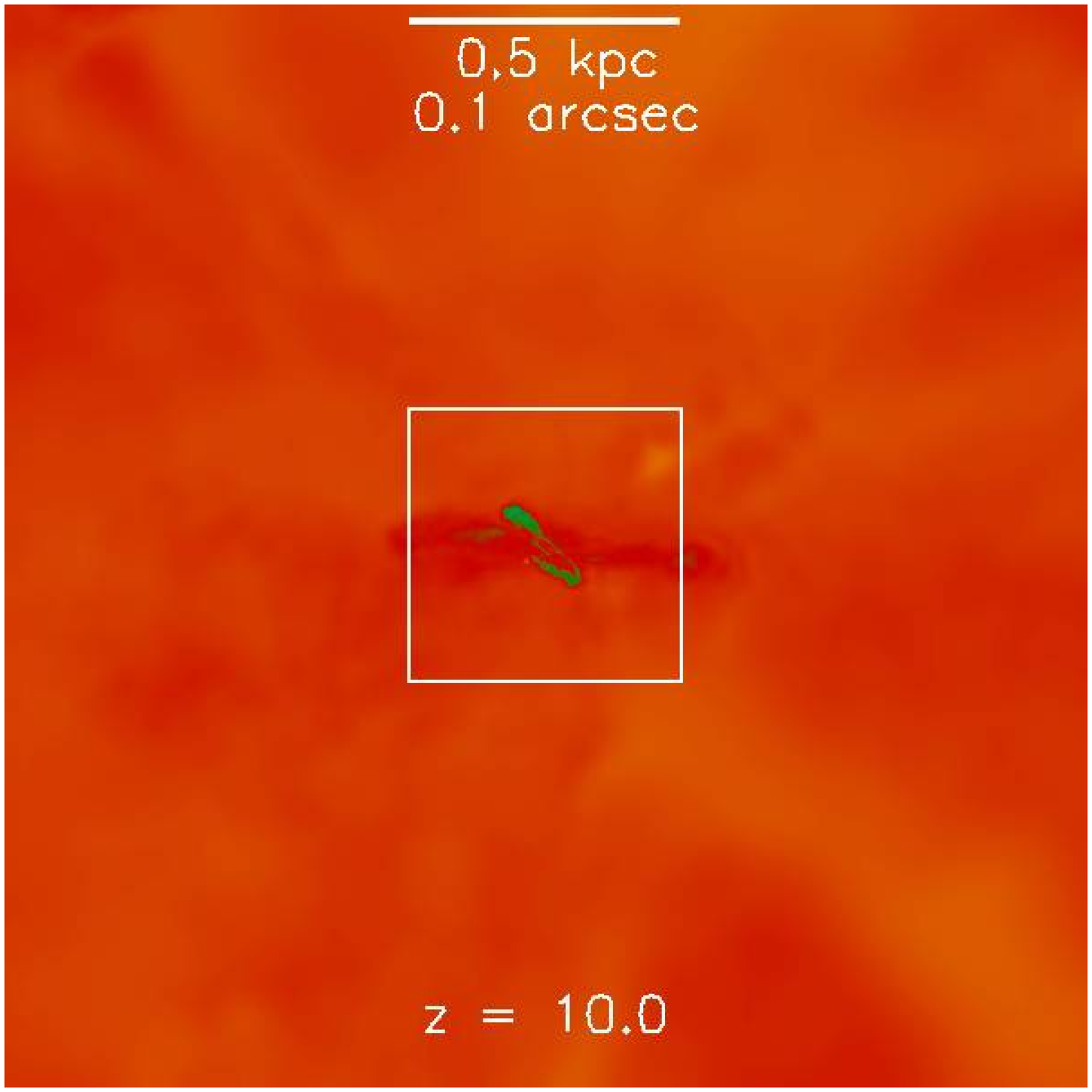}
\includegraphics[width = 0.29\textwidth]{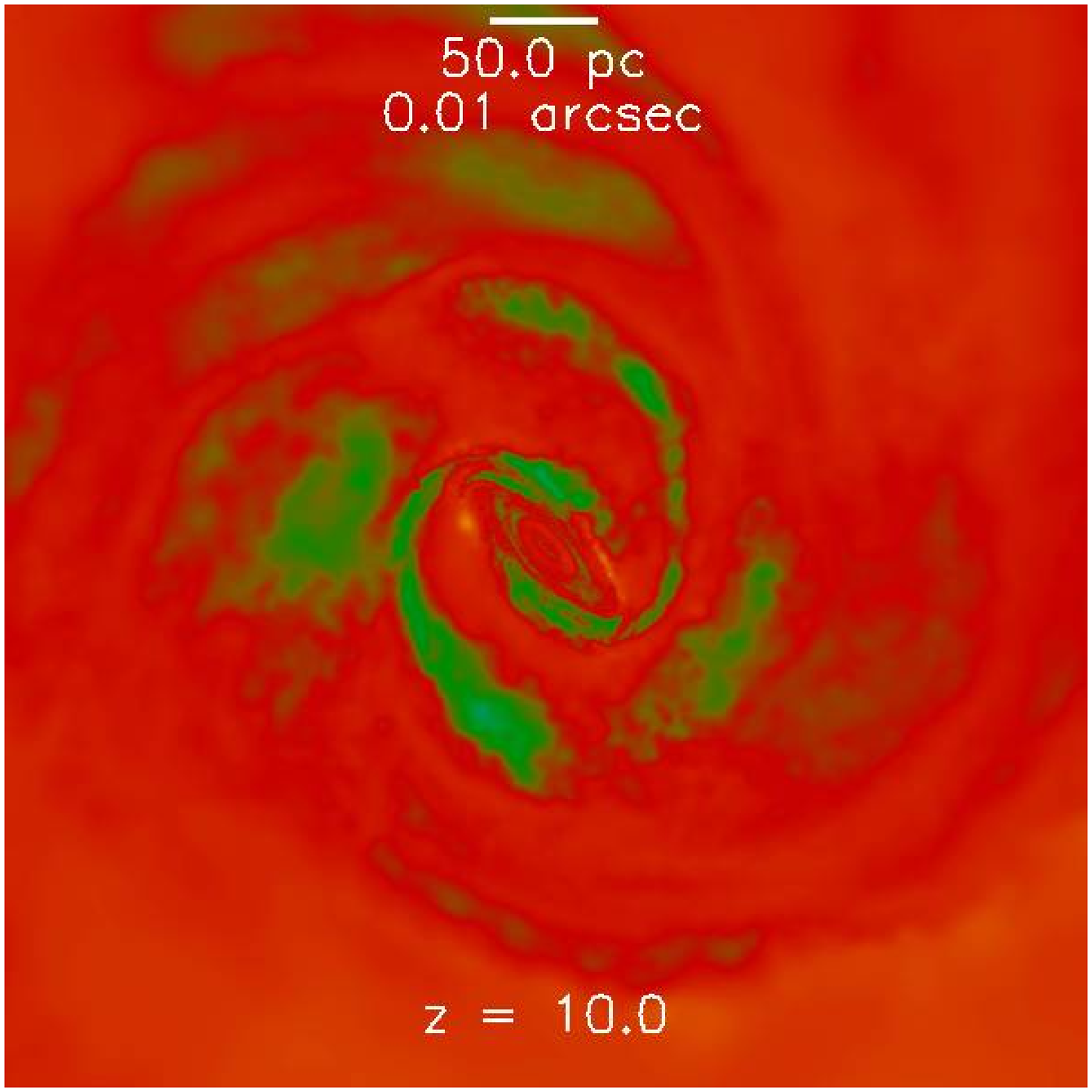}
\includegraphics[clip=true, trim = 320 0 50 0,height = 0.3\textwidth, width = 0.1\textwidth]{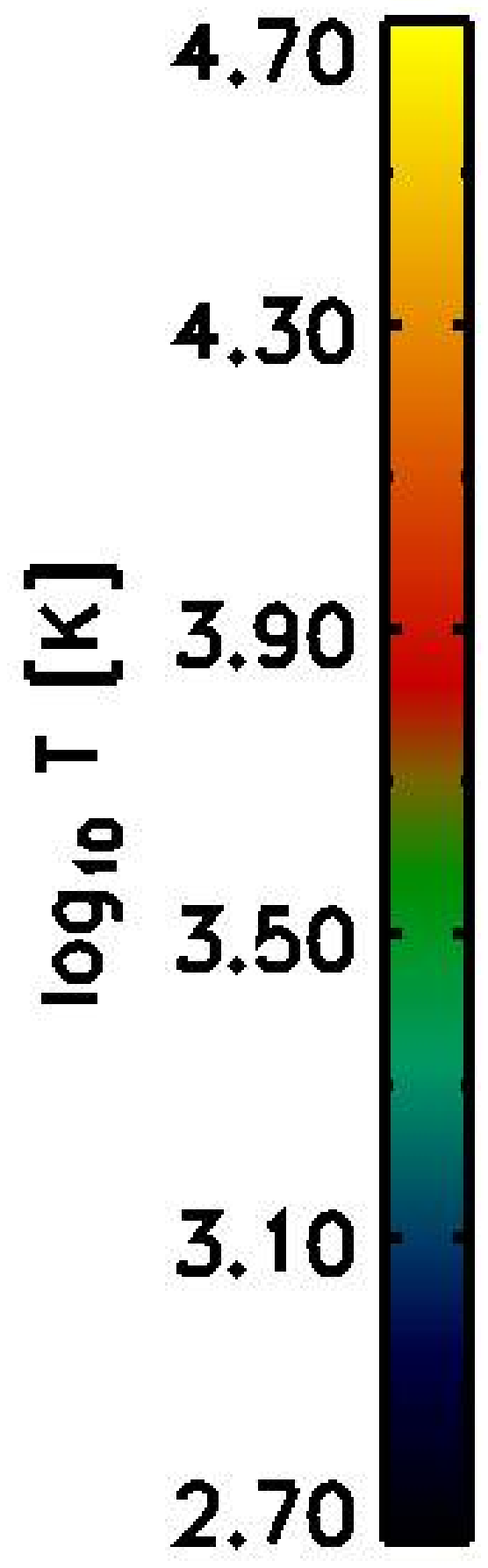}\\
\end{minipage}
\caption{Same as Figure~\ref{Fig:Images:Densities:Z4}, but for
  simulation {\it Z4NOMOL}. As in
  Figure~\ref{Fig:Images:Densities:Z4}, the left-hand and middle
  panels present edge-on views of the galaxy, and the right-hand
  panels show the galaxy face-on. Because of the lack of efficient
  low-temperature coolants and in contrast to simulation {\it Z4}
  (Figure~\ref{Fig:Images:Densities:Z4}), the gas inside the filaments
  is at about the same temperatures as the diffuse gas in between the
  filaments. The temperature of the underresolved gas core is
  artificially elevated because of the use of a density-dependent
  temperature floor to prevent artificial
  fragmentation. \label{Fig:Images:Densities:Z4NOMOL}}
\end{center}
\end{figure*}

\begin{figure}
  \includegraphics[width = 0.35\textwidth]{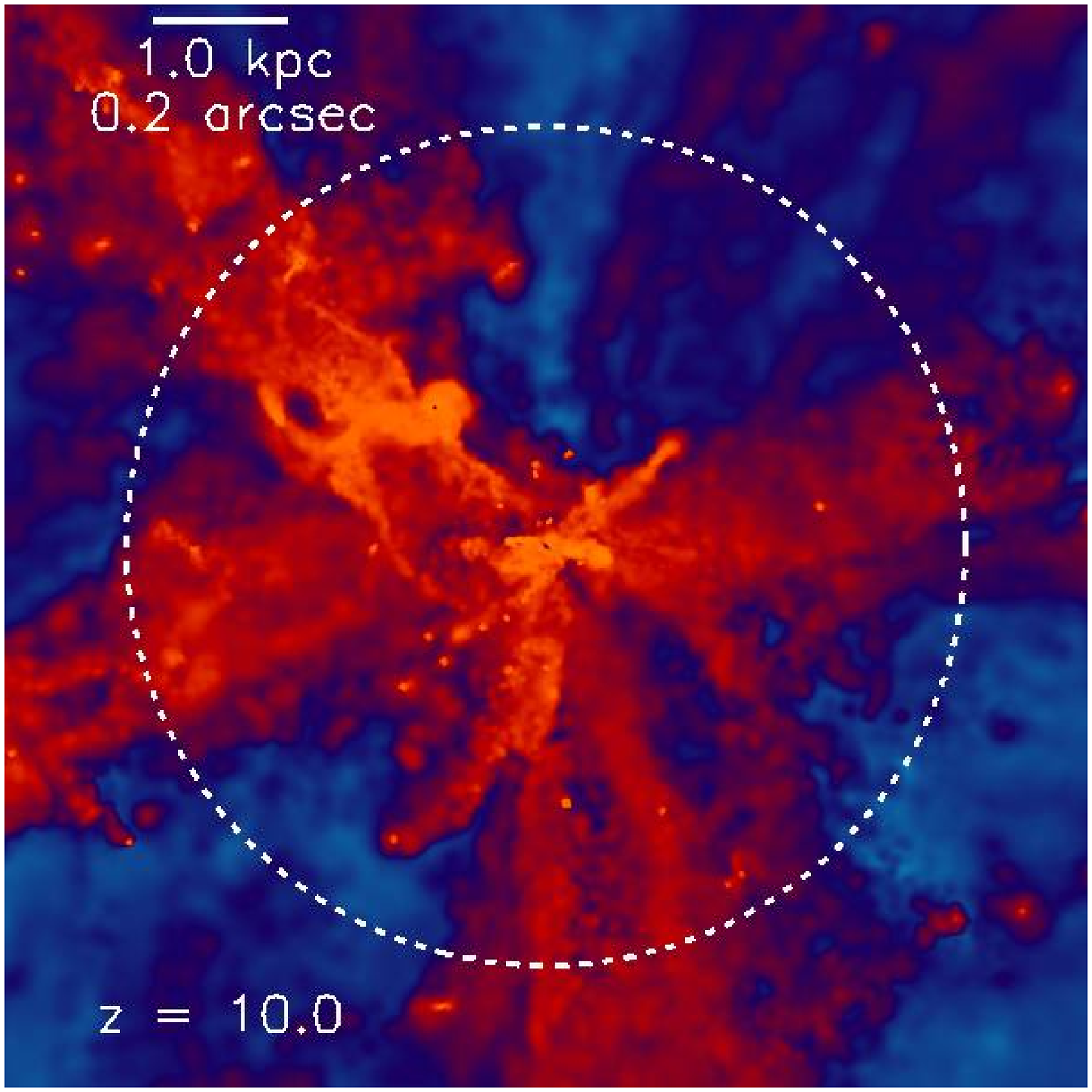}
\includegraphics[clip=true, trim = 320 0 50 0,height = 0.35\textwidth, width = 0.1\textwidth]{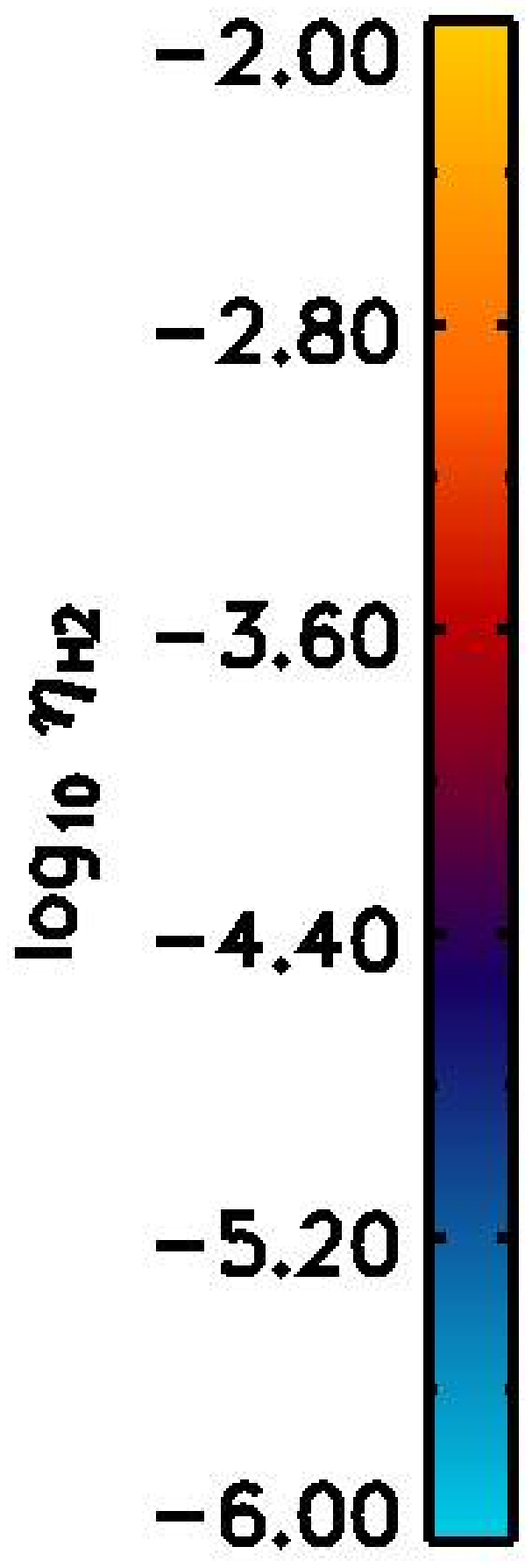} 
\caption{Molecular hydrogen fraction $\eta_{\rm H_2} = n_{\rm H_2} /
n_{\rm H}$ at $z = 10$ in simulation \textit{Z4}. The image shows
the same cubical region as the left panels in
Figure~\ref{Fig:Images:Densities:Z4}. The dashed circle marks the
virial radius $r_{\rm vir} \approx 3.1 \kpc$. The gas inside filaments
and subhalos has large molecular fractions $\eta_{\rm H_2} \sim 10^{-3.5} - 10^{-3}$ which
enables efficient low-temperature radiative cooling. \label{Fig:Images:H2}}
\end{figure}

\par
The left panel of Figure~\ref{Fig:Profile:Density} shows 
density profiles spherically averaged around the most bound halo
particle obtained from simulation \textit{Z4} at $z = 10$. The dark matter density profile (red dash-dotted
curve; obtained by summing particle masses inside spherical shells and
dividing by the shell volume) follows an isothermal shape $\rho
\propto r^{-2}$ for radii $\epsilon \lesssim r \lesssim r_{\rm
vir}$. The gas density profile does not follow the shape of the dark matter 
profile but shows significant small-scale structure. We have
computed the gas density profile both by averaging the SPH particle
densities inside spherical shells (green dotted curve) and by summing
particle masses inside spherical shells and dividing by the shell
volume (blue dashed curve). The two methods for computing the density
profile yield different results because the spatial distribution of the gas mass is highly
non-uniform, as will be discussed below. We also show the gas density profile in simulation
{\it Z4NOMOL} (black dotted curve). The dark matter density profile obtained in
simulation \textit{Z4NOMOL} is nearly identical to that from
simulation \textit{Z4} and hence is not shown.
\par
The middle panel of Figure~\ref{Fig:Profile:Density} shows the
enclosed mass as a function of distance $r$ from the halo center for
simulation \textit{Z4}. The cumulative mass of dark matter dominates
the significantly more centrally concentrated cumulative mass of gas
for radii $r \gtrsim 0.4 \kpc$. The build-up of the final gas mass
profile shown in the middle panel is illustrated in the right panel of
Figure~\ref{Fig:Profile:Density}. There is a large rapid increase in
the mass of the central region ($r \lesssim \epsilon$) around redshift
$z \approx 12$. In little more than $40 \Myr$ ($11.5 \lesssim z
\lesssim 12.5$), the central gas mass grows from $\sim 2 \times 10^7
\Msun$ to $\sim 5 \times 10^7 \Msun$. In Section~\ref{Sec:JWST} we
will assume that this rapid collapse of large gas masses triggers a
massive burst of star formation in the central core. The cumulative
mass profiles from simulation \textit{Z4NOMOL} exhibit a nearly
identical behavior and again are not shown.
\par
Figures~\ref{Fig:Images:Densities:Z4} and
\ref{Fig:Images:Densities:Z4NOMOL} give further impressions of the baryonic
structure of the simulated halos at the final simulation time, i.e., at
redshift $z = 10$. All quantities are shown for both simulation
$\textit{Z4}$ (Figure~\ref{Fig:Images:Densities:Z4}) and simulation
$\textit{Z4NOMOL}$ in which molecular hydrogen formation was suppressed
(Figure~\ref{Fig:Images:Densities:Z4NOMOL}).  The panels in the
left columns of Figures~\ref{Fig:Images:Densities:Z4} and
\ref{Fig:Images:Densities:Z4NOMOL} show the hydrogen number densities
and temperatures, mapped to a three-dimensional grid using standard
mass-conserving SPH interpolation and mass-weighted averaging along
the line of sight, within a cubical volume encompassing the virial
region. 
\par
The panels show that independent of the inclusion of
molecular cooling, gas is organized in four geometrically distinct
components: diffuse low-density ($n_{\rm H} \lesssim
10^{-2} \cmci$) gas, collimated streams, or filaments, of smooth dense
($n_{\rm H} \lesssim 10^{-1} \cmci$) gas that penetrate deep into the
virialized region, dense ($n_{\rm H} \gtrsim 10^{-1} \cmci$) gas-rich
clumps with mostly-spherical appearance and a central dense ($n_{\rm
H} \gtrsim 10^{1} \cmci$) gaseous flattened object seen edge-on. The
dense clumps inside the virial radius are associated with low-mass
($\lesssim 10^7 \Msun$) halos that entered the virial region before $z
= 10$. In the following we refer to these halos as subhalos.
\par
The panels in the middle columns of Figures~\ref{Fig:Images:Densities:Z4} 
and \ref{Fig:Images:Densities:Z4NOMOL} are
zooms into the cubical regions marked by the white solid rectangles in
the panels of the left columns. The panels in the right
columns are zooms into the cubical
region marked in the panels of the middle columns, but with their coordinate axes
rotated. The zooms resolve the central flattened object into two
nested disks whose orientations are tilted with respect to each
other. The disks cause the steepenings of the gas density profile at
$r \approx 0.07 \kpc$ and $r \approx 0.3 \kpc$ seen in
the left panel of Figure~\ref{Fig:Profile:Density}. We let these 
radii define the sizes of the disks. Note that the disks surround a 
central unresolved core of radius $r \lesssim \epsilon$. 
The disks will be discussed in more detail in Section~\ref{Sec:Disks}.
\par
The images of the gas temperatures reveal a qualitative difference
between simulation \textit{Z4} and simulation \textit{Z4NOMOL} in which molecular
hydrogen formation was suppressed. While in both simulations the tenuous gas in between the
filaments is at temperatures close to the virial temperature of the
halo, the temperatures of the gas inside filaments, subhalos and 
the disks are up to an order of magnitude lower
in \textit{Z4} than in \textit{Z4NOMOL}.  Figure~\ref{Fig:Images:H2} shows
the molecular hydrogen fraction $\eta_{\rm H_2} \equiv  n_{\rm H_2} / n_{\rm H}$ 
inside the same cubical region as shown in the left panels of 
Figure~\ref{Fig:Images:Densities:Z4}. The molecular fraction is
greatly increased up to $\eta_{\rm H_2} \sim 10^{-3}$ in gas with densities $n_{\rm H} \gtrsim 1 \cmci$ and
temperatures $T \lesssim 1000 \K$, which resides mostly in the
filaments and subhalos and in the central halo region.
\par
Without sufficient molecular hydrogen, the intra-halo gas cannot cool
efficiently below temperatures $T \sim 10^4 \K$ as atomic cooling is
exponentially suppressed because of the lack of thermal excitation of
bound electrons. Gas that is shock-heated to $T \gtrsim 10^4 \K$ upon entry in the halo 
then remains hot at $T \approx 10^4 \K$ until it is incorporated into the
dense disks where it cools to slightly lower temperatures. In the
presence of a sufficient amount of molecular hydrogen, on the other
hand, radiative cooling counters virial heating in the dense gas
inside filaments down to temperatures $T \lesssim 10^3 \K$. The accretion along
the filaments then occurs in a cold mode (\citealp{Wise:2007}; \citealp{Greif:2008};
see also \citealp{Birnboim:2003}; \citealp{Keres:2005}; \citealp{Brooks:2009}; \citealp{Freeke:2010} for cold accretion 
of gas inside more massive halos).
\par

\subsection{Dynamics}
\label{Sec:Dynamics}
\begin{figure*}
\includegraphics[trim = 30mm 0mm 30mm 0mm, width = 0.33\textwidth]{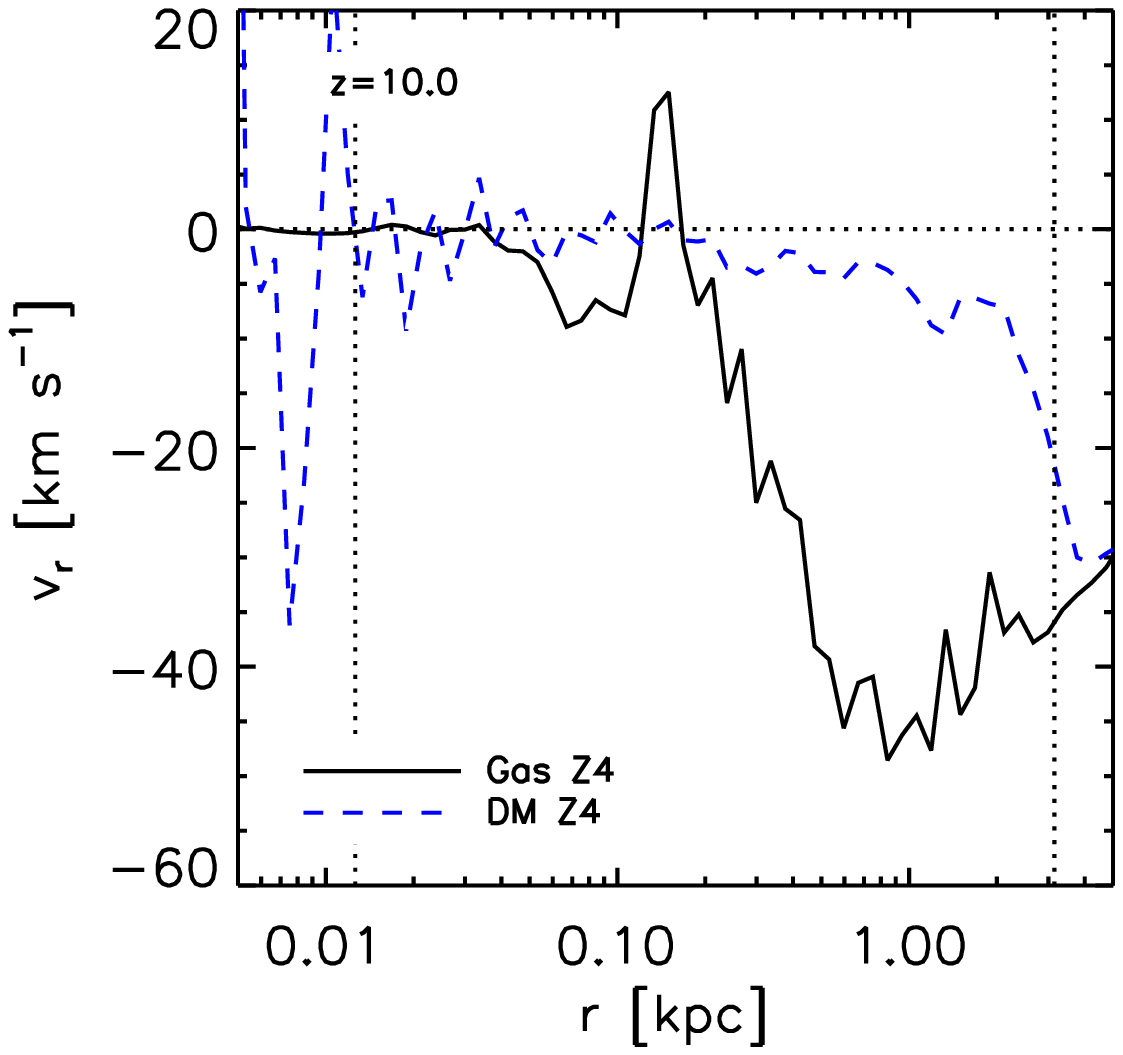}
\includegraphics[trim = 30mm 0mm 30mm 0mm, width = 0.33\textwidth]{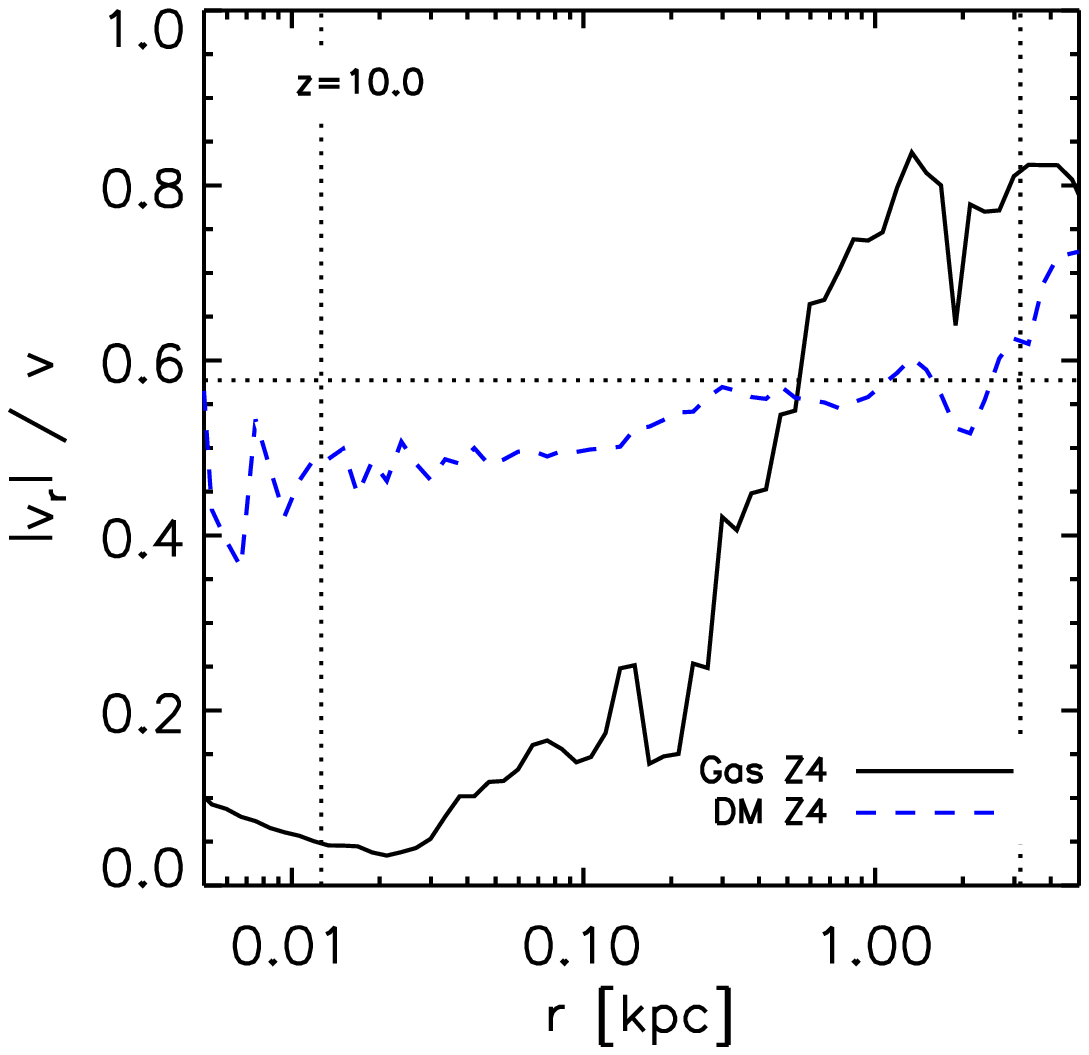}
\includegraphics[trim = 30mm 0mm 30mm 0mm, width = 0.33\textwidth]{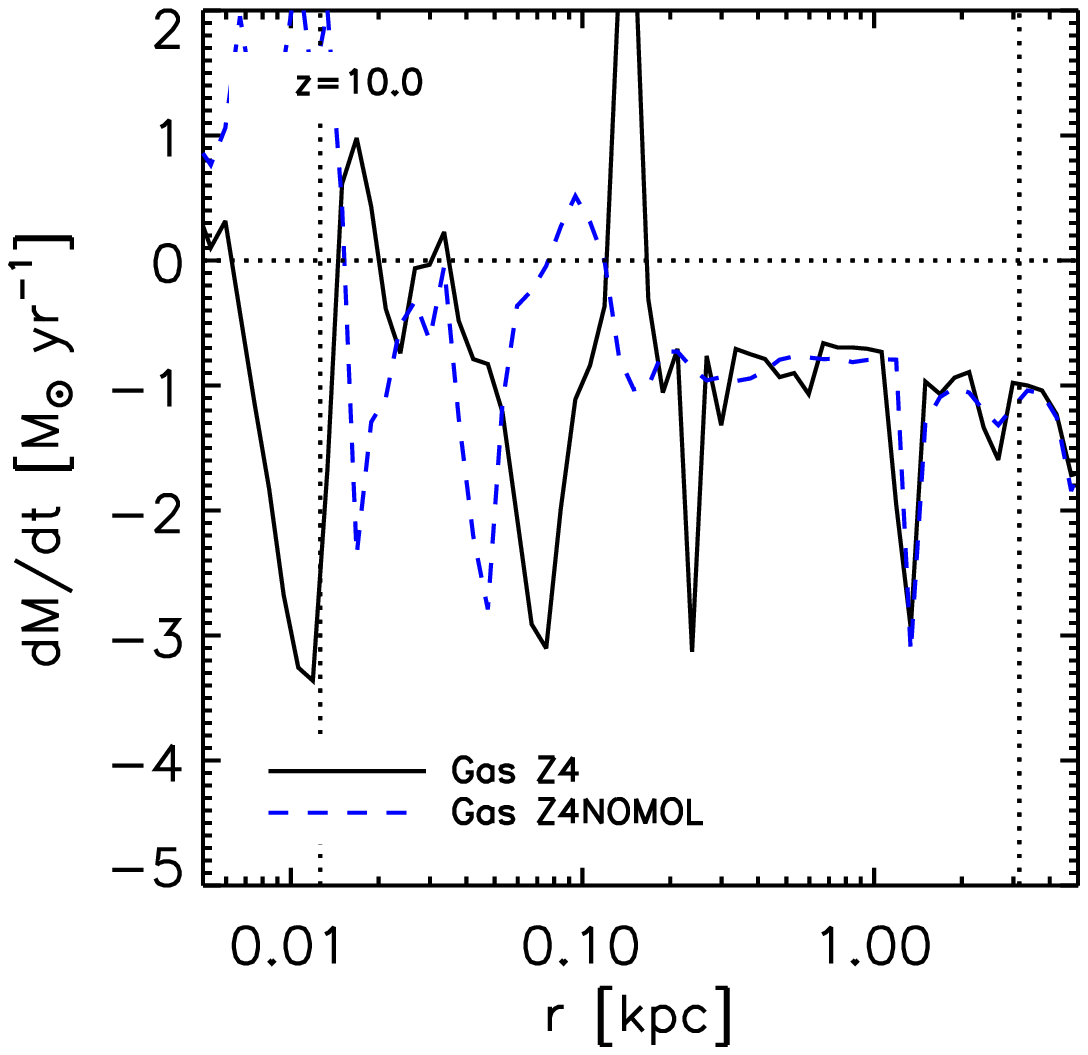}
\caption{Gas flow at $z = 10$ around the most bound halo particle in
simulation \textit{Z4}.  In all panels, the vertical line
on the left shows the gravitational softening radius $\epsilon$ and the 
vertical line on the right marks the virial radius $r_{\rm vir}$. {\it Left panel}: 
Spherically averaged radial particle
velocities for dark matter (blue dashed curve) and gas (black solid
curve). The horizontal line marks zero radial velocity. {\it Middle panel:}
Same as left panel, but with particle radial velocities divided
by the particle total velocities. The horizontal line marks the ratio of radial
and total velocity expected for an isotropic velocity distribution. While the
dark matter isotropizes after entering the virial region, the gas
keeps falling in along mostly radial orbits until it reaches the outer
disk at $r \lesssim 0.3 \kpc$. {\it Right panel}: Gas accretion
rates. For comparison, we also show the gas accretion rates in
simulation \textit{Z4MOMOL}. The spikes are associated with gas-rich
subhalos, a majority of which accretes along filaments. Note that the scale of the vertical axis is linear.
\label{Fig:Profiles:Velocities}}
\end{figure*}

\begin{figure*}
\begin{center}
\begin{minipage}[c]{0.9\linewidth}
\includegraphics[width = 0.29\textwidth]{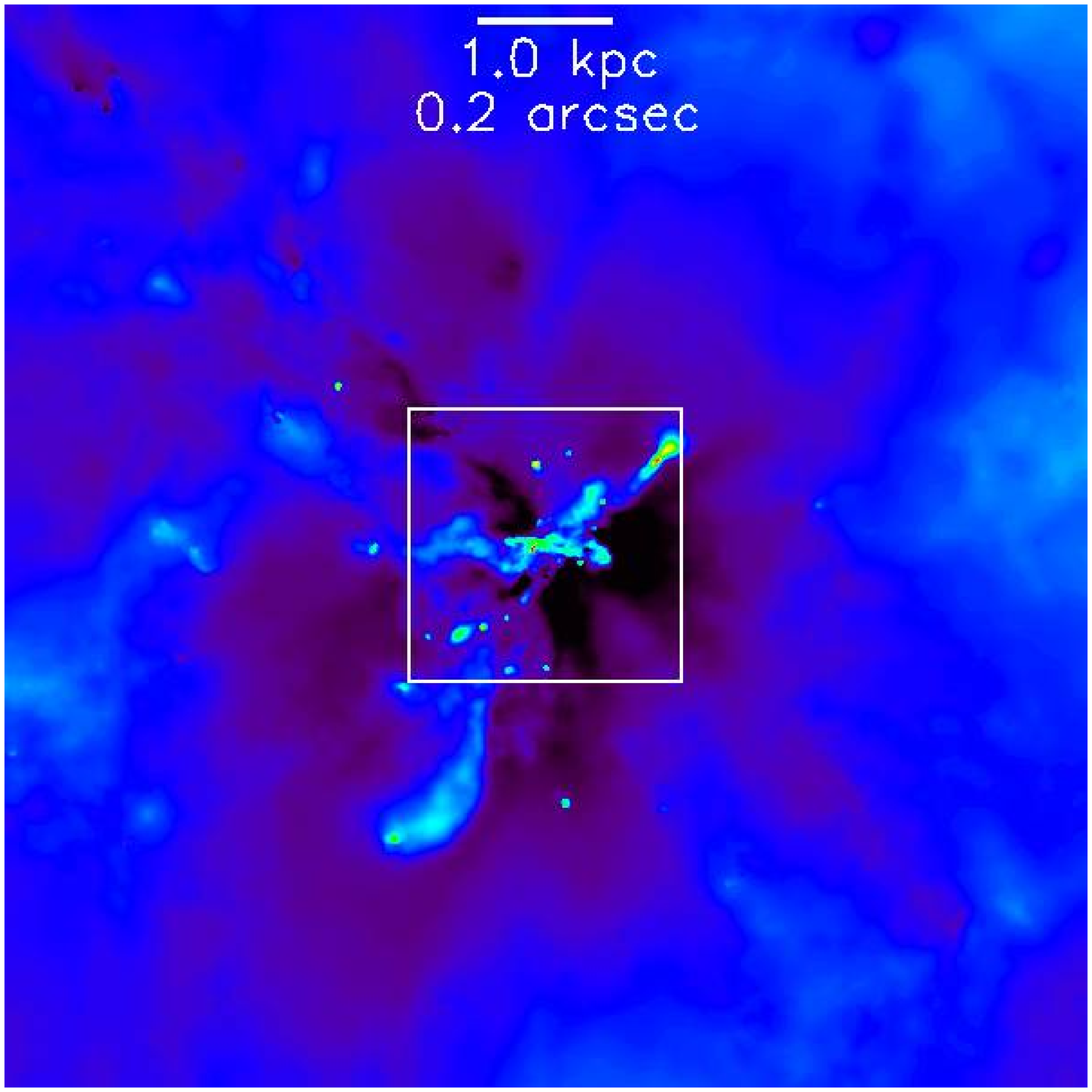}
\includegraphics[width = 0.29\textwidth]{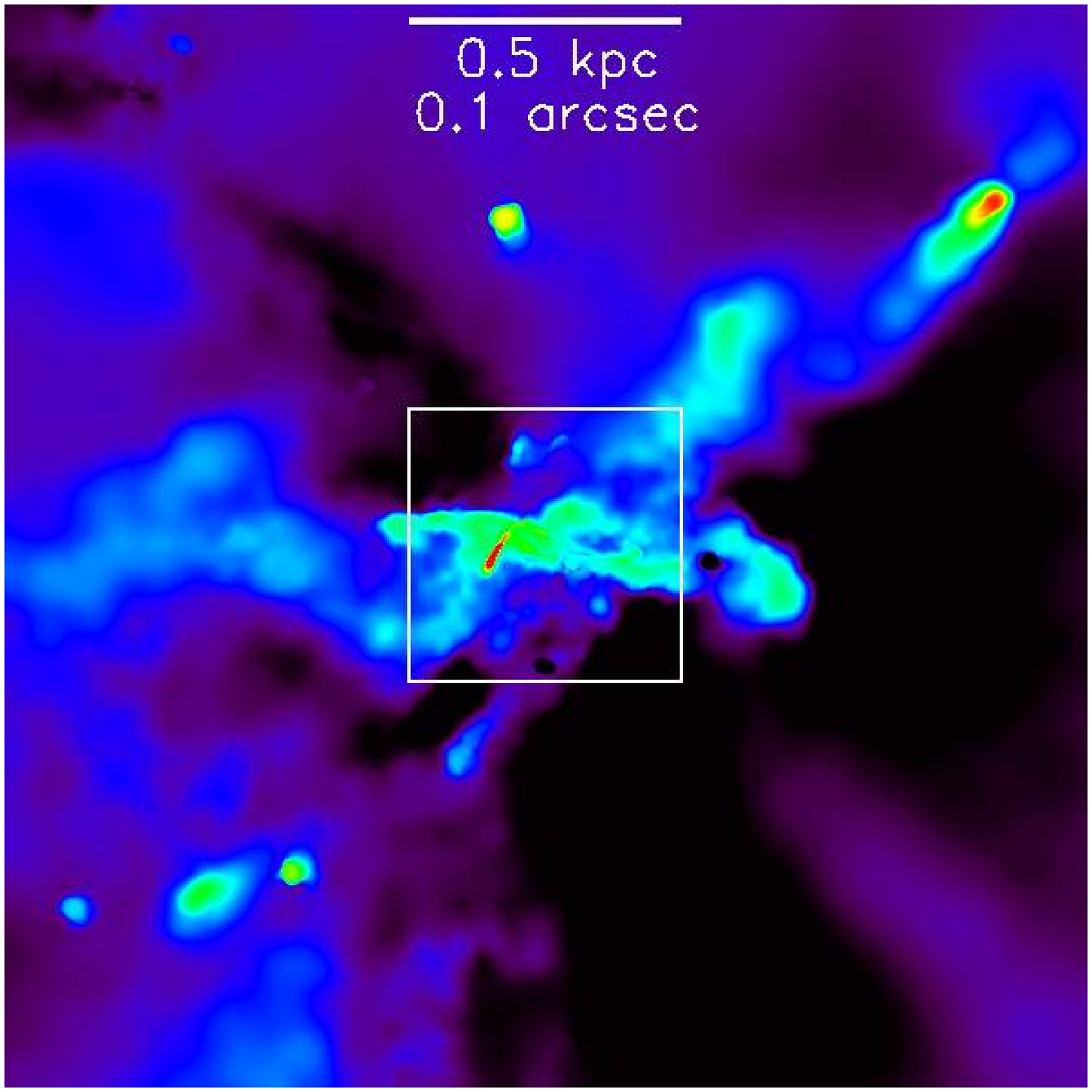}
\includegraphics[width = 0.29\textwidth]{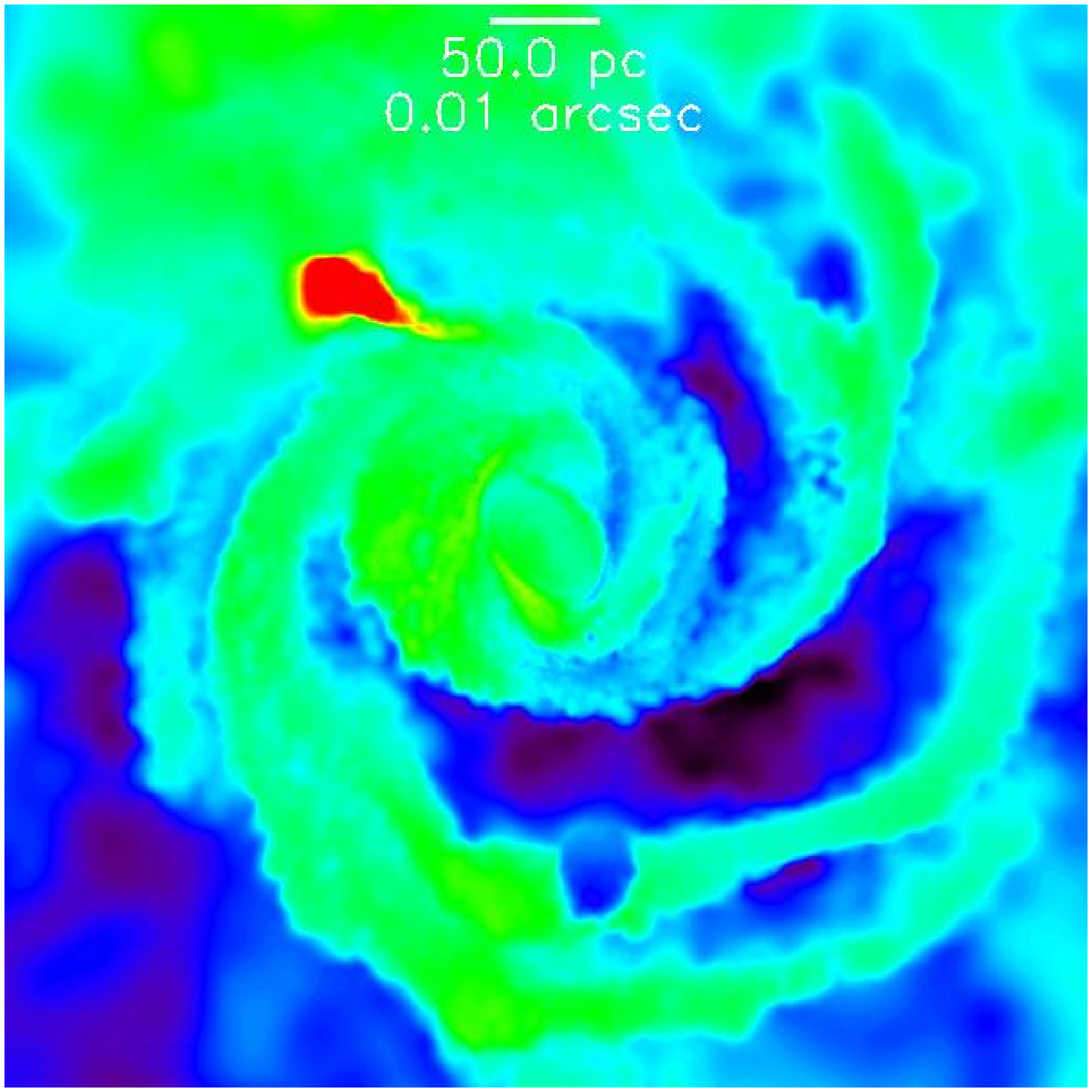}
\includegraphics[clip=true, trim = 310 0 50 0,height = 0.3\textwidth, width = 0.1\textwidth]{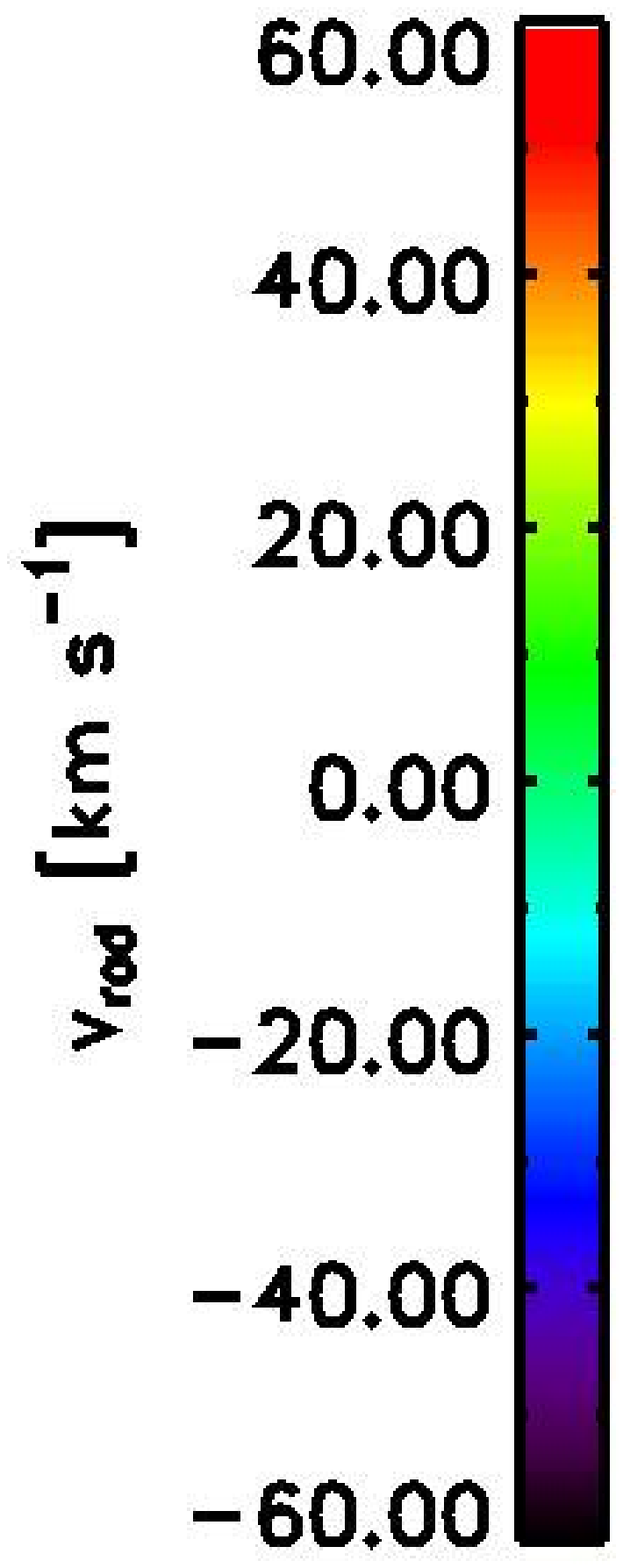}
\\
\includegraphics[width = 0.29\textwidth]{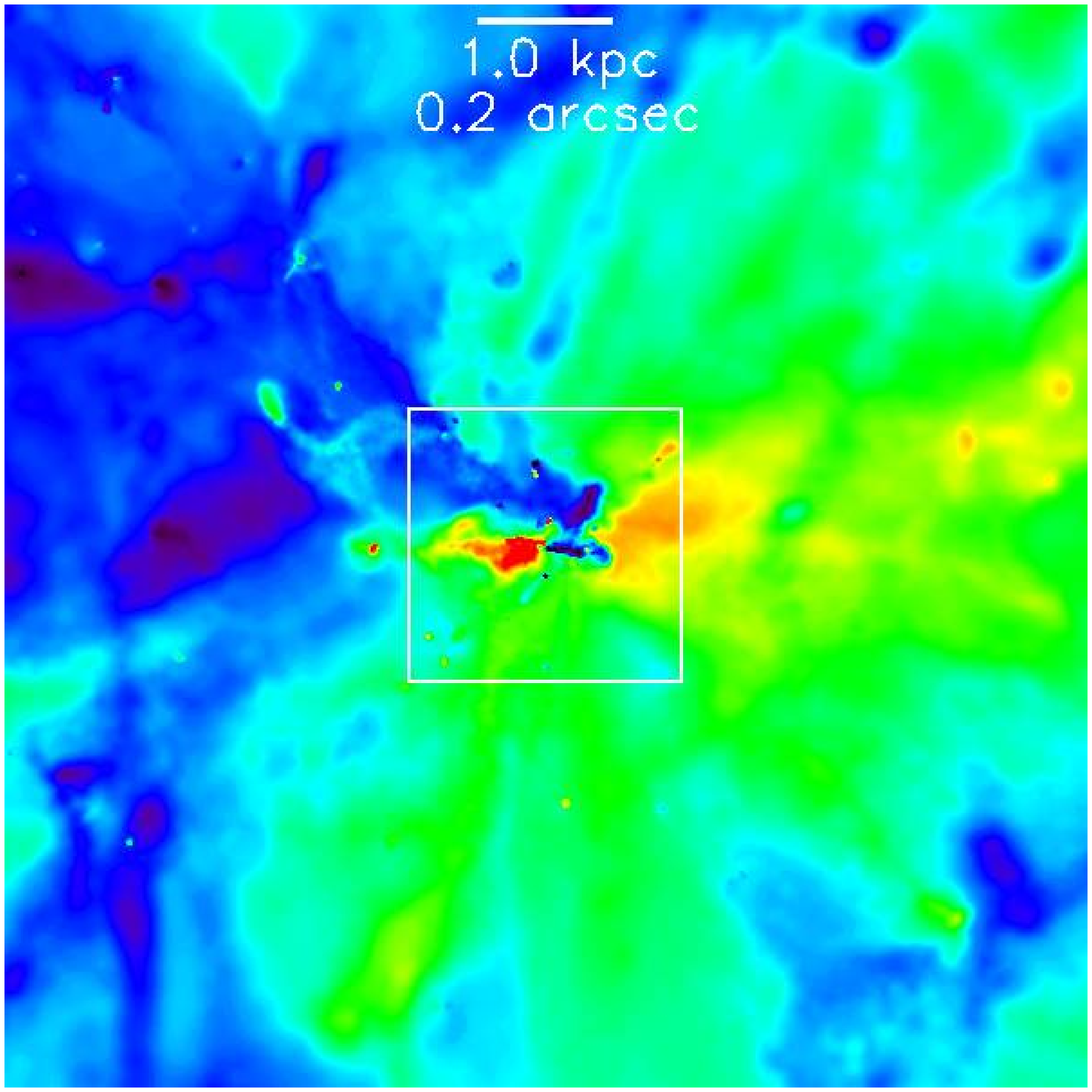}
\includegraphics[width = 0.29\textwidth]{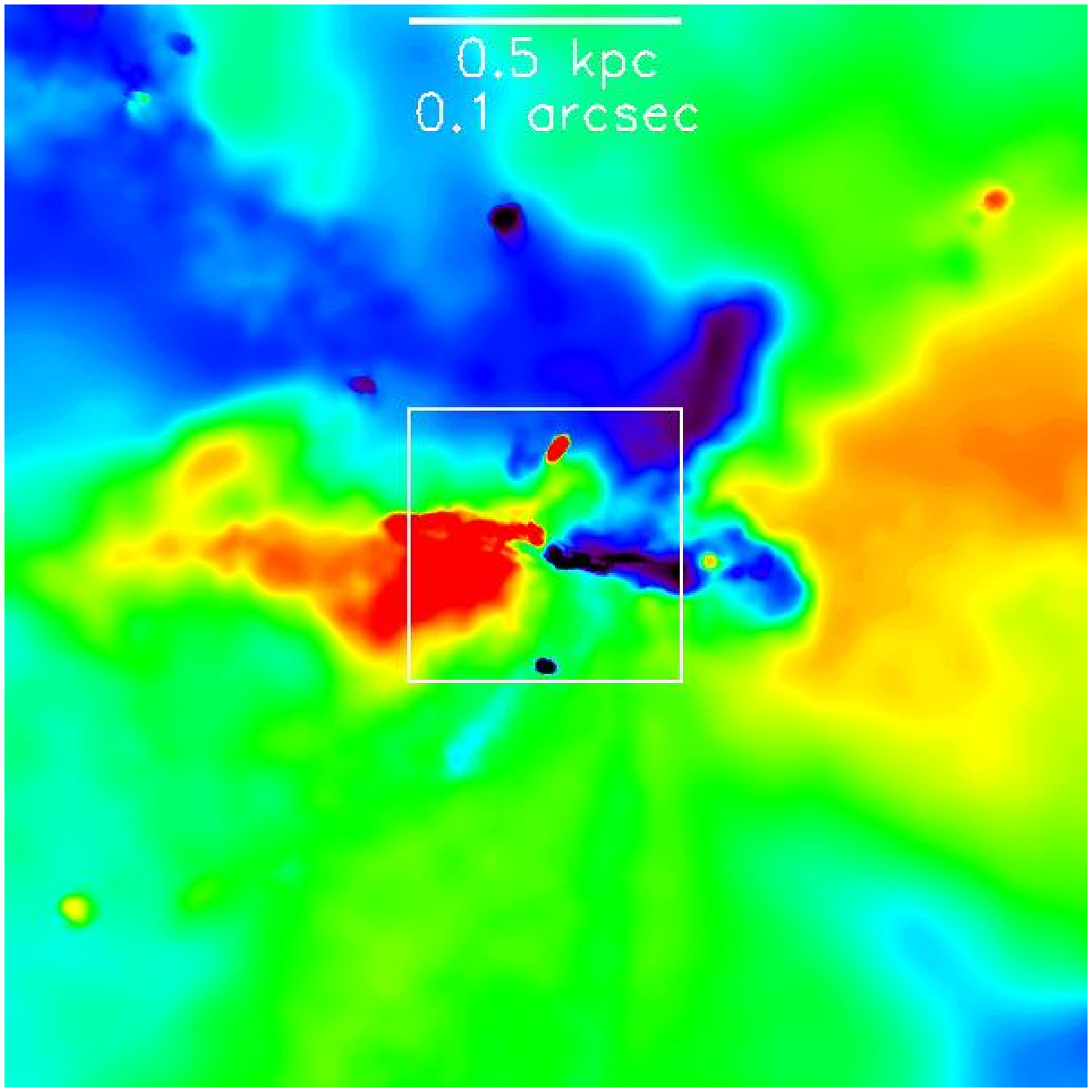}
\includegraphics[width = 0.29\textwidth]{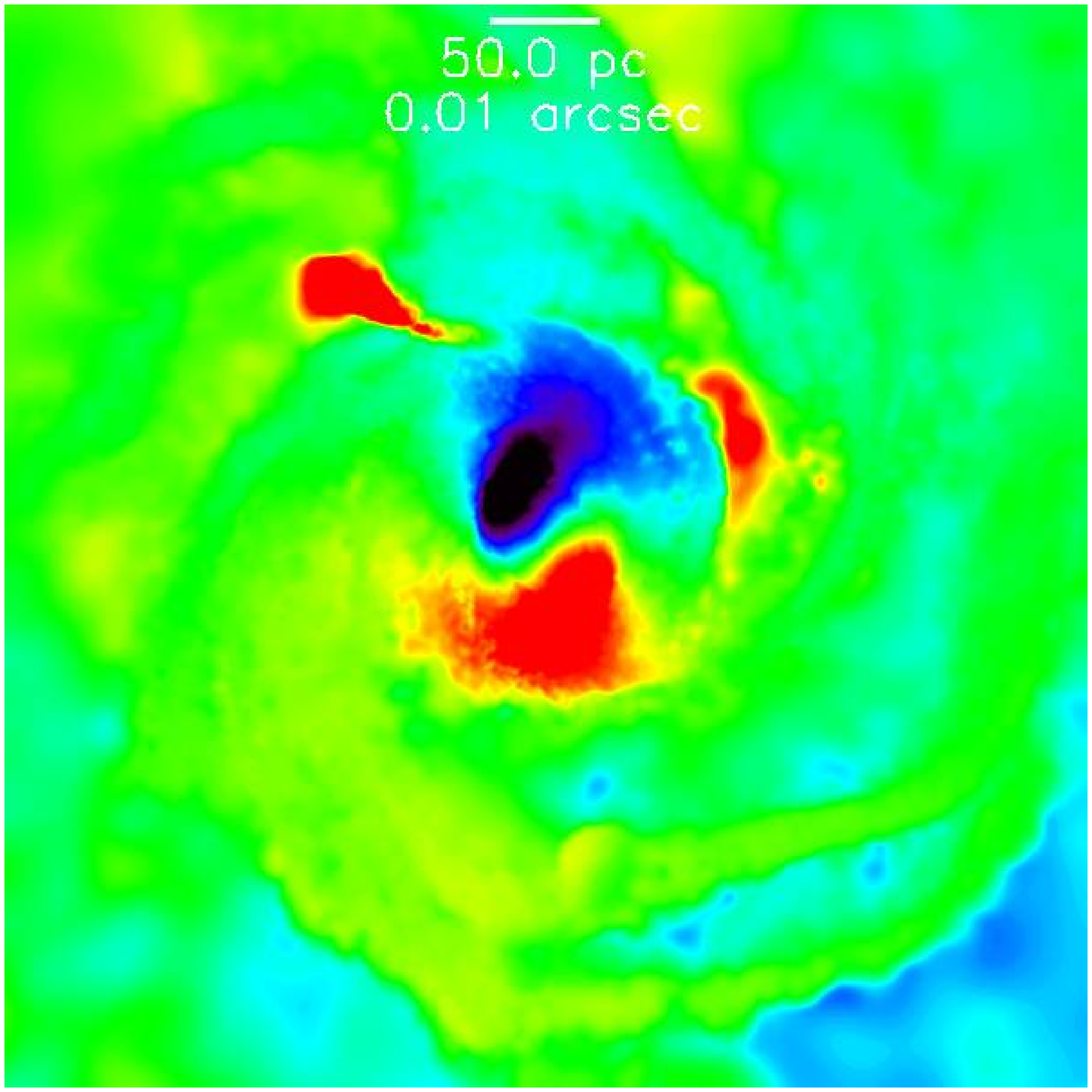}
\includegraphics[clip=true, trim = 310 0 50 0,height = 0.3\textwidth, width = 0.1\textwidth]{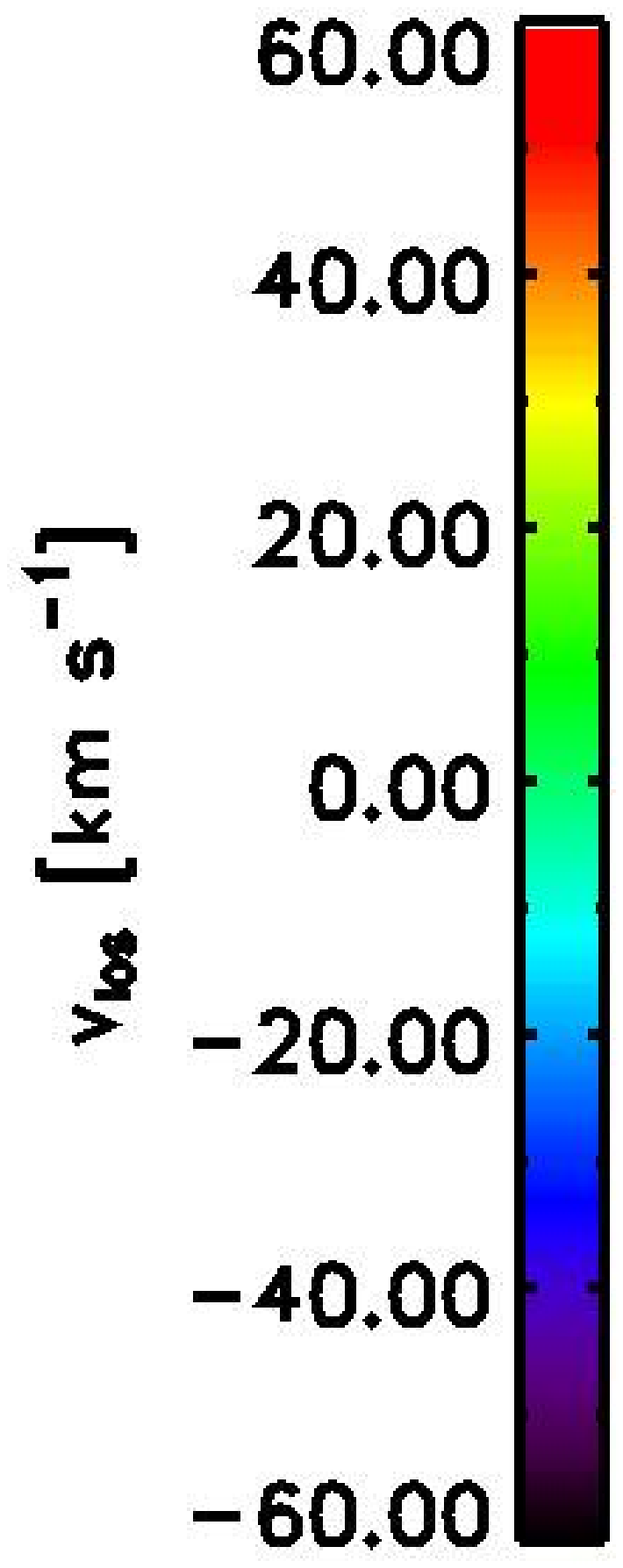}
\end{minipage}
\caption{Radial (top row) and line-of-sight (i.e., along the axis of
  projection; bottom row) gas velocities at $z = 10$ in simulation
  {\it Z4}. The images correspond in size and orientation to the
  images in Figure~\ref{Fig:Images:Densities:Z4} which allows the
  matching of features in velocity and real space. As in
  Figure~\ref{Fig:Images:Densities:Z4}, the left-hand and middle
  panels present edge-on views of the simulated galaxy, and the right-hand
  panels show the galaxy face-on. The top row panels show that the
  central disks grow through infall of dilute gas, channeling of both
  dense smooth gas and gas-rich subhalos along filaments and merging
  with gas-rich subhalos from outside filaments. The bottom row panels
  show the kinematic signature of the
  disks. \label{Fig:Images:Velocities}}
\end{center}
\end{figure*}
\par
Figure~\ref{Fig:Profiles:Velocities} shows spherically averaged
(mass-weighted) profiles of the gas radial velocities $v_{\rm g,r}$
(left panel) and fractional radial velocities $v_{\rm g,r }/
v_{\rm g}$ (right panel) at $z = 10$ in simulation \textit{Z4}, 
where $v_{\rm g} = |\mathbf{v}_{\rm g}|$ and $\mathbf{v}_{\rm g}$
is the gas velocity. For comparison, the corresponding velocities for the dark matter 
are also shown.  The velocities were corrected for the
bulk halo motion by subtracting the velocity of the center of mass of
all gas particles within the virial region  and were calculated
relative to the location of the most bound particle. 
Figure~\ref{Fig:Profiles:Velocities} shows that both the dark matter and
the gas approach the virial region along mostly radial orbits with 
similar infall velocities consistent with the halo circular
velocity (Section~\ref{Sec:Structure}).  Inside it, their velocity
distributions, however, differ significantly.
\par
The dark matter isotropizes just upon entry in the virial region,
i.e. at $r \approx r_{\rm vir}$ (vertical line on the right), which is
reflected in a sharp drop of the ratio of radial to total velocity to
$3^{-1/2}$ expected for an isotropic velocity distribution (horizontal
dotted line in the middle panel of
Figure~\ref{Fig:Profiles:Velocities}).  The gas, being able to
radiatively cool and lose gravitational energy, keeps streaming with
radial velocities that increase towards the halo center and reach
maximum values $-v_{\rm g,r }\lesssim 60 \kms$. The gas eventually
hits the central disk at $r \approx 0.3\kpc$ where it
circularizes. The right panel in
Figure~\ref{Fig:Profiles:Velocities} shows that at redshift $z = 10$,
in both simulation \textit{Z4} and simulation \textit{Z4NOMOL}, the
spherically integrated gas accretion rates are, on average, $-r^2 \int
\rho_{\rm g} v_{\rm g,r}d\Omega \approx 1 \Msun \invyr$, independent
of radius $r \gtrsim 0.07 \kpc$. The prominent spikes exhibited by the
gas accretion rates are due to the infall of gas-rich subhalos along
the radial filaments.
\par
The complexity of the gas dynamics within the virial region in
simulation \textit{Z4} is revealed by the radial and line-of-sight
(i.e., along the axis of projection) velocities that are shown,
respectively, in the top and bottom panels of
Figure~\ref{Fig:Images:Velocities}. The velocity structure of the gas
in simulation \textit{Z4NOMOL} is very similar.  The images show views
that correspond in size and orientation to the views shown in
Figure~\ref{Fig:Images:Densities:Z4}.  This allows to identify
structures in the velocity images with those in the images of density
and temperature.  The velocity images were obtained by mapping
particle radial and line-of-sight velocities to a three-dimensional
grid using standard SPH interpolation and performing a mass-weighted
average along the line of sight to project the grid into a
two-dimensional plane.
\par
The line-of-sight velocities (bottom panels in
Figure~\ref{Fig:Images:Velocities}) show the kinematic
signatures of two rotating disks that are centered on a spatially
unresolved core. The inner disk continues to grow in
mass not only through accretion of gas from outside the disk plane but
also through accretion of gas from within the outer disk to which it
is kinematically connected (bottom middle panel in
Figure~\ref{Fig:Images:Velocities}). The radial velocity (top panels in
Figure~\ref{Fig:Images:Velocities}) shows a complex inflow pattern. The
gas stream that points to the top right corner in the top middle panel
of Figure~\ref{Fig:Images:Velocities} is gas ejected by the passage of
a gas-rich subhalo from beneath the disk immediately before $z = 10$. 
A majority of subhalos appears to be associated with
filaments, which resembles the picture of anisotropic accretion of satellites in
simulations of massive galaxies and clusters at low redshift 
(e.g., \citealp{Libeskind:2010}; \citealp{Knebe:2004}).
\par
\subsection{The Disks}
\label{Sec:Disks}
\par
\begin{figure*}
  \begin{center}
\begin{minipage}[c]{1\linewidth}
\includegraphics[width = 0.24\textwidth]{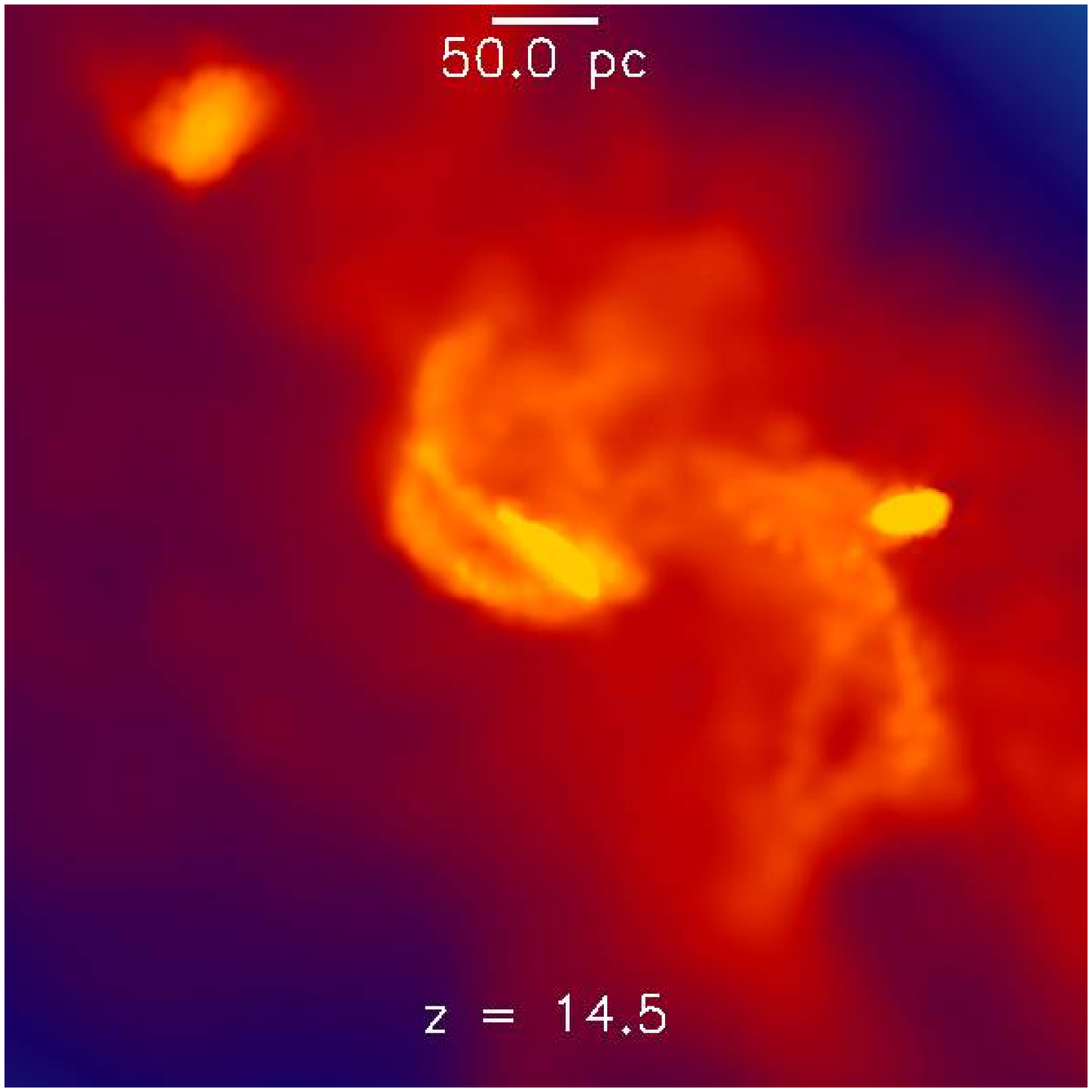}
\includegraphics[width = 0.24\textwidth]{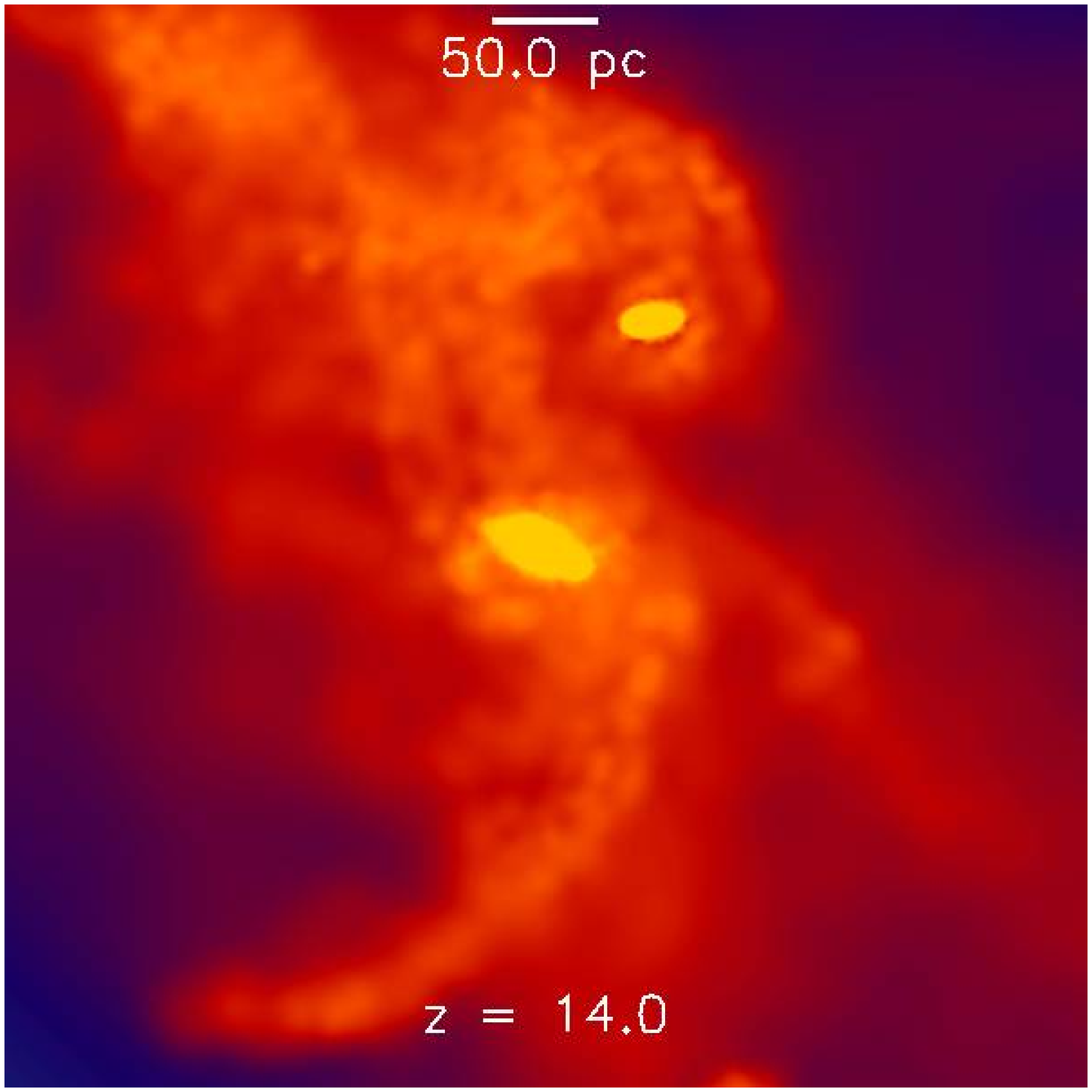}
\includegraphics[width = 0.24\textwidth]{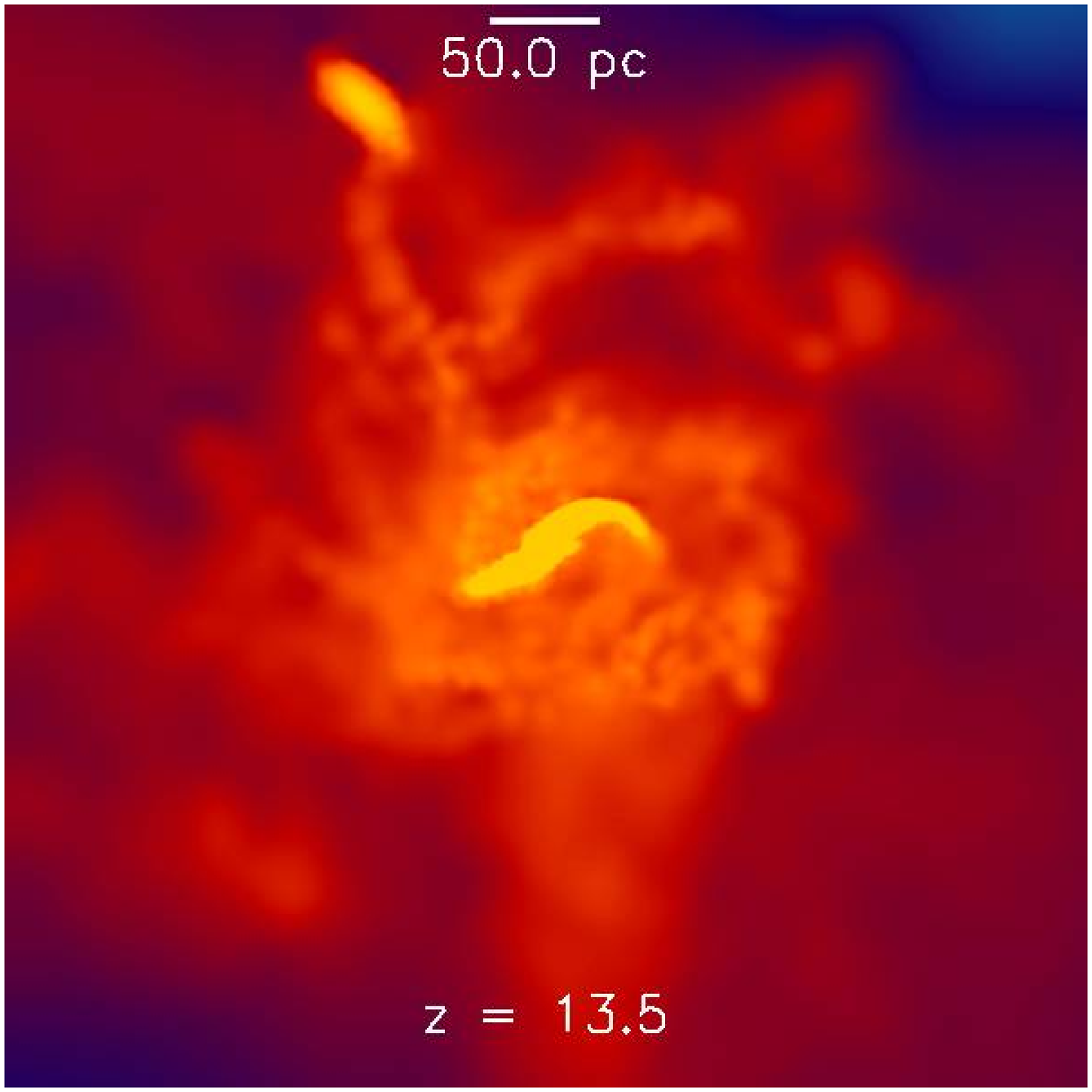}
\includegraphics[width = 0.24\textwidth]{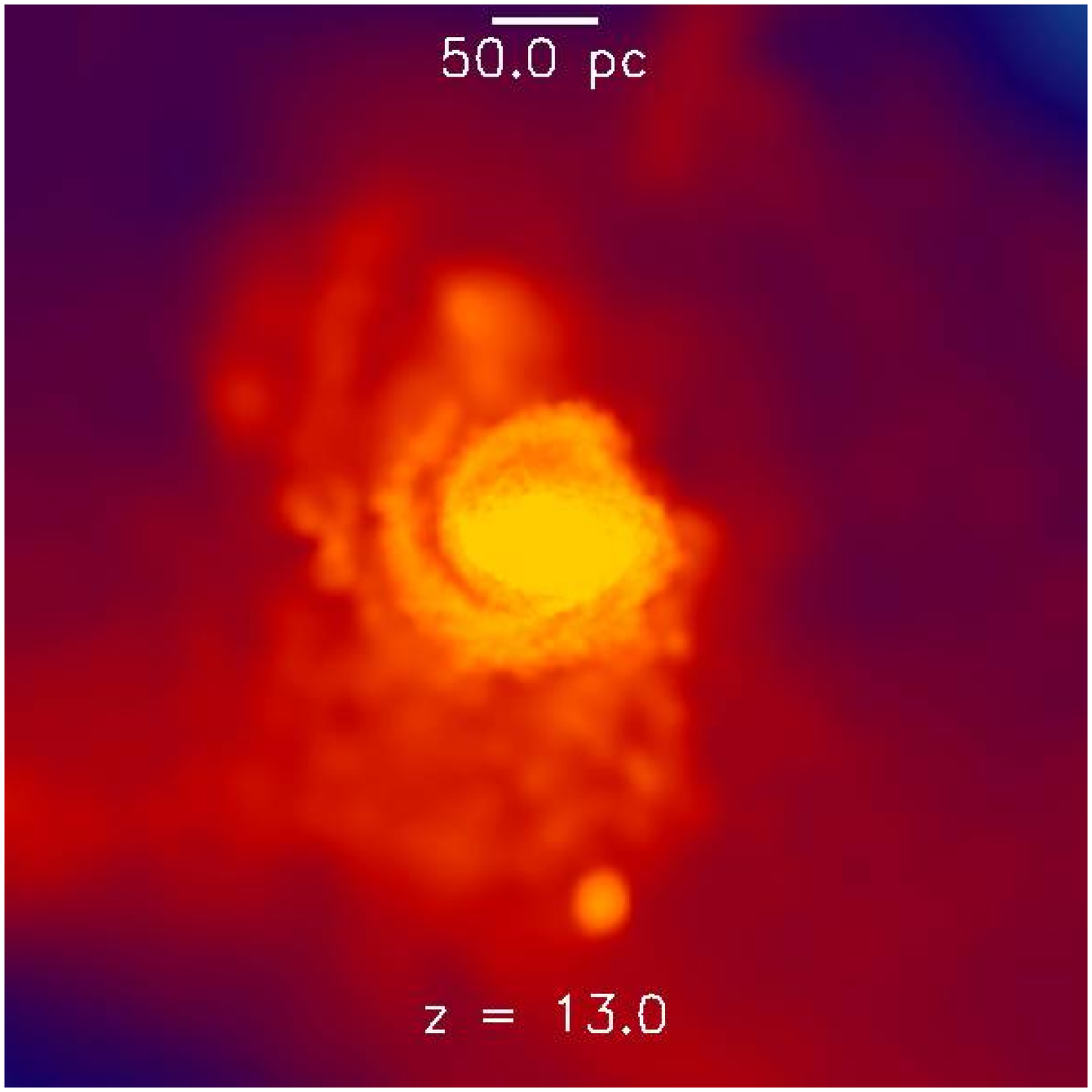}
\\
\includegraphics[width = 0.24\textwidth]{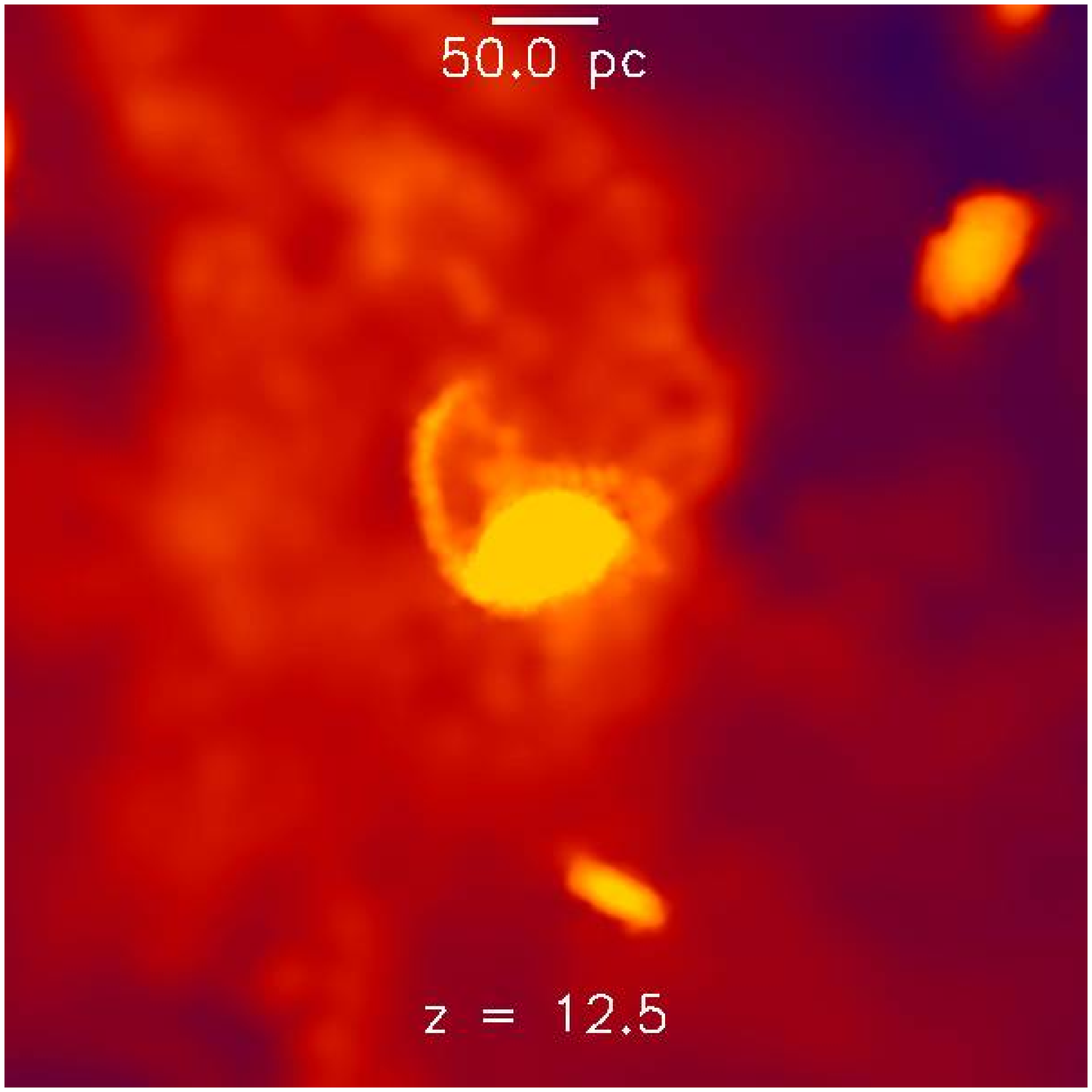}
\includegraphics[width = 0.24\textwidth]{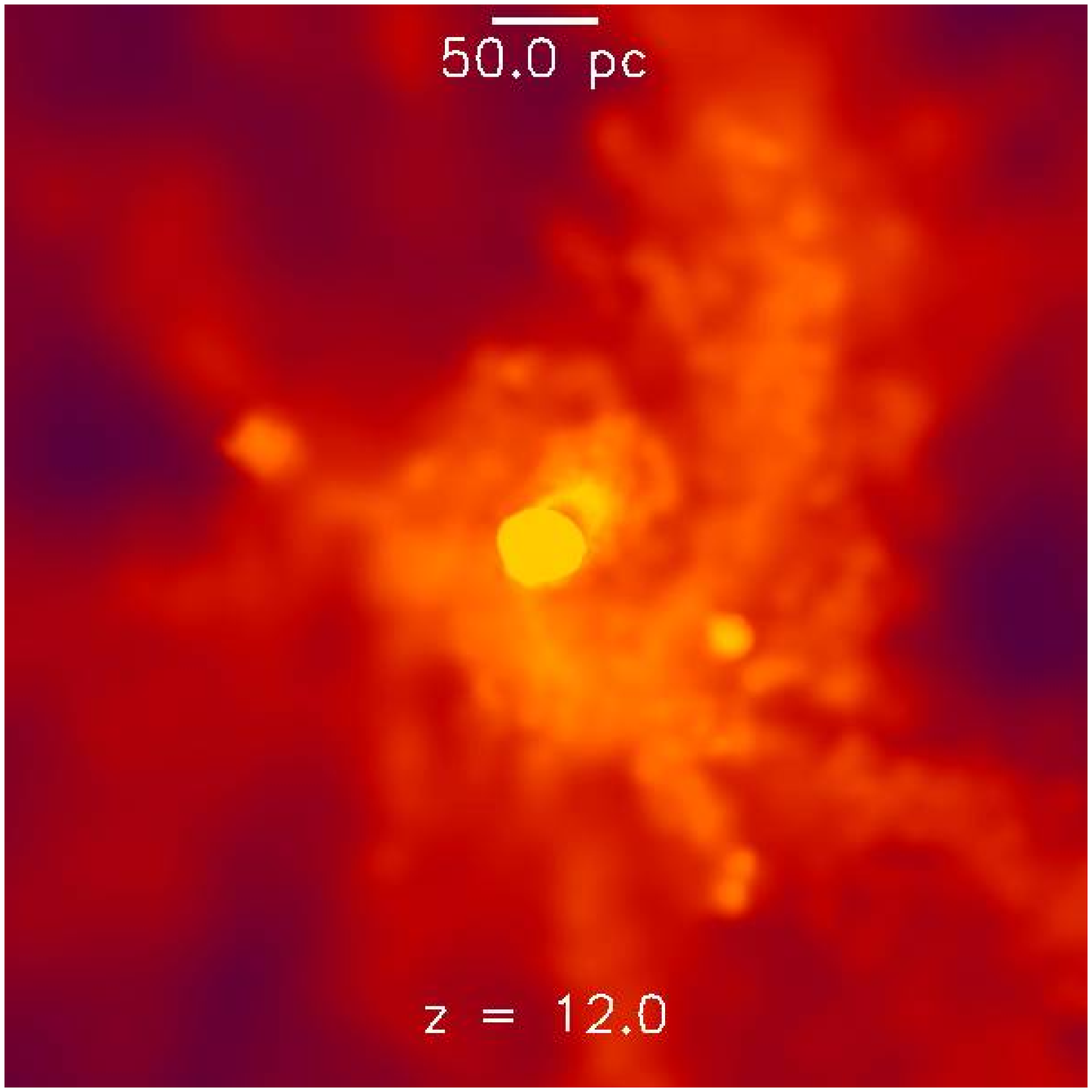}
\includegraphics[width = 0.24\textwidth]{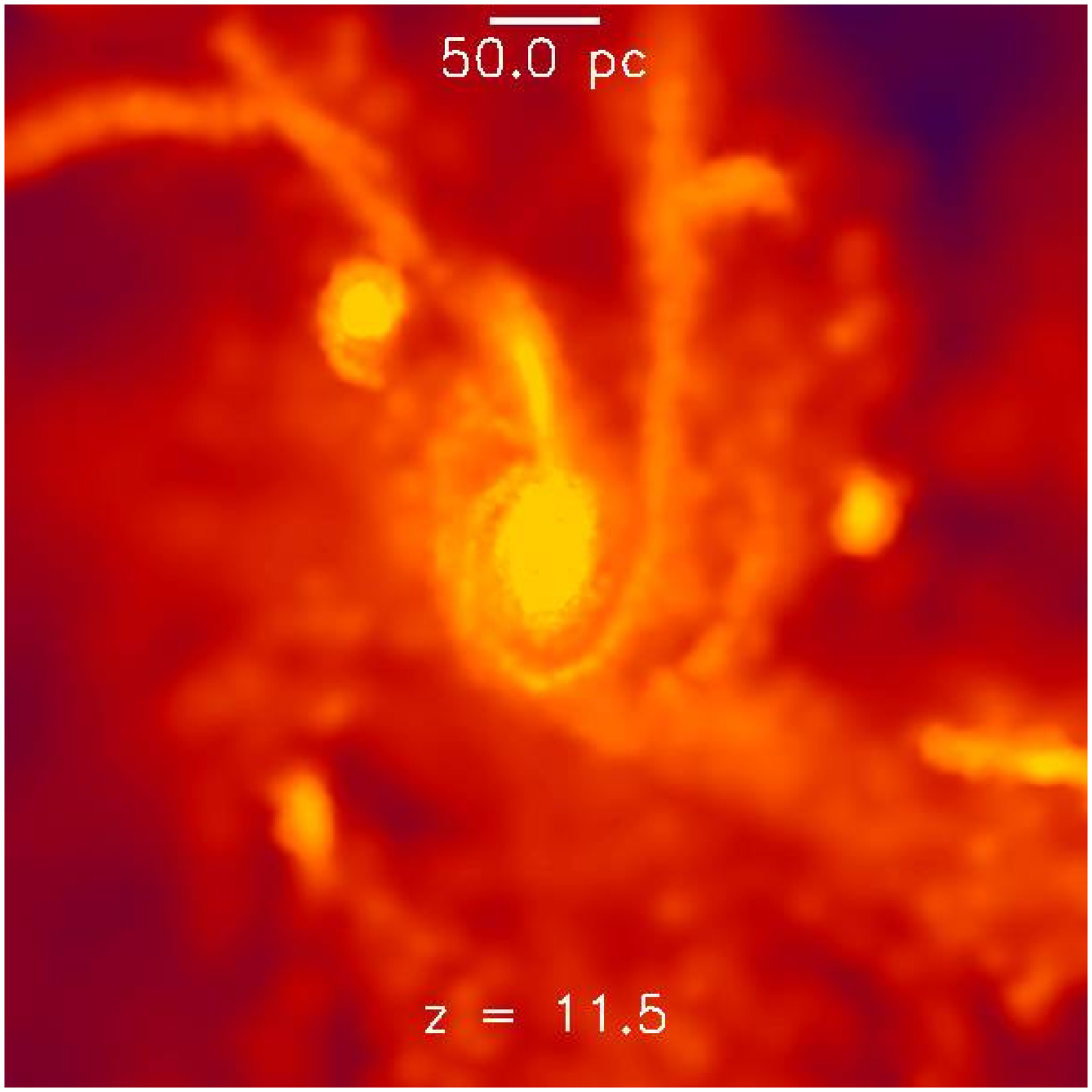}
\includegraphics[width = 0.24\textwidth]{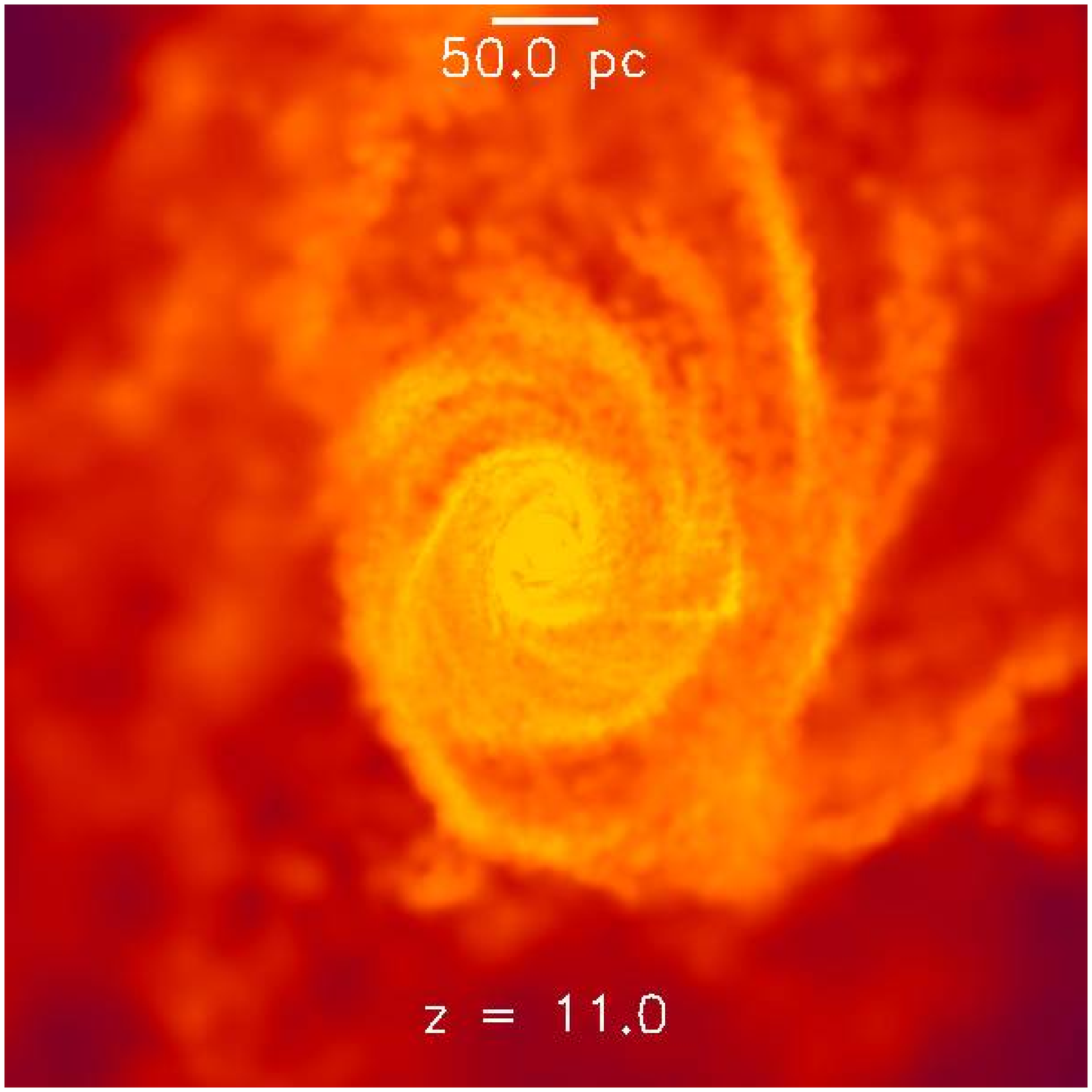}
\end{minipage}

\caption{Assembly of the inner and the outer gas disks in
simulation {\it Z4}. The panels present face-on views of the 
simulated galaxy and show the evolution of 
gas densities in the redshift range $11.0 \le z \le 14.5$ 
in cubical slices of linear size $0.5 \kpc$. They can
be compared to the top right panel of
Figure~\ref{Fig:Images:Densities:Z4} which shows a face-on view of 
the gas density at the final simulation redshift in the same cubical
slice. The color coding is identical to that in
Figure~\ref{Fig:Images:Densities:Z4}. The outer disk forms at $z
\approx 11.5$, roughly $100 \Myr$ after the assembly of the inner
disk at $z \approx 13.5$.\label{Fig:Images:Disks}}
\end{center}
\end{figure*}

\par
\begin{figure*}
\includegraphics[trim = 30mm 0mm 30mm 0mm, width = 0.33\textwidth]{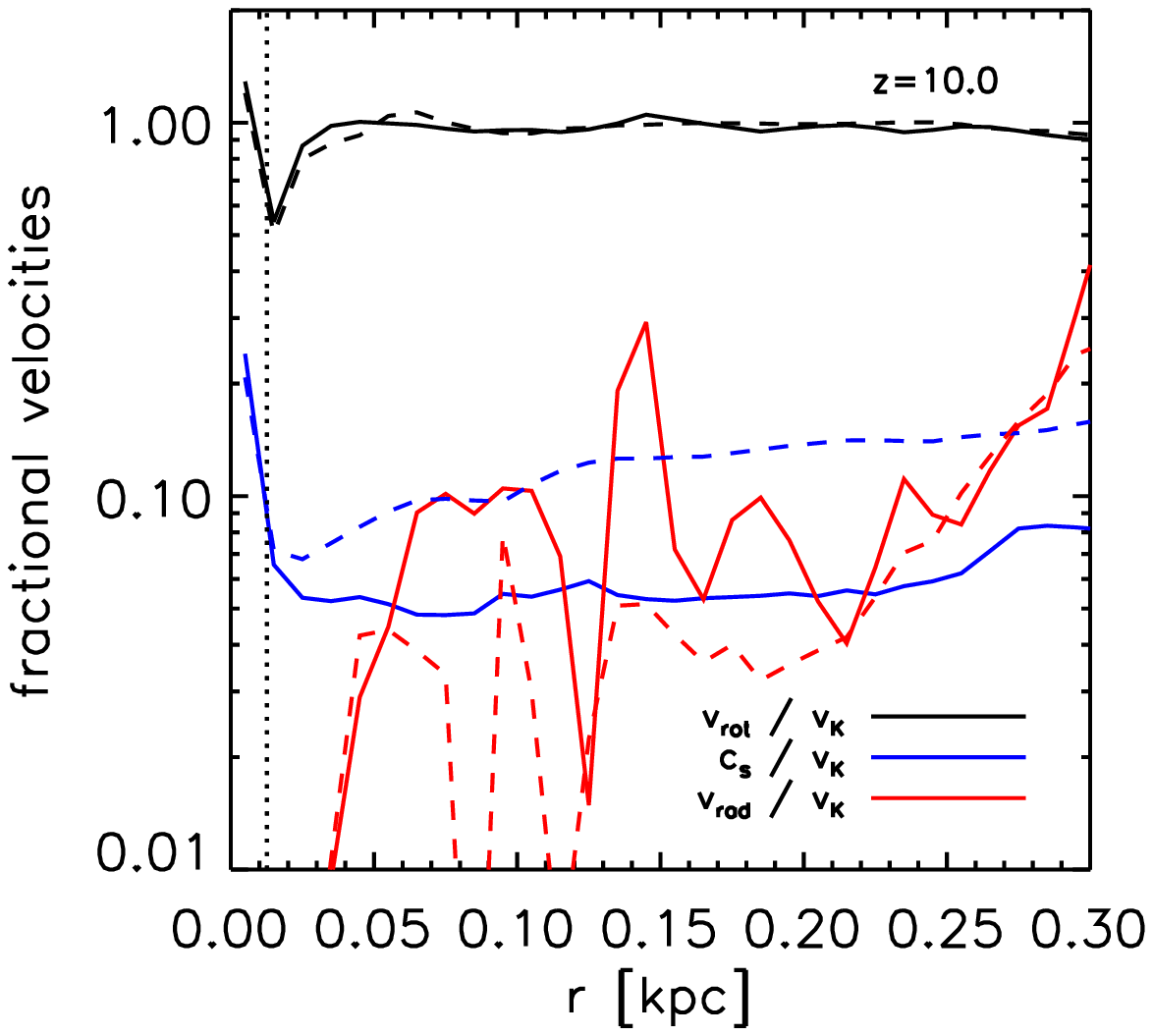}
\includegraphics[trim = 30mm 0mm 30mm 0mm, width = 0.33\textwidth]{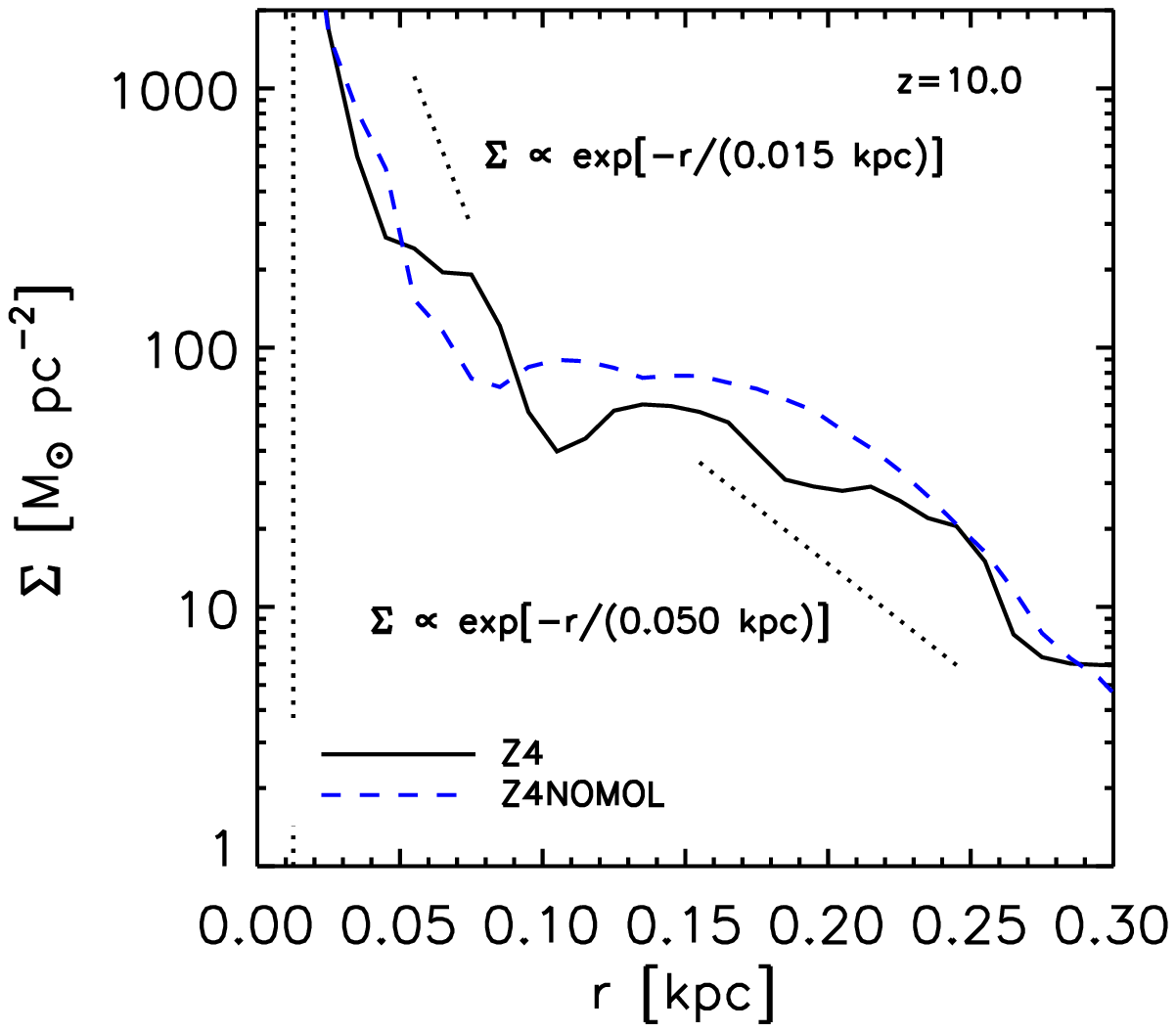}
\includegraphics[trim = 30mm 0mm 30mm 0mm, width = 0.33\textwidth]{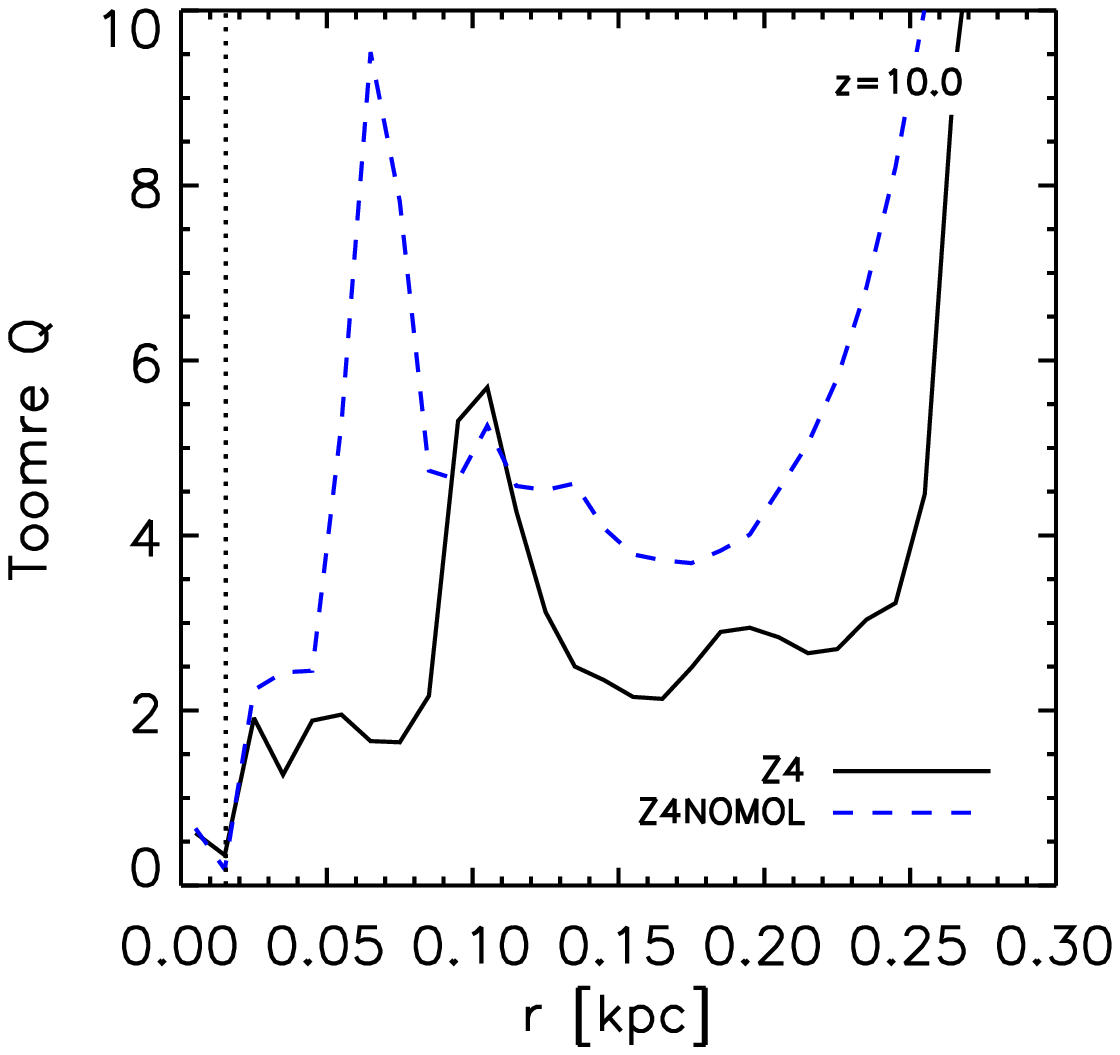}
\caption{Properties of the disks at $z = 10$.  {\it Left panel}:
Rotational gas velocities $ v_{\rm g,rot}$ (black curves), adiabatic
sound speeds $c_{\rm s}$ (blue curves) and radial gas velocities $v_{\rm g,r}$ 
(red curves) with respect to Keplerian velocities in
simulations \textit{Z4} (solid curves) and \textit{Z4NOMOL}
(dashed curves). {\it Middle panel}: Gas surface density profiles in
simulations \textit{Z4} (black solid curve) and \textit{Z4NOMOL} (blue
dashed curve).  {\it Right panel}: Toomre $Q$ parameter in
simulations \textit{Z4} (black solid curve) and \textit{Z4NOMOL} (blue
dashed curve). In each panel, the vertical line marks the gravitational softening
radius $\epsilon$. The gas in the outer disk ($0.1 \kpc <r< 0.3
\kpc$) is on nearly Keplerian orbits. Both the inner ($\epsilon
<r<0.07 \kpc$) and the outer disk show surface density profiles that
are roughly exponential with scale lengths as indicated in the middle
panel (dotted lines). Toomre parameters $Q > 1$ indicate stable disks. \label{Fig:Profiles:Disk}}
\end{figure*}
Figure~\ref{Fig:Images:Disks} shows the assembly of the inner and the
outer disk in simulation \textit{Z4}. The disk assembly times and
histories in simulation \textit{Z4NOMOL} are very similar.  The inner
disk forms at $z \approx 13.5$ as a result of a major merger. 
Initially, it is relatively thin and extended and also
shows marked spiral structure. Its gas, however, 
collapses into the highly concentrated disk that is seen at $z
= 10$ in Figure~\ref{Fig:Images:Densities:Z4} rather quickly, within a redshift
interval $\Delta z \lesssim 1$, corresponding to $\lesssim 40 \Myr$. 
The outer disk is assembled from the halo gas at $z \approx 11.5$,
i.e., $\lesssim 100 \Myr$ after the assembly of the inner disk. It 
grows in size and develops increasingly pronounced spiral structure. In none 
of our simulations do the disks show signs of fragmentation. 
\par
The right panels in Figures~\ref{Fig:Images:Densities:Z4} and
\ref{Fig:Images:Densities:Z4NOMOL} present face-on views of the disks
at $z = 10$.  In simulation \textit{Z4}, the outer disk shows much
more developed spiral structure.  When seen edge-on (middle
panels of Figures~\ref{Fig:Images:Densities:Z4} and
\ref{Fig:Images:Densities:Z4NOMOL}), the outer disk in \textit{Z4}
looks somewhat thinner and more perturbed than the outer disk in
\textit{Z4NOMOL}. The latter may be explained in part by the fact that in
simulation \textit{Z4}, thanks to the efficient low-temperature
molecular cooling, subhalos have larger gas fractions than in
simulation \textit{Z4NOMOL}, which increases the chance for possibly
violent subhalo-disk interactions.
\par
In simulation \textit{Z4NOMOL}, the disk gas reaches minimum
temperatures $T \gtrsim 8000 \K$ slightly below the temperatures in
the diffuse gas and filaments because the increased densities in the
disks imply shorter cooling times. In simulation \textit{Z4}, on the
other hand, the disk gas can cool to temperatures $T \lesssim 1000 \K$
thanks to the presence of molecular hydrogen.  Note that the disk
temperatures in simulation {\it Z4} are also determined by the
temperature floor enforced to prevent artificial fragmentation
(Section~\ref{Sec:Simulations:Fragmention}). The temperature floor
affects the evolution of the gas for densities above $n_{\rm H}
\gtrsim 10 \cmci$ in simulation \textit{Z4} and densities above
$n_{\rm H} \gtrsim 10^6 \cmci$ in simulation \textit{Z4NOMOL} (see
equation~[\ref{Eq:Floor}]). These densities correspond, respectively,
to radii $r \lesssim 0.3 \kpc$ and $r \lesssim 0.03 \kpc$ (see
Figure~\ref{Fig:Profile:Density}).
\par
The final mass distributions in the disk region in the simulations
{\it Z4} and {\it Z4NOMOL} are very similar. In both simulations, the
volume inside $r \le 0.3 \kpc$ contains gas and total masses of
$\approx 1.4 \times 10^8 \Msun$ and $\approx 2.5 \times 10^8 \Msun$,
respectively (see Figure~\ref{Fig:Profile:Density}). These masses
amount to $\approx 65 \%$ of the gas mass and to $\approx 25\%$ of the
total mass inside the virial region. In simulation
{\it Z4}, roughly $20\%$ ($35\%$) of the total (gas) mass within $r
\le 0.3 \kpc$ is in the central unresolved core, about $30\%$ ($40\%$)
is in the inner disks and about $50\%$ ($25\%$) is in the outer
disks. In simulation \textit{Z4NOMOL}, roughly $\lesssim 17\%$
($\lesssim 28\%$) of the total (gas) mass out to radii $r \le 0.3
\kpc$ is in the central unresolved core, $\lesssim 32\%$ ($\lesssim
40\%$) is in the inner disk and about $51\%$ ($\lesssim 32\%$) is in
the outer disk.
\par
Figure~\ref{Fig:Profiles:Disk} shows several azimuthally averaged properties of 
the disks at $z = 10$ in simulation \textit{Z4} and \textit{Z4NOMOL}. 
The left panel of Figure~\ref{Fig:Profiles:Disk} shows rotational
gas velocities $ v_{\rm g,rot}$ (black curves), adiabatic sound speeds
$c_{\rm s} = [\gamma k_{\rm B} T / (\mu m_{\rm H})]^{1/2}$ (blue curves) and radial
velocities $v_{\rm g,r}$ (red curves), all with respect to the
Keplerian velocities $v_{\rm K} = [G M (<r) / r]^{1/2}$ and for both
simulation \textit{Z4} (solid curves) and simulation \textit{Z4NOMOL}
(dashed curves).  The rotational velocities were computed using
$ v_{\rm g,rot} = (\mathbf{v}_{\rm g}^2 - \mathbf{v}_{\rm g,r}^2)^{1/2}$. 
For computing the sound speed we have assumed a ratio of
specific heats of $\gamma = 5/3$ appropriate for a mostly atomic gas
and assumed that the cold disk gas is mostly neutral, i.e., $\mu = 1.2$. 
Both the inner and the outer disks exhibit a high degree of rotational
support with the outer disk showing nearly Keplerian
motion. The rotational velocities are larger than the sound velocities
by factors of $\lesssim 5-10$. Radial velocities are small compared to rotational
velocities at all disk radii and drop sharply to zero once the gas
hits the central core at $r \lesssim \epsilon$.  
\par 
The middle panel of Figure~\ref{Fig:Profiles:Disk} shows the gas
surface density profiles. In both \textit{Z4} and \textit{Z4NOMOL},
the inner disk, which is resolved with $\gtrsim 5$ gravitational
softening radii, is characterized by an exponential surface density
profile with scale length $0.015 \kpc$ that extends over several scale
lengths.  In \textit{Z4NOMOL}, the outer disk follows an exponential
profile over several scale lengths of $0.05 \kpc$. In \textit{Z4}, the
surface density profile shows significant deviations from an
exponential profile. The nearly exponential density profiles of the
outer disks in our simulations are only gradually built up and at
higher redshifts the density profile in simulation {\it Z4} shows
pronounced ripples due to the presence of thick spiral structure. 
\par
The right panel of Figure~\ref{Fig:Profiles:Disk} shows the Toomre
parameter $Q = c_{\rm s} \kappa/ (\pi G \Sigma)$, where $c_{\rm s}$ is the 
velocity dispersion, $\kappa = (4 \Omega^2 + r d\Omega^2/dr)^{1/2}$ is
the epicyclic frequency and $\Omega = v_{\rm g,rot} / r$ is the angular
velocity (\citealp{Toomre:1964}). A standard linear theory instability
analysis (\citealp{Toomre:1964}; \citealp{Goldreich:1965};
\citealp{Binney:2008}) shows that a gas disk becomes unstable to
axisymmetric perturbations if $Q \lesssim 1$; the precise
threshold for instability depends on the disk thickness. The linear
analysis is confirmed with detailed simulations of disk instability, which also
show that a similar criterion applies to the discussion of non-axisymmetric 
perturbations (see, e.g., the review by \citealp{Durisen:2007}). 
In computing $Q$ we identify the velocity dispersion $c_{\rm s}$ with the
adiabatic sound speed and we set $\kappa = v_{\rm K} / r$ appropriate
for Keplerian motion. Both in simulation \textit{Z4} and in
simulation \textit{Z4NOMOL} the Toomre parameter $Q \gtrsim 1$ for all
radii $\epsilon \lesssim r < 0.3 \kpc$ that cover the two disks. In
both simulations the inner disk is characterized by $Q \lesssim 2$
while the outer disk is characterized by $Q \gtrsim 2$.  
\par
The measured Toomre $Q$ parameters imply that the disks in simulation
\textit{Z4} are somewhat less stable than the disks in simulation
\textit{Z4NOMOL}. This is consistent with the observation that spiral
arms in \textit{Z4} are more distinct than in \textit{Z4NOMOL} (right
panels in Figures~\ref{Fig:Images:Densities:Z4} and
\ref{Fig:Images:Densities:Z4NOMOL}) and is likely a direct consequence
of the fact that in simulation \textit{Z4} the disk temperatures are
significantly lower than in simulation \textit{Z4NOMOL} thanks to
efficient low-temperature cooling by molecular hydrogen. Note that the
stability of the outer and of the inner disk in simulation \textit{Z4}
and the stabilitiy of the inner disk in simulation \textit{Z4NOMOL}
may be artificially increased due to the imposed Jeans floor as
mentioned above. Note also that in equating the velocity dispersion
with the sound speed, we may have underestimated the true velocity
dispersion and hence the stability of the disks.
\par
\section{Detecting the First Galaxies with JWST}
\label{Sec:JWST}
\begin{figure*}
\includegraphics[trim = 20mm 0mm 20mm 0mm, width = 0.33\textwidth]{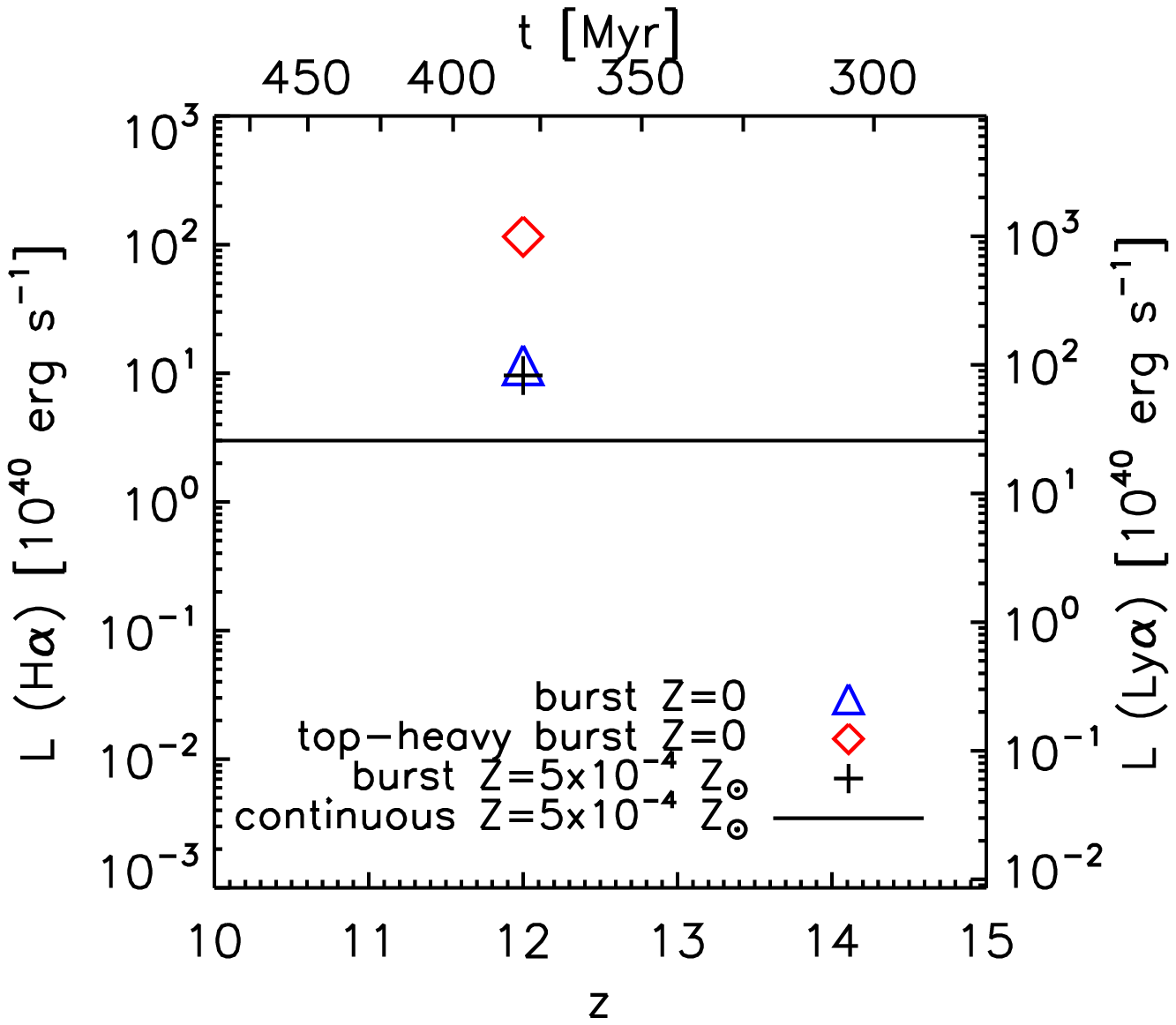}
\includegraphics[trim = 20mm 0mm 20mm 0mm, width = 0.33\textwidth]{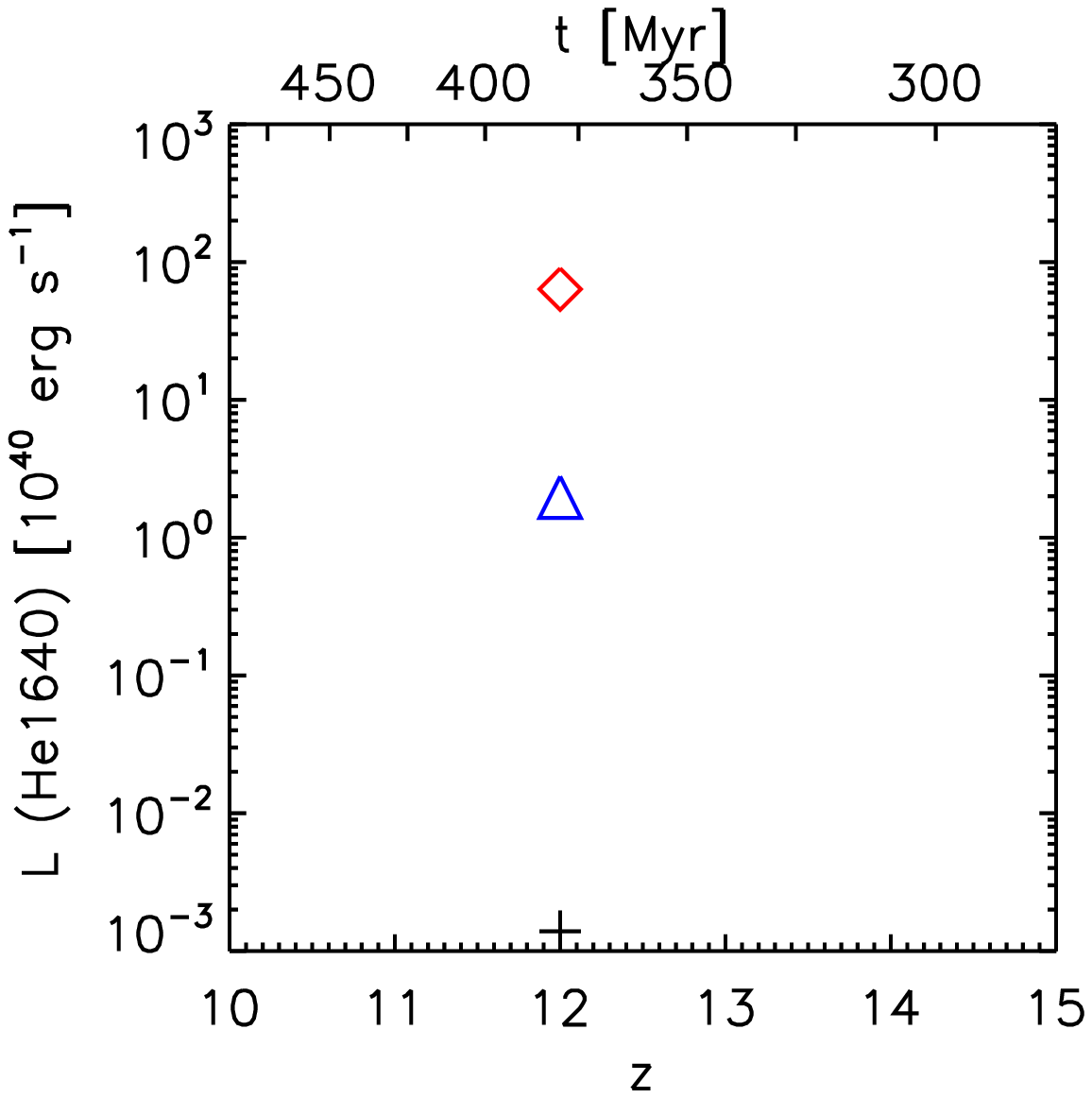}
\includegraphics[trim = 20mm 0mm 20mm 0mm, width = 0.33\textwidth]{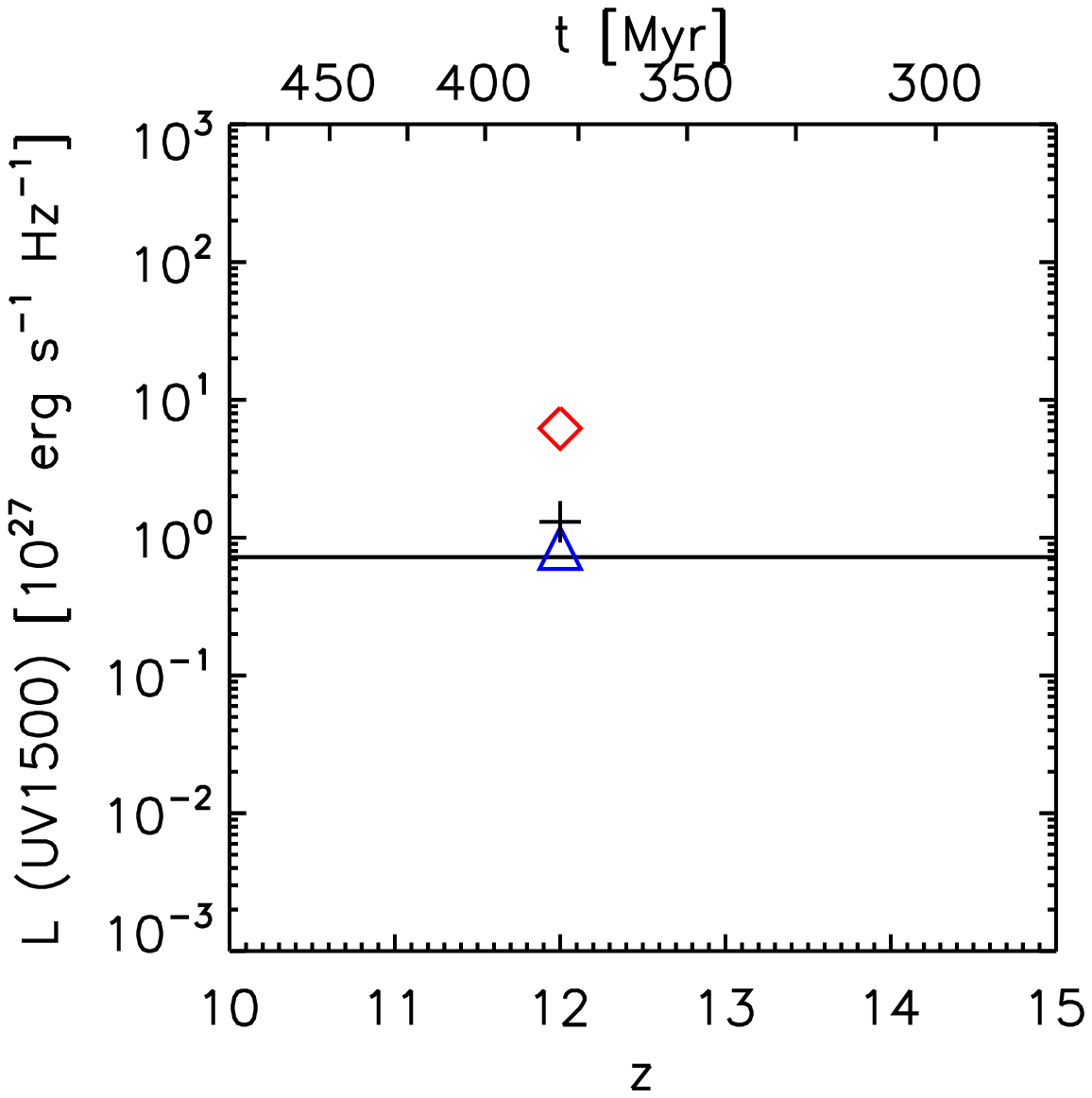}
\caption{Luminosities in the H$\alpha$ (left panel, left axis)
and Ly$\alpha$ (left panel, right axis) and He1640 (middle panel)
nebular recombination lines and the intensity of the combined stellar
and nebular UV1500 continuum (right panel) for galaxies inside
halos with mass $M_{\rm vir} = 10^9 \Msun$. The estimates are based on the
\cite{Schaerer:2003} stellar population synthesis models and assume
Case B recombination theory and zero escape fractions. Red diamonds assume the formation of
metal-free stars with top-heavy IMF in an instantaneous burst of total
stellar mass $M_{\star} = 10^6 \Msun$ at $z = 12$.  We also show the
luminosities expected for identical bursts but assuming metal-free
stars with a normal IMF (blue triangles) and stars with metallicities
$Z=5\times 10^{-4} \Zsun$ and normal IMF (black crosses). The lines assume the
continuous formation of stars with metallicities $Z=5\times 10^{-4} \Zsun$ and
normal IMF at a rate $0.05\Msun\invyr$. All results scale linearly
with the star formation efficiencies of $f_{\star} = 0.1$ and
$0.05$ assumed for, respectively, the starburst
(equation~[\ref{Eq:Starburst}]) and the continuous star formation
scenario (equation~[\ref{Eq:Continuous}]). Note the large dependence
of the He1640 flux on metallicity and IMF. The He1640 line
luminosities estimated for the continuous star formation scenario are
not shown because they are too low to fall inside the plot range.
\label{Fig:IntrinsicLuminosities}}
\end{figure*}
One of the main science goals of the upcoming {\it JWST} is the detection of light from the first
galaxies (\citealp{Gardner:2006}). Here we present an estimate of the expected flux from the
first galaxies based on our simulations of galaxies with halo masses
$M_{\rm vir} \sim 10^9 \Msun$ at $z \gtrsim 10$ and investigate their detectability with the
instruments aboard {\it JWST}.
\par
{\it JWST} will observe the first galaxies using deep field imaging
and spectroscopy with the Near Infrared Camera
(NIRCam) and the Mid Infrared Instrument (MIRI) and
using spectroscopy with the Near Infrared Spectrograph
(NIRSpec) and also MIRI (for an overview of these instruments
see http://www.stsci.edu/jwst; see also \citealp{Gardner:2006}).
NIRCam will allow imaging and low resolution ($R \equiv
\lambda/\Delta\lambda \lesssim 100$) spectroscopy within a field of
view of $2.2'\times 4.4'$ and an angular resolution of $\sim
0.03''-0.06''$ in the range of observed wavelengths $\lambda_{\rm o} =
0.6-5 \mum$.  The multi-object spectrograph NIRSpec will enable
medium resolution ($R \sim 100 - 3000$) spectroscopy of up to $\sim
100$ objects simultaneously within a field of view of $3.4'\times
3.4'$. NIRSpec will operate in the same wavelength range as 
NIRCam but at lower angular resolution ($\sim 0.1''$). Finally, 
MIRI will complement NIRCam and NIRSpec by providing
imaging, low and medium resolution spectroscopy within the range of
observed wavelengths $\lambda_{\rm o} = 5-28.8 \mum$ and fields of
view and angular resolutions of, respectively, $\sim 2'\times 2'$ and
$\sim 0.1''-0.6''$.
\par
\subsection{Intrinsic Luminosities}
Our estimates of the observability of the first galaxies are based 
on our simulations of galaxies in halos with masses $\sim
10^9 \Msun$ at $z \gtrsim 10$. We examine and compare two 
idealized scenarios for star formation derived from the gas accretion
rates observed in our simulations, chosen to bracket the range of
likely scenarios and parametrized such as to enable the straightforward rescaling 
and extrapolation of our results. We combine assumptions about the nature of the
forming stellar populations with population synthesis
models to estimate the luminosities in the Ly$\alpha$, H$\alpha$ and
HeII (restframe wavelength $\lambda_{\rm e} = 1640 \Ang$; hereafter
He1640) nebular recombination lines and the intensities of the
non-ionizing (combined stellar and nebular) UV continuum (rest-frame
wavelength $\lambda_{\rm e} = 1500\Ang$; hereafter UV1500). We
translate line luminosities and UV continuum intensities
into observed fluxes and compare them with the expected flux limits
for observations with {\it JWST}.  Based on extrapolation of our
results to galaxies with both lower and larger halo masses, we
estimate the number of high-redshift halos {\it JWST} will detect.
\par
The first of the two star formation scenarios explored here 
assumes that stars form in a single central
instantaneous burst with total stellar mass
\begin{equation}
M_\star = 10^6 \Msun \left(\frac{f_{\star}}{ 0.1}\right) \left(\frac{f_{\rm cool}}{0.01}\right) \left(\frac{M_{\rm vir}}{ 10^9 \Msun}\right), 
\label{Eq:Starburst}
\end{equation}
where $f_{\rm cool}$ is a conversion factor that determines the amount of
gas mass available for starbursts inside halos with virial 
masses $M_{\rm vir}$, and $f_\star$ is the star formation efficiency,
i.e., the fraction of the available gas mass that is turned into
stars.  Setting $f_{\rm cool} = 0.01$, this scenario is motivated by the
rapid accretion of large
gas masses ($M_{\rm g} \gtrsim 10^7 \Msun$) onto the central
unresolved core observed in our simulations of halos with
virial masses $M_{\rm vir} \sim 10^9 \Msun$ at around $z \lesssim 12$
(see the right panel of Figure~\ref{Fig:Profile:Density}). We
choose a relatively high star formation efficiency, $f_\star = 0.1$, 
expected for the initial bursts (e.g, \citealp{Wise:2009}; \citealp{Johnson:2009}). 
The adopted conversion factor $f_{\rm cool} = 0.01$ between 
the gas mass available for star formation and the halo virial mass is consistent with 
gas collapse fractions in previous simulations of the first 
galaxies (e.g., \citealp{Wise:2008}; \citealp{Regan:2009a}).
\par
The second scenario for star formation considered here assumes that
stars form continuously at a rate 
\begin{equation}
\dot{M}_\star(z) = 0.05 \Msun\invyr
\left(\frac{f_\star}{0.05}\right)\left(\frac{\dot{M}_{\rm g}(z)}{ 1 \Msun\invyr}\right)
\label{Eq:Continuous}
\end{equation}
 in proportion to the rate $\dot{M}_{\rm g}$ at which gas is accreted.
Indeed, galaxies with masses $\gtrsim 10^9 \Msun$ may be sufficiently
massive to sustain a moderate level of near-continuous star
formation despite ongoing feedback (e.g., \citealp{Wise:2009}; we 
discuss the effects of feedback in more detail in
Section~\ref{Sec:Limitations} below). We adopt a gas
accretion rate $\dot{M}_{\rm g} \sim 1\Msun\invyr$ that describes the
rate of accretion of gas onto the central region with radius $r
\lesssim 0.1 r_{\rm vir}$ for $z \lesssim 15$ in our simulations of
halos with virial masses $\sim 10^9 \Msun$ (see the right panel
of Fig.~\ref{Fig:Profiles:Velocities}). At this radius the gas surface
density is roughly in agreement with the critical surface density 
$\gtrsim 10-100 \Msun\invpcsq$ for star formation in the low-redshift universe
(see the middle panel of Figure~\ref{Fig:Profiles:Disk}); this critical surface density 
is further discussed in Section~\ref{Sec:Implications} below. We approximately include the effects
of feedback from star formation by employing a lower star formation
efficiency $f_\star = 0.05$ than used for the starburst. The implied
star formation rates $\dot{M}_\star(z) = 0.05 \Msun\invyr$ are
consistent with star formation rates found in recent feedback simulations 
of galaxies inside halos with masses $\sim 10^9 \Msun$ (e.g., \citealp{Wise:2009}; \citealp{Razoumov:2010}).
\par
To compute the luminosities of the stellar populations that form in
the two scenarios we must specify the metallicities of the stars, the stellar
ages, and the distribution of stellar masses at the time of 
formation, i.e., the initial mass function (IMF). We first assume that
starbursts occur in metal-free gas and form clusters of
zero-metallicity stars.  We adopt a top-heavy IMF, i.e., an IMF biased
towards massive ($M_\star \sim 100\Msun$) stars. Such an IMF is
expected to characterize the first, metal-free 
generation of stars which form by radiative cooling from collisionally 
excited molecular hydrogen (e.g., the review by \citealp{Bromm:2009}).
\par
The IMF of metal-free stars, however, is still subject to large
theoretical uncertainty. Stars forming out of gas with an elevated
electron fraction, such as produced behind structure formation or
supernova (SN) shocks or as inside and near ionized regions, could have
characteristic masses substantially less than $\sim 100
\Msun$. This is because the increased electron abundance boosts the
production of HD which enables gas to cool to
much lower temperatures than is possible with molecular hydrogen alone
(e.g., \citealp{Nakamura:2002}; \citealp{Nagakura:2005};
\citealp{Johnson:2006}; \citealp{Stacy:2007}; see also
\citealp{Shapiro:1987} and \citealp{Clark:2010}). We therefore repeat our analysis assuming the
formation of metal-free stars with a normal IMF, similar to the one used
to describe star formation in the nearby universe.
\par
Enrichment to critical metallicities as low as
$Z_{\rm c} \lesssim 10^{-6} - 10^{-3.5} \Zsun$, where we set $\Zsun \equiv 0.02$, 
will also imply the transition from a
top-heavy IMF to a normal IMF (e.g., \citealp{Bromm:2001};
\citealp{BrommLoeb:2003}; \citealp{Schneider:2006};
\citealp{Smith:2009}). We therefore complement our study of metal-free
starbursts with the study of a burst of stars with above-critical but low metallicities $Z \gtrsim 3 \times 10^{-4} \Zsun$ 
and normal IMF. Note that a few massive star SN explosions may already be sufficient to enrich the first 
galaxies to metallicities $Z \gtrsim Z_{\rm c}$
(e.g., \citealp{Scannapieco:2003}; \citealp{Tornatore:2007}; \citealp{Wise:2008a}; \citealp{Karlsson:2008};
\citealp{Greif:2010}; \citealp{Maio:2010}). We therefore always adopt above-critical (but low) metallicities 
$Z \gtrsim 3 \times 10^{-4} \Zsun$ and a normal IMF in the continuous star formation scenario.
\par
We compute the stellar ionizing luminosities expected for
the two star formation scenarios using the population synthesis models
for zero-age instantaneous bursts and continuous star formation from
\citet[][some of these models have been previously published in
\citealp{Schaerer:2002}]{Schaerer:2003}.  The \cite{Schaerer:2003}
models assume a power-law IMF $p(m_\star) \propto m_\star^{-\alpha}$
with \cite{Salpeter:1955} exponent $\alpha = 2.35$ but allow for
different ranges for the masses $m_\star$ of individual stars. We use
the \cite{Schaerer:2003} zero metallicity models for instantaneous
starbursts with initial masses in the range $50-500\Msun$ and
$1-100\Msun$ to describe metal-free stars with a top-heavy and normal
IMF, respectively.  We describe the populations of low-metallicity
stars using the \cite{Schaerer:2003} models with initial masses in the
range $1-100\Msun$ and metallicities\footnote{Our conclusions are
insensitive to the precise choice for the metallicity $Z > 0$ in the
models. } $Z=5 \times 10^{-4} \Zsun$.  Following \cite{Schaerer:2003},
we use case B recombination theory (e.g., \citealp{Osterbrock:1989})
to relate the ionizing luminosities of the stellar populations of
specific age, mass, and metallicity to the nebular luminosities of the
surrounding gas, assuming that all ionizing photons are absorbed,
i.e., that the fraction of ionizing photons escaping the galaxy,
$f_{\rm esc}$, is zero.\footnote{We obtain the luminosities in the
nebular recombination lines using equations (7) and (8) in
\cite{Schaerer:2003} together with the data in their Tables 1, 3, and
4. We obtain the combined stellar and nebular UV1500 continuum
intensity, averaged within a $20\Ang$ band centered on $1500\Ang$,
from the corresponding online data sets provided at
http://obswww.unige.ch/sfr.} We will discuss these assumptions at 
the end of this section. 
\par
Figure~\ref{Fig:IntrinsicLuminosities} shows the luminosities of the
H$\alpha$ and Ly$\alpha$ (left panel) and He1640 (middle panel)
nebular lines and the intensities of the non-ionizing stellar and
nebular UV1500 continuum (right panel) expected for the two star
formation scenarios described above.  Note that the 
luminosities in the Ly$\alpha$ line are related to those in the
H$\alpha$ line by the simple scaling $L({\rm Ly}\alpha) \approx 8.6
L({\rm H}\alpha)$ (Tables~1 and 4 in \citealp{Schaerer:2003}). For 
low metallicity and normal IMF, the starburst scenario implies line
luminosities and continuum intensities that are roughly twice as large
as those implied by the continuous star formation scenario.    
\par
At
fixed IMF, the luminosities in H$\alpha$ and Ly$\alpha$ and the UV1500
continuum intensities of the starbursts are insensitive to the stellar
metallicity. The zero-metallicity starburst models with top-heavy IMF
imply H$\alpha$ and Ly$\alpha$ line luminosities and UV1500
continuum intensities larger by about one order of magnitude than
those implied by the zero-metallicity starburst model with a normal
IMF. In contrast, the starburst luminosities in the He1640 line depend
strongly on both the IMF and the stellar metallicity. At fixed normal IMF, a
change from low to zero stellar metallicity implies an increase in the
He1640 line luminosity by about three orders of magnitude. This is
because the exceptionally hot atmospheres of zero-metallicity stars
turn them into strong emitters of HeII ionizing radiation (e.g.,
\citealp{Tumlinson:2000}; \citealp{Bromm:2001};
\citealp{Schaerer:2003}). An additional change from normal to
top-heavy IMF increases the luminosity in the He1640 line by another
order of magnitude. 
\par
The large differences in He1640 line luminosities offer the prospect
of distinguishing observationally between stellar populations made of
metal-free and metal-enriched stars and of constraining their IMFs
(e.g., \citealp{Tumlinson:2000}; \citealp{Bromm:2001};
\citealp{Oh:2001}; \citealp{Johnson:2009}). The He1640 recombination
line will also be excited due to the emission of ionizing radiation from a
central accreting black hole, if present (e.g., \citealp{Oh:2001}; 
\citealp{Tumlinson:2001}; \citealp{Johnson:2010}).  Observationally, accreting black holes
could be distinguished from metal-free stellar populations through the
detection of their X-ray emission (e.g, \citealp{Haiman:1999}). Note though that an evolved stellar
population may also contribute to the X-ray 
emissivity (e.g., \citealp{Oh:2001b}; \citealp{Power:2009}). X-ray sources may ionize the
gas in a larger region than stellar sources, implying a spatially more
extended emission of recombination radiation. 
\par
\subsection{Observed Fluxes}
We translate the line luminosities and UV continuum
intensities into observed fluxes to investigate the detectability with
{\it JWST}. The flux density from a spatially unresolved object emitted in a
spectrally unresolved line with rest-frame wavelength $\lambda_{\rm e}$
and line luminosity $L$ is given by (e.g.,
\citealp{Oh:1999}; \citealp{Johnson:2009})
\begin{eqnarray}
f_\nu(\lambda_{\rm o}) &=& \frac{L \lambda_{\rm e} (1+z) R}{4 \pi c d_{\rm L}^2(z)} \label{Eq:LineFlux} \\
&\sim& 3 \nJy \left(\frac{L}{10^{40} \erg \invs}\right) \nonumber \\
&\times& \left( \frac{\lambda_{\rm e}}{1216 \Ang}\right)  \left(\frac{R}{1000}\right)\left(\frac{1+z}{11}\right)^{-1} \nonumber,
\end{eqnarray}
where $\lambda_{\rm e}$ is the rest-frame wavelength, $\lambda_{\rm o}
= (1+z)\lambda_{\rm e}$ the observed wavelength, and $d_{\rm L} \sim
100 [(1+z) / 10] \Gpc$ the luminosity distance. 
The UV continuum intensity $L_\nu$ of a spatially unresolved
object implies an observed flux density (e.g., \citealp{Oh:1999};
\citealp{Bromm:2001})
\begin{eqnarray}
f_\nu (\lambda_{\rm o}) &=& \frac{L_{\nu}(\lambda_{\rm e})}{4 \pi d_{\rm L}^2(z)} (1+z) \label{Eq:ContFlux}\\
 &\sim & 1 \nJy \left(\frac{L_{\nu}(\lambda_{\rm e})}{10^{27} \erg \invs \invHz}\right) \left(\frac{1+z}{11}\right)^{-1}\nonumber.
\end{eqnarray}
Flux densities, $f_\nu$, are related to 
AB magnitudes, $m_{\rm AB}$, via (\citealp{Oke:1974}; \citealp{Oke:1983})
\begin{equation}
m_{\rm AB} = -2.5 \log_{\rm 10} \left ( \frac{f_\nu}{\nJy} \right) + 31.4.
\end{equation}
\par
\begin{figure}
\includegraphics[trim = 20mm 0mm 20mm 0mm, width = 0.5\textwidth]{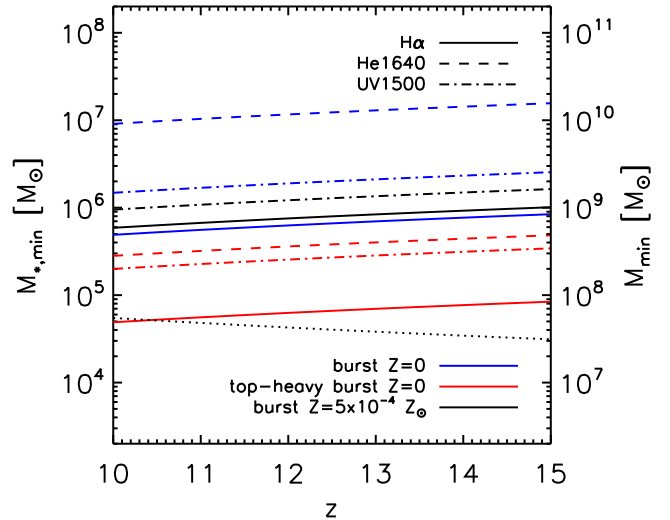}
\caption{Stellar masses $M_{\star,\rm min}$ of the lowest mass
starburst observable through the detection of the H$\alpha$ line (solid
curves) or the He1640 line (dashed curves) or the UV1500 continuum
(dash-dotted curves) with {\it JWST}, assuming an exposure of $t_{\rm
exp} = 10^6 \s$ and S/N = 10. The masses scale with $t_{\rm
exp}^{-1/2}$. Stellar masses derived from the \cite{Schaerer:2003} zero-metallicity starbursts with normal IMF,
the zero-metallicity starbursts with top-heavy IMF, and the low-metallicity starburst, are shown, respectively, in 
blue, red, and black. The right axis shows the masses $M_{\rm min} = 10^3
M_{\star,\rm min}$ of halos expected to host a starburst with stellar mass
$M_{\star,\rm min}$. The conversion between stellar and halo masses
is based on equation~(\ref{Eq:Starburst}) with $f_{\rm cool} = 0.01$ and $f_\star = 0.1$.  
For reference, the dotted curve shows the virial mass (with corresponding labels on the right
axis) for a halo with virial temperature $T_{\rm vir}=10^4 \K$. The black
dashed curve is not shown because it exceeds the plot range.
\label{Fig:MinObsMass}}
\end{figure}
\begin{figure*}
\includegraphics[trim = 20mm 0mm 20mm 0mm, width = 0.33\textwidth]{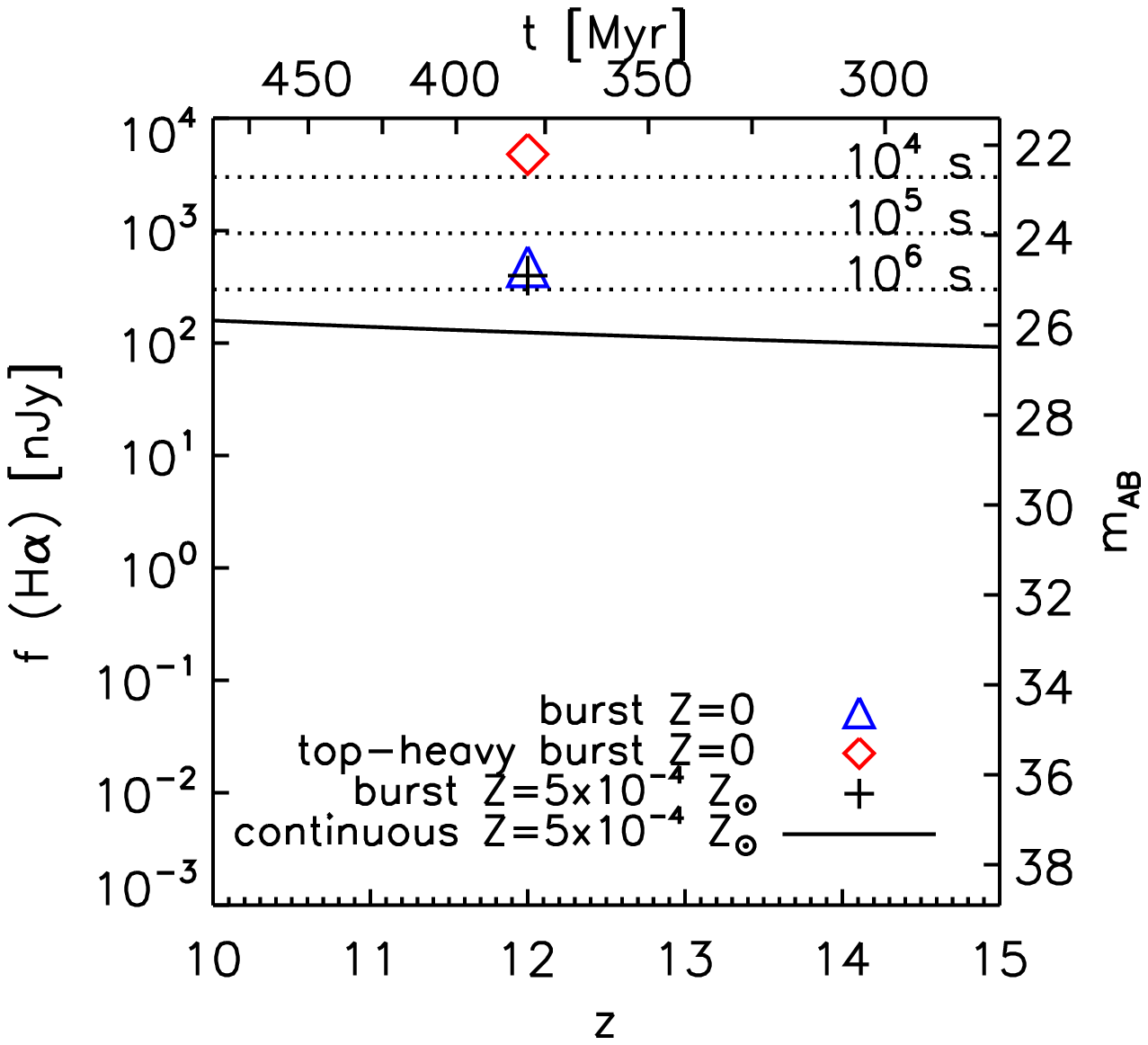}
\includegraphics[trim = 20mm 0mm 20mm 0mm, width = 0.33\textwidth]{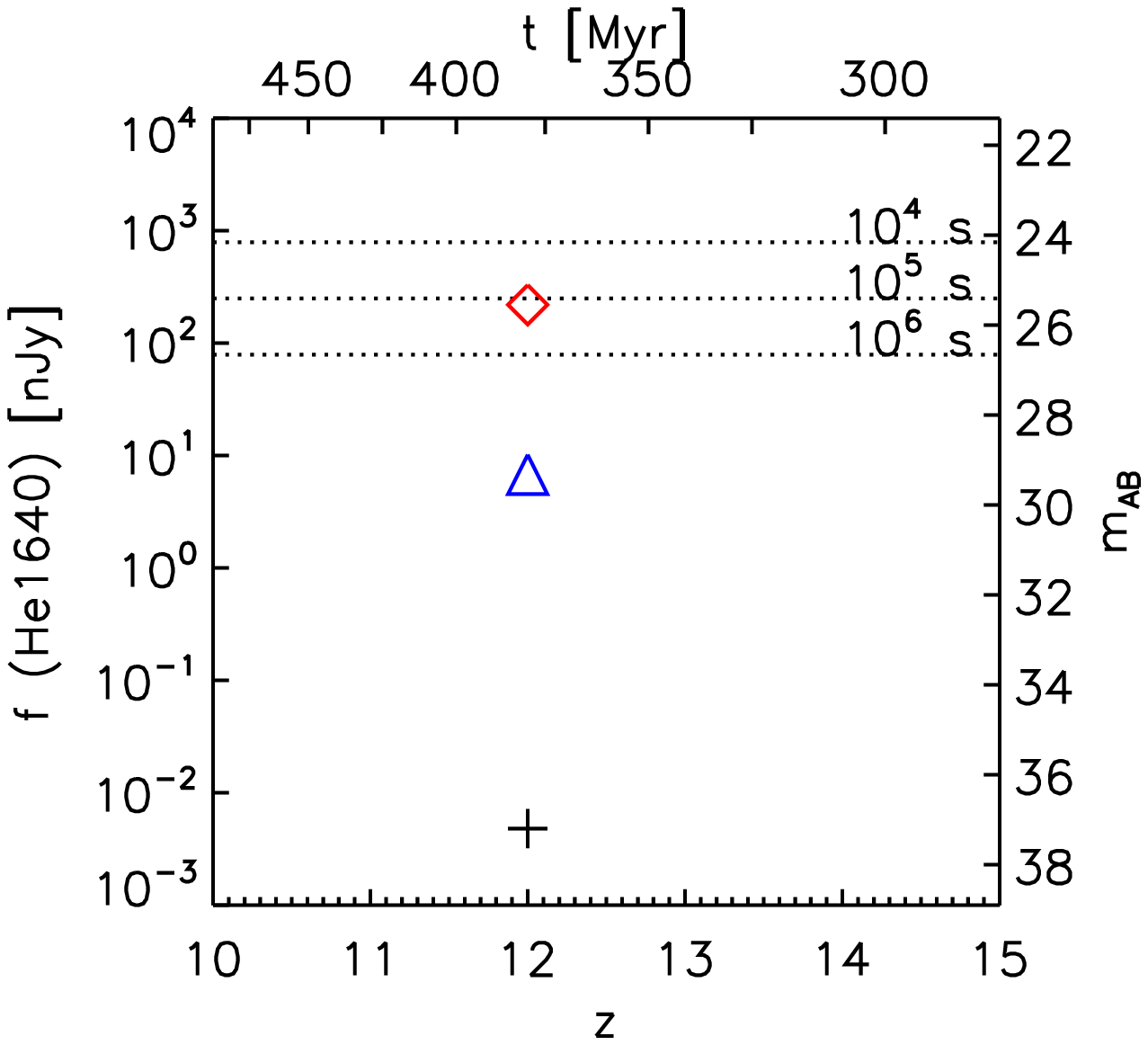}
\includegraphics[trim = 20mm 0mm 20mm 0mm, width = 0.33\textwidth]{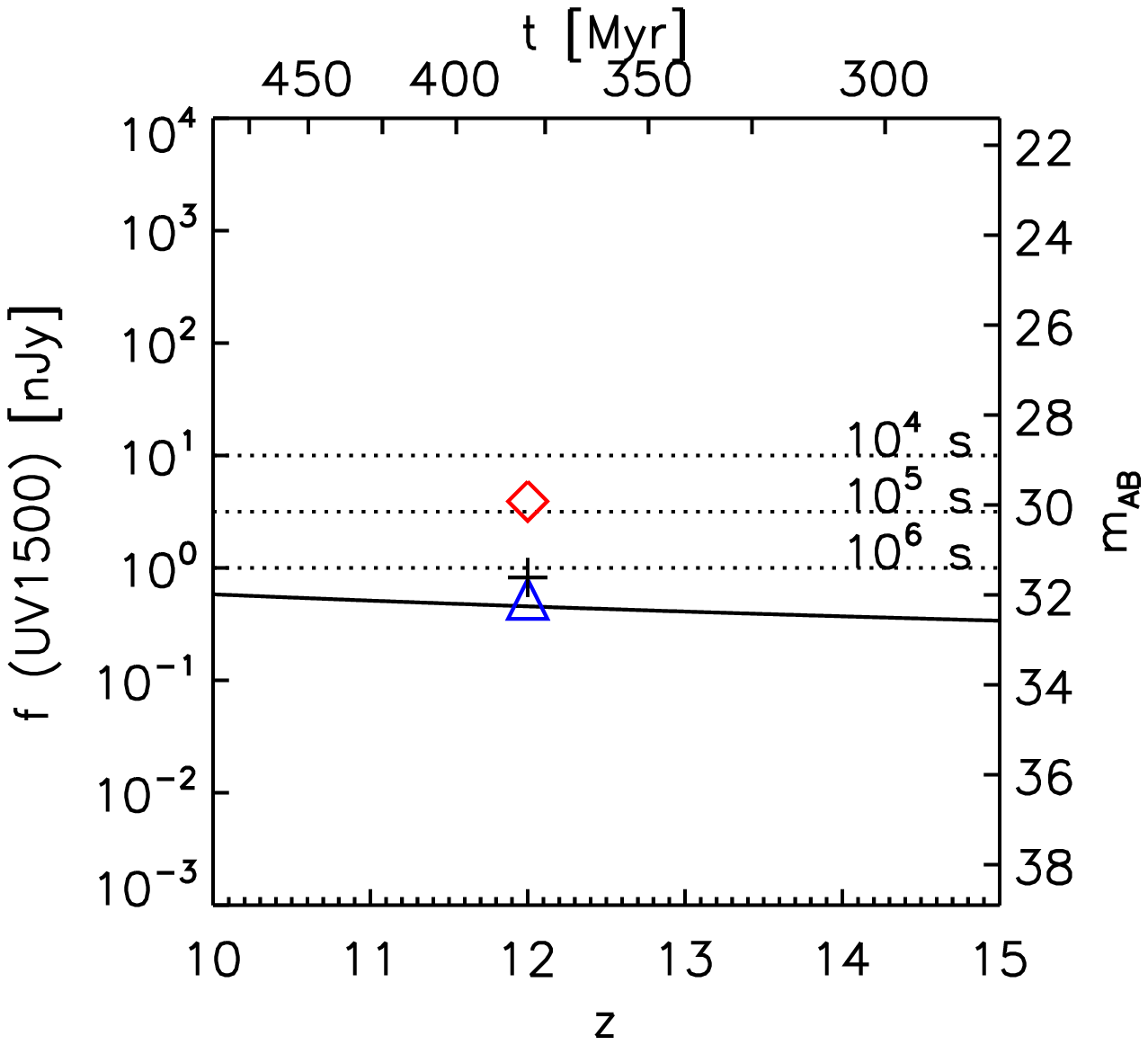}
\caption{Observed flux densities in the H$\alpha$ and He1640
recombination lines and of the combined stellar and nebular UV1500
continuum derived using the line luminosities and continuum
intensities shown in Figure~\ref{Fig:IntrinsicLuminosities} and using
equations~(\ref{Eq:LineFlux}) and (\ref{Eq:ContFlux}).  The flux
densities scale linearly with the star formation efficiencies
of $f_{\star} = 0.05$ and $0.1$ for, respectively, the continuous star
formation and the starburst scenario.  Dotted lines show the
sensitivity limits for observations with {\it JWST}, assuming exposures 
of $10^4$, $10^5$, and $10^6\s$ (top to bottom) and S/N=10. The
He1640 line fluxes obtained for the continuous star formation
scenario are not shown because they are too low to fall inside the
plot range. With exposures $t_{\rm exp} \lesssim 10^6 \s$,
{\it JWST} will have the sensitivity to distinguish between
metal-free starbursts with top-heavy IMF and metal-free or
metal-enriched starbursts with normal IMF inside $\sim
10^9 \Msun$ halos based on the detection of the He1640 line.
\label{Fig:ObservedLuminosities}}
\end{figure*}
The assumption that the lines are spectrally unresolved is excellent
for both H$\alpha$ and He1640, whose line widths $\Delta \lambda /
\lambda \lesssim 10^{-4} (T/10^4\K)^{1/2}$ are set by thermal Doppler broadening at
temperature $T \lesssim 10^4 \K$ (e.g., \citealp{Oh:1999}).  We also
note that at redshifts $z \gtrsim 10$ a transverse physical scale
$\Delta l$ corresponds to an observed angle $\Delta \theta = \Delta l
/ d_{\rm A} \sim 0.1'' (\Delta l / 0.5 \kpc) [(1+z)/10]$, where
$d_{\rm A} = (1+z)^{-2} d_{\rm L}$ is the angular diameter
distance. Hence, if most of the nebular emission originates from
within the vicinity of the stellar populations, which we assumed to be
concentrated in the halo centers, i.e., at $r \lesssim 0.1 r_{\rm vir}$,
the assumption that the emitting regions are spatially unresolved is
good for both the H$\alpha$ and the He1640 line and it applies equally
well to the UV continuum.
\par
In contrast, the Ly$\alpha$ line radiation undergoes resonant
scattering. Hence, it will likely be additionally spectrally
broadened (e.g., \citealp{Neufeld:1990}), and spatially 
extended with typical angular size $\Delta \theta
\sim 15''$ (\citealp{Loeb:1999}). It will be damped due to
absorption by intergalactic neutral hydrogen (e.g.,
\citealp{Miralda:1998}; \citealp{Santos:2004}). Note that Ly$\alpha$ radiation from galaxies
at redshifts $z \gtrsim 10 $ may be particularly strongly damped
because the reionization of the universe was probably only
accomplished at much lower redshifts (e.g., \citealp{Fan:2006}). On
the other hand, scattering off outflowing 
interstellar gas may help the Ly$\alpha$ radiation to
escape (e.g., \citealp{Dijkstra:2010}), and galaxies may reside in a ionized bubble 
sufficiently large for Ly$\alpha$ photons to redshift away 
in the expanding universe (e.g., \citealp{Cen:2000}; \citealp{Haiman:2002}; 
\citealp{Loeb:2005}; \citealp{Wyithe:2005}; \citealp{Lehnert:2010}). 
Since we do not treat radiative transfer effects here, Ly$\alpha$ line
fluxes implied by equation~(\ref{Eq:LineFlux}) must be considered
upper limits. In the following we therefore mostly discuss the
observability of the H$\alpha$ and He1640 lines and of the UV1500
continuum.
\par
\par
\par
\begin{figure*}
\includegraphics[trim = 20mm 0mm 20mm 0mm, width = 0.5\textwidth]{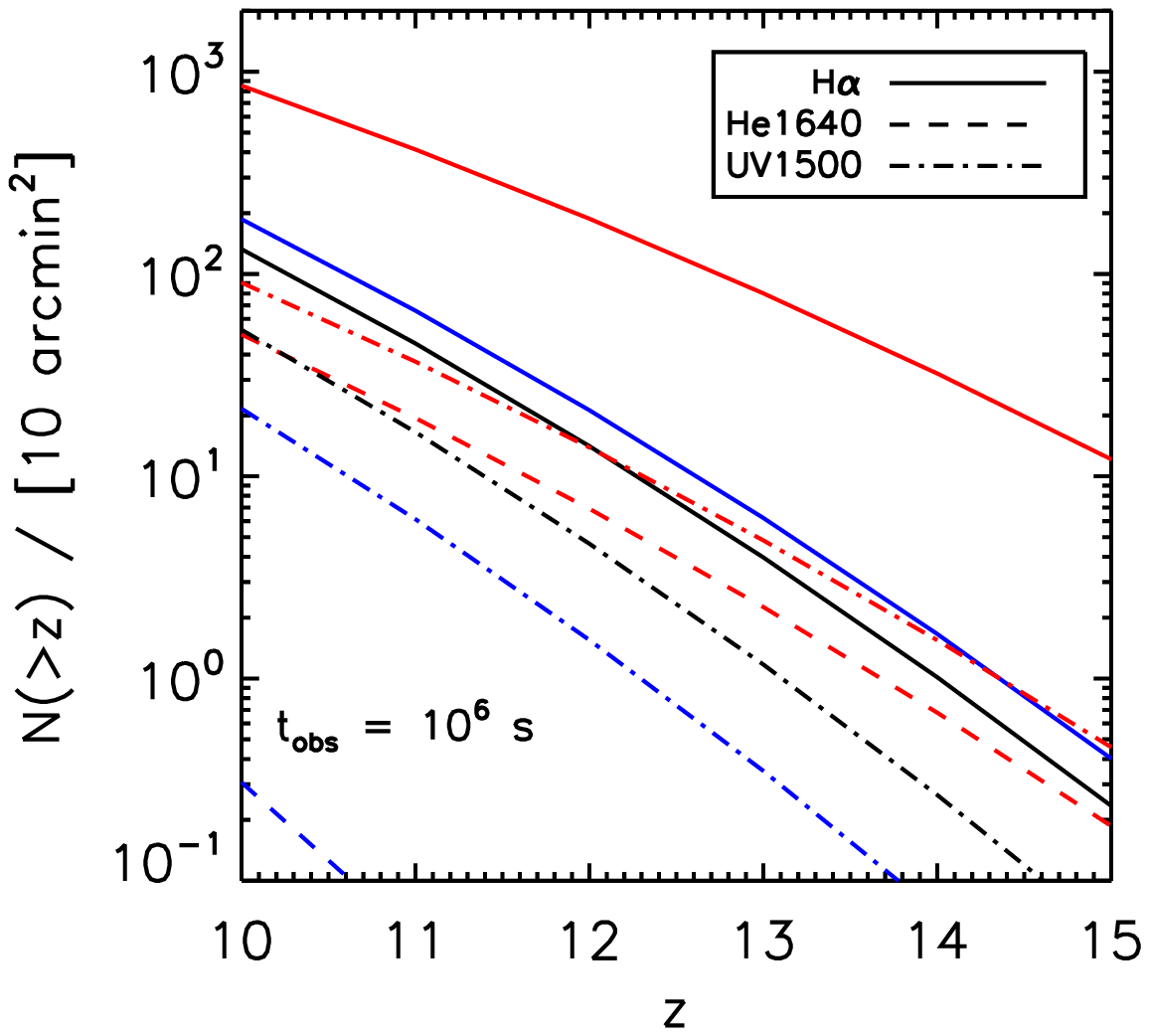}
\includegraphics[trim = 20mm 0mm 20mm 0mm, width = 0.5\textwidth]{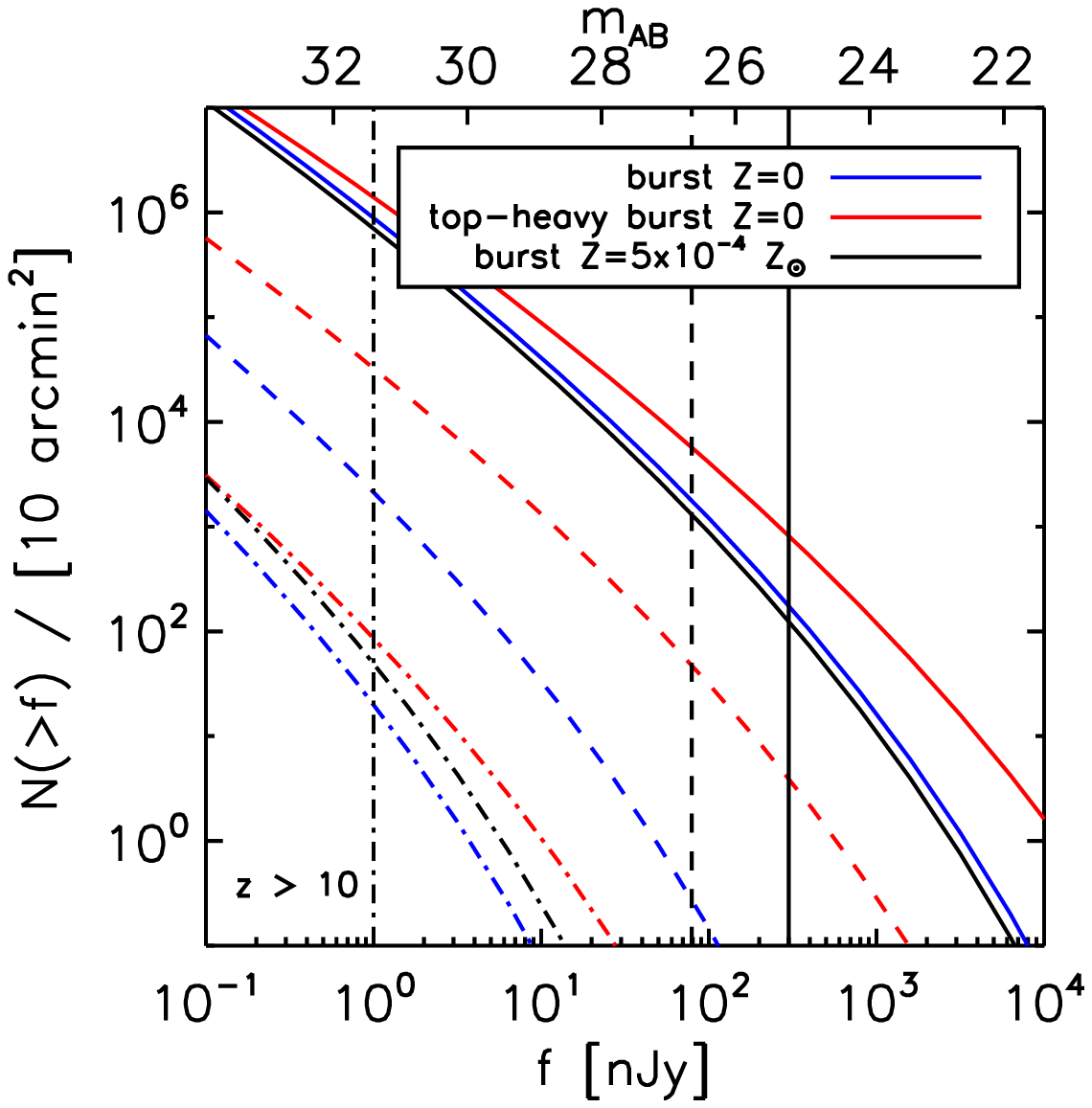}
\caption{{\it JWST} starburst counts. {\it Left panel}: Number of
halos $N(>z)$ with redshifts $>z$ and masses $> M_{\rm min}$, where
$M_{\rm min}$ is the lowest mass halo capable of hosting a starburst
observable through the detection of the H$\alpha$ line (solid curves) or
the He1640 line (dashed curves) or the UV1500 continuum (dash-dotted
curves) with {\it JWST} (Figure~\ref{Fig:MinObsMass}, right axis).
We have assumed an exposure of $t_{\rm exp} = 10^6 \s$ and S/N = 10. Counts
for the model starbursts of zero metallicity and
normal IMF, zero metallicity and top-heavy IMF, and low non-zero
metallicity are shown, respectively, in blue, red, and black. The
black dashed curve is not shown because it falls below the
plot range. {\it Right panel}: Number of halos $N(>f)$ above $z
> 10$ with observed fluxes $>f$. The vertical lines
show the {\it JWST} flux limits $f_{\rm lim} \propto t_{\rm
exp}^{-1/2}$ for observations of the H$\alpha$ line (solid), the He1640 line (dashed)
and the UV1500 continuum (dot-dashed), assuming exposures $t_{\rm exp} = 10^6 \s$
and S/N = 10. The black dashed curve is not shown because it falls below the
plot range. {\it JWST} may detect a few tens (for nonzero metallicities and
normal IMFs) up to a thousand (for zero metallicity and top-heavy
IMFs) starbursts with redshifts $z > 10$ in its field of view of $\sim 10~\mbox{arcmin}^2$. Our estimates for the number 
counts scale linearly with the assumed starburst durations of $\tau_{\rm sb} = 3 \Myr$ and $30 \Myr$ for, respectively, 
the starbursts with top-heavy and normal IMFs.
\label{Fig:NumberCounts}}
\end{figure*}

The He1640 recombination line ($\lambda_{\rm
e} = 1640 \Ang$), as well as the Ly$\alpha$
line ($\lambda_{\rm e} = 1216 \Ang$) not further discussed here, will
be detected by {\it JWST} with NIRSpec at a spectral resolution 
$R \sim 1000$, while the  H$\alpha$ line ($\lambda_{\rm
e} = 6563 \Ang$) will be detected with MIRI at a spectral
resolution $R \sim 3000$.  {\it JWST} will detect the 
UV1500 ($\lambda_{\rm e} = 1500 \Ang$) continuum using 
NIRCam. As an illustration, Figure~\ref{Fig:MinObsMass} shows the
minimum stellar mass required for the \cite{Schaerer:2003} model
starbursts employed here to be observable with {\it JWST}. We assume
exposures with signal to noise ratio S/N = 10 and duration $t_{\rm
exp} =10^6\s$ and the currently expected flux limits\footnote{Flux
limits $f_{\rm lim}$ reported in \cite{Gardner:2006} assume S/N = 10
and $t_{\rm exp} =10^4 \s$.  Here we rescale these limits to other
exposure times using $f_{\rm lim}\propto t_{\rm
exp}^{-1/2}$. Flux limits for $t_{\rm exp} = 10^6 \s$ and S/N=10 can be read 
from Figures \ref{Fig:ObservedLuminosities} and \ref{Fig:NumberCounts}.} for observations with {\it JWST}
(Table~10 in \citealp{Gardner:2006}; see \citealp{Panagia:2005} for a
graphical presentation and \citealp{Johnson:2009} for a useful
summary). Figure~\ref{Fig:MinObsMass} demonstrates that even for
exposure times as long as $10^6 \s$, {\it JWST} will not have
sufficient sensitivity to detect stellar populations with masses below
$\sim 10^5-10^6 \Msun$. {\it
JWST} will thus not be able to see stellar light from individual first stars
(e.g., \citealp{Oh:1999}; \citealp{Oh:2001}; \citealp{Gardner:2006}). 
\par
Figure~\ref{Fig:ObservedLuminosities} shows the H$\alpha$ (left panel) 
and He1640 (middle panel) recombination line fluxes and the
non-ionizing UV1500 continuum fluxes (right panel) for the models
presented in Figure~\ref{Fig:IntrinsicLuminosities}.  The {\it JWST}
flux limits for exposure times $t_{\rm exp} = 10^4, 10^5$ and $10^6\s$
are indicated by the dotted lines in each panel. The figure reveals
that the scenario of continuous star formation implies line and
continuum fluxes too low to be observable, even when assuming exposure times as large as $10^6 \s$.  
{\it JWST}, however, may see starbursts similar to those
modelled here. In exposures with duration $\sim 10^6 \s$, MIRI will
detect such starbursts in H$\alpha$ for all metallicities and IMFs
explored. 
\par
Figure~\ref{Fig:ObservedLuminosities} also reveals that {\it JWST} has
the potential to constrain the properties of starbursts in galaxies
with halo masses as low as $\sim 10^9 \Msun$, based on the detection
of the He1640 line. Indeed, only the zero-metallicity starburst with a
top-heavy IMF and observed with an exposure of $\lesssim 10^6 \s$ is
detected in He1640. Starbursts inside $\gtrsim 10-100$ times more
massive halos would be detected in the He1640 line independent of
whether their IMFs are top-heavy.  Their nature could then be further
constrained by measuring the ratio of the H$\alpha$ and He1640 line
strengths. 
\par
Note, finally, that if scattering by the intergalactic gas can be
ignored, the total (i.e., integrated over the line) flux in the
unresolved Ly$\alpha$ line would be a factor $\approx 8.6$ larger than
the total flux in the unresolved H$\alpha$ line, based on the relation
$L({\rm Ly}\alpha) \approx 8.6 L({\rm H}\alpha)$ noted above and shown
in Figure~\ref{Fig:IntrinsicLuminosities}.  For a galaxy at redshift
$z \approx 10$, {\it JWST}'s NIRSpec is a factor of $\approx 3$ more
sensitive to the detection of the redshifted Ly$\alpha$ line than MIRI
is to the detection of the redshifted H$\alpha$ line\footnote{See the
sensitivity limits for detection of narrow unresolved line fluxes
quoted at
http://www.stsci.edu/jwst/science/data\_simulation\_resources/sensitivity}.
That is, unless radiative transfer effects cause {\it JWST} to see
less than $1/(8.6 \times 3) \approx 4\%$ of the Ly$\alpha$ line flux,
the Ly$\alpha$ line will be easier to detect than the H$\alpha$
line. Hence, despite the large uncertainties arising from its resonant
nature, the Ly$\alpha$ line remains a powerful probe of high-redshift
galaxy formation (\citealp{Partridge:1967}).
\par
\par
\subsection{JWST Number Counts}
How many star-forming galaxies can we expect {\it JWST} to detect?  For simplicity 
and brevity of the presentation we ignore that galaxies may form stars in a continuous mode and assume
that starbursts shine at constant luminosity over a time interval $\tau_{\rm
sb}$. The number of 
galaxies, per unit solid angle, above redshift $z$ that
{\it JWST} will detect is then obtained using
\begin{equation}
\frac{dN}{d \Omega}(>z) = \int_z^\infty dz'\ \frac{dV}{dz'd\Omega}
\frac{\tau_{\rm sb}}{t_{\rm H}(z')}\int_{M_{\rm min} (z')}^\infty dM\ \frac{dn(M,
z')}{dM},
\end{equation}
where $t_{\rm H}(z)$ is the age of the universe at redshift $z$, $dV =
c H^{-1}(z) d_{\rm L}^2 (1+z)^{-2}$ is the comoving volume element,
$H(z) = H_0 [\Omega_{\rm m} (1+z)^{3} + \Omega_{\Lambda}]^{1/2}$, and $H_0
= 100 h \kms \invMpc$ is the Hubble constant. We approximate
the comoving number density $n(M,z)$ of halos with mass $M$ at
redshift $z$ by the Press-Schechter halo abundance 
(\citealp{Press:1974}; \citealp{Bond:1991}; see, e.g., \citealp{Zentner:2007} for a recent review). We
set $M_{\rm min}(z) = f_{\rm cool}^{-1}f_\star^{-1} M_{\star,\rm
min}(z)$, where $M_{\star,\rm min} (z)$ is the smallest stellar mass
observable at redshift $z$ (Figure~\ref{Fig:MinObsMass}), and we use $f_{\rm cool} = 0.01$
and $f_\star = 0.1$ (see equation~[\ref{Eq:Starburst}]). 
\par
Figure~\ref{Fig:NumberCounts} shows our estimates of the number of
observable starbursts in exposures of $t_{\rm exp} = 10^6 \s$ and
S/N=10. The estimates scale linearly with the assumed durations
$\tau_{\rm sb}$ of the starbursts. We set $\tau_{\rm sb} = 3 \Myr$ for
the zero-metallicity starbursts with top-heavy IMF, which
approximately corresponds to the time it takes for its massive $\sim
100 \Msun$ stars to age and explode, upon which further star
formation, if not suppressed by SN feedback, will more likely occur
inside metal-enriched gas, hence ceasing the zero metallicity
burst. We set $\tau_{\rm sb} = 30 \Myr$ for the zero-metallicity and
low-metallicity starbursts with normal IMF. Our choice for this longer
duration reflects the longer time it takes, on average, for massive stars forming 
inside bursts with normal IMFs to evolve and explode in SNe. Starburst
durations up to $\sim 30 \Myr$ are consistent with the gas dynamics
and the amount of gas available for star formation in our simulations 
(see Section~\ref{Sec:Structure} and Figure~\ref{Fig:Profile:Density}).
\par
Figure~\ref{Fig:NumberCounts} demonstrates that {\it JWST} will enable
the detection of a few tens up to a thousand star-bursting galaxies
with redshifts $z \gtrsim 10$ in its field of view of $\sim
10~\mbox{arcmin}^2$. {\it JWST} will allow to constrain the
predominant nature of the first starbursts as the He1640 recombination
line is only detected in significant numbers for the case of
zero-metallicity starbursts with top-heavy IMF. Intriguingly, our
estimates imply that the first galaxies may be more readily detectable
in H$\alpha$ spectroscopic searches than in UV continuum surveys 
(see also Figure~\ref{Fig:ObservedLuminosities}). The
expected number of star-bursting galaxies with redshifts $z > 10$ to
be detected with {\it JWST} in exposures other than $10^6 \s$ can be
read from the right panel in Figure~\ref{Fig:NumberCounts}. Our
estimates are consistent with previous estimates of {\it JWST}
starburst counts for similar assumptions about the conversion between
stellar and halo mass (e.g., \citealp{Haiman:1998};
\citealp{Oh:1999}).
\par
\par
The estimates of the observability of the first galaxies are
uncertain due to the assumptions underlying the computation of UV line
and continuum emission. We followed \cite{Schaerer:2003} and computed
the nebular contribution to the stellar luminosities of the first
galaxies using Case B recombination theory. \cite{Raiter:2010}
point out that the Case B approximation ignores photoionizations from
excited states and hence underestimates the photoionization rate. At
low nebular metallicities and for hot stellar sources, their detailed
photoionization models imply Ly$\alpha$ line luminosities and
nebular UV continuum intensities larger by factors of $2-3$ (see their Figure~10) than expected under the Case B
assumption. \cite{Raiter:2010} also find that the line luminosities
in H$\alpha$ are insensitive to whether Case B is
assumed. Hence, Ly$\alpha$ line and UV continuum emission may provide relatively 
stronger signatures of star formation inside the first galaxies than suggested
here.
\par
Finally, we have assumed that a negligible fraction $f_{\rm esc}$ of
stellar ionizing photons escapes the star-forming regions without
being converted into recombination radiation by the surrounding gas,
i.e., $f_{\rm esc} = 0$. Recent numerical work has emphasized that the
escape fraction may depend strongly on the structural properties of
galaxies as determined by their masses and internal processes like
star formation and feedback (e.g., \citealp{Fujita:2003};
\citealp{Gnedin:2008b}; \citealp{Wise:2009}; \citealp{Johnson:2009};
\citealp{Razoumov:2010}; \citealp{Yajima:2010}). While escape
fractions of order unity $f_{\rm esc} \sim 0.5$ are possible for
low-mass ($\lesssim 10^9 \Msun$) halos with turbulent gas dynamics and
amorphous morphology, significantly smaller escape fractions are
expected for galaxies that form most of their stars inside dense
rotationally-supported disks (e.g., \citealp{Gnedin:2008b};
\citealp{Wise:2009}; \citealp{Razoumov:2010}). The 
luminosities implied by our assumption of a zero escape fraction can be rescaled 
to account for non-zero escape
fractions by multiplication with the factor $(1 - f_{\rm esc})$.
\par
\section{Discussion}
\label{Sec:Implications}
An interesting outcome of our simulations is the collapse of the halo
gas into two extended rotationally supported disks. When and how the
first galaxy-scale disks were formed is currently not well
understood. Extended disks are found in large-scale hydrodynamical
cosmological simulations and in cosmological simulations of individual
massive ($\gtrsim 10^{10} \Msun$) halos or halos at low redshifts (e.g.,
\citealp{Kaufmann:2007}; \citealp{Mashchenko:2008}; \citealp{Levine:2008};
\citealp{Sawala:2010}; \citealp{Governato:2010}; \citealp{Schaye:2010}; 
\citealp{Sales:2010}). Cosmological
simulations of the first minihalos and low-mass ($\lesssim 10^8
\Msun$) halos, however, have not yielded such disks
(e.g., \citealp{Wise:2008}; \citealp{Greif:2008}; \citealp{Wise:2009}; 
\citealp{Regan:2009a}; \citealp{Stacy:2010}). The gas dynamics inside these 
halos is dominated by turbulent motions instead. This morphological 
bimodality suggests that mass is an
important factor in determining whether a given halo may host an
extended disk.  Evidence for the suppression of disk
formation in low-mass galaxies comes from observations of dwarf
galaxies in the local universe which suggest a critical stellar mass
below which stellar disks become systematically thicker (e.g.,
\citealp{Sanchez:2010}; \citealp{Roychowdhury:2010}). 
\par
Our simulations show that the formation of extended gas disks is
possible in halos with masses as low as $M_{\rm vir} \sim 10^9 \Msun$
at redshifts as high as $z \gtrsim 10$ (but see
\citealp{Latif:2010}). However, we acknowledge that the assembly of
the disks may be determined in part by the imposed Jeans floor. The
Jeans floor artificially heats the gas and increases the sound speed,
and this reduces the Mach numbers at which accretion shocks supply
fresh gas to the central high-density regions. Our simulations may
therefore potentially underestimate the ability of accretion flows to
stir up the gas and to channel energy into turbulent motions, which
could otherwise impede the formation of thin extended disks by
transporting angular momentum and driving material that would without
turbulence have circularized in a thin disk rapidly inward. Indeed,
high Mach number accretion flows have been identified as major drivers
of the turbulent gas motions and morphologies seen in simulations of
high-redshift low-mass galaxies (\citealp{Wise:2007};
\citealp{Wise:2008}; \citealp{Greif:2008}). Note that in the
simulation without molecular cooling ({\it Z4NOMOL}), the pressure
floor affects only the inner disk. The formation of the outer disk
thus is a robust outcome of this simulation.
\par
The unresolved gas cores embedded within the disks in our simulations
have masses ($\sim 5 \times 10^7 \Msun$, see
Figure~\ref{Fig:Profile:Density}) and compact sizes ($\lesssim 10
\pc$) to potentially evolve into nuclear star clusters or massive
black holes in the centers of dwarf galaxies. There is an observed
relation between the masses $M_{\rm gal}$ of galaxies and the masses
$M_{\rm cent}$ of the nuclear star clusters in spheroidal galaxies or of massive
black holes in ellipticals and bulges, both of which are $M_{\rm cent} \sim 2
\times 10^{-3} M_{\rm gal}$ (e.g., \citealp{Ferrarese:2006};
\citealp{Wehner:2006}, and references therein). It is not
straightforward to define the masses of the galaxies in our
simulations in a manner consistent with the observational definitions (see,
e.g., the discussion in Li et al. 2007). However, we may
conservatively assume that galaxy masses $M_{\rm gal}$ are fractions
$< 1$ of the associated halo virial masses $M_{\rm vir}$. Then, the
masses of the central cores are too large by factors $ > (M_{\rm cent} / M_{\rm vir}) / (2\times 10^{-3}) = (5\times 10^7
\Msun / 10^9 \Msun) / (2\times 10^{-3}) = 25$ to fit the observed
relation.  Feedback from a central source, which was ignored here, may
be crucial for establishing this relation (e.g.,
\citealp{McLaughlin:2006};
\citealp{Narayanan:2008}; \citealp{Johnson:2010}; but see, e.g.,
\citealp{Li:2007}; \citealp{Larson:2010}). However, it is not
known if the locally observed relation is already established at the
high redshifts of interest and if it extends to the low-mass halo
regime considered here.
\par
We can combine the surface density profiles of the disks obtained in
our simulations with assumptions about a threshold surface density for
star formation to speculate on the stellar radii of the simulated
galaxies. A threshold of $\sim 10 \Msun \invpcsq$ is implied by
observations at kiloparsec scales of star formation in nearby disk galaxies
(e.g., \citealp{Kennicutt:1989}; \citealp{Kennicutt:1998};
\citealp{Bigiel:2008}) and is supported by semi-analytical and numerical
work (e.g., \citealp{Elmegreen:1994}; \citealp{Elmegreen:2002};
\citealp{Schaye:2004}; \citealp{GnedinKravtsov:2010}).  
At high redshifts this threshold could be larger, $\lesssim 100
\Msun \invpcsq$, mostly because of the low dust abundances (implying less 
shielding from the supposed UV background) at these epochs (\citealp{GnedinKravtsov:2010}; 
see also, e.g., \citealp{Schaye:2004}; \citealp{Krumholz:2009}). Our
simulations then imply stellar radii $\lesssim 0.1 \kpc$ (see
Figure~\ref{Fig:Profiles:Disk}). Such small stellar radii are
characteristic of dwarf-globular transition objects and small dwarf
spheroidals around the Milky Way and other members of the Local Group
(see Figure~8 in \citealp{Belokurov:2007}). We, however,
caution that the relatively massive dwarf galaxies simulated here may continue their
stellar growth well below $z \lesssim 10$ as they may accrete gas also
after reionization has raised the Jeans mass in the intergalactic
medium to $\sim 10^8 \Msun$ (see Section~\ref{Sec:Limitations} below). The possibility 
remains that the central gas core forms
stars but the disks do not, in which case the galaxies
would, upon gas loss, evolve into a massive, compact star cluster.
\par
We have shown that the detection of recombination lines and UV
continuum radiation emitted by gas surrounding the first stellar
populations in deep exposures with the upcoming {\it JWST} will likely
only probe galaxies inside halos with masses $\gtrsim 10^9 \Msun$.
Detection of stellar light and recombination radiation from smaller
galaxies may be possible if these galaxies are gravitationally lensed
(e.g., \citealp{Johnson:2009}). Note that recombination radiation may
also be produced by halo gas that does not join the central disks
smoothly but comes to a halt in a shock. The infall energy of the
shocked gas would be transformed into radiation, ionize the disk
environment and be re-emitted as
recombination lines (\citealp{Birnboim:2003}). We have ignored this potentially significant
contribution to the recombination line luminosities in our study of
the observability of the first galaxies presented here.
\par
In addition to detecting recombination radiation from the interstellar gas, {\it JWST} may
observe the first galaxies through the detection of 
cooling radiation emitted during their assembly. {\it JWST} will probably not have the
sensitivity to detect Ly$\alpha$ cooling radiation from galaxies with
halo virial masses $M_{\rm vir} \lesssim 10^{10} \Msun$ (e.g.,
\citealp{Haiman:2000}; \citealp{Dijkstra:2009b}). However, {\it JWST}
and other telescopes such as the {\it Atacama Large Millimeter Array} 
or the proposed {\it Single Aparture Far-Infrared
Observatory}\footnote{http://safir.jpl.nasa.gov} and {\it Space
Infrared Telescope for Cosmology and
Astrophysics}\footnote{http://www.ir.isas.jaxa.jp/SPICA/} 
may detect these low-mass galaxies through the cooling radiation
emitted by molecular hydrogen (for an overview see, e.g.,
\citealp{Appleton:2009}). The luminosity in cooling
radiation from molecular hydrogen will be strongly boosted if emitted by gas
inside SN shells (\citealp{Ciardi:2001}) or by gas powered by X-ray
irradiation from a central black hole (\citealp{Spaans:2008}).
\par
\section{Limitations and Future Work}
\label{Sec:Limitations}
In this work we presented our first steps towards self-consistent simulations
of the formation and evolution of the first galaxies. As such, we have
limited ourselves to the study of important aspects of the
gravitational assembly of individual dwarf galaxy halos and of the
gas-dynamical processes inside their virial regions. Our goal is to
build, step by step, ever more realistic simulations that will allow
us to draw an increasingly detailed picture of high-redshift dwarf
galaxy formation. The most important challenges for future work 
concern effects related to the formation of stars and
associated feedback that we have ignored here. Processes that are
known to strongly affect the assembly and evolution of galaxies
include photodissociation, photoionization, SN explosions
and associated chemical enrichment and radiation pressure from stellar
clusters or black holes (see \citealp{Ciardi:2005} for a review).
\par
Our metal-free atomic cooling simulation is consistent with
scenarios outlined in previous works in which molecular hydrogen
formation and star formation and feedback are suppressed in progenitors
of the assembling dwarf galaxy due to the presence of a 
photodissociating Lyman-Werner background (e.g., \citealp{Oh:2002};
\citealp{Johnson:2008}; \citealp{Regan:2009a}; \citealp{Regan:2009b}; \citealp{Shang:2010}). Star formation in the
progenitors may also be efficiently suppressed due to photoionization
from early (local) reionization, which affects the gas fractions in
low-mass halos primarily by boiling the gas out of the shallow halo
potential wells (e.g., \citealp{Thoul:1996}; \citealp{Barkana:1999};
\citealp{Kitayama:2000}; \citealp{Gnedin:2000}; \citealp{Dijkstra:2004};
\citealp{Shapiro:2004}; \citealp{Whalen:2008a}; \citealp{Petkova:2010}). Reionization also
raises the cosmological Jeans mass in the ionized intergalactic gas
(e.g., \citealp{Shapiro:1994}; \citealp{Gnedin:1998};
\citealp{Gnedin:2000}; \citealp{Hoeft:2006}; \citealp{Okamoto:2008}; 
\citealp{Petkova:2010}), and it affects the rate at which gas can cool and sink towards the
halo centers (\citealp{Efstathiou:1992}; see also \citealp{Wiersma:2009}). Both effects lower the star formation
efficiency because they prevent or impede the replenishing of the gas in 
photoevaporated low-mass halos as well as the accretion of gas by subsequent low-mass
halo generations.
\par
The effects of photodissociation and photoionization on the final
state of our simulated galaxies are difficult to assess without
detailed radiative transfer simulations, also because internal
radiation sources may play an important role (\citealp{Miralda:2005};
\citealp{Schaye:2006}; \citealp{Gnedin:2010}). Preliminary numerical
experiments based on simulations that include the effects of
photodissociation and photoionization from a UV background in the
optically thin approximation\footnote{We have performed a simulation
identical to {\it Z4} but assuming equilibrium cooling in the presence of a uniform
photodissociating and photoionizing \cite{Haardt:2001} background in
the optically thin approximation (and with UV intensities equal to
their $z=9$ values for all $z > 9$; see \citealp{Pawlik:2009} for a
detailed description of the simulation technique) and using only 
3 instead of 4 zoom levels, corresponding to 8 times lower mass resolution.} indicate that UV
radiation is unlikely to prevent the formation of disks in our
simulations. However, photoionization may affect the disk structure, e.g., 
because it may determine the local star formation
efficiency through its effects on the Jeans mass and by producing free
electrons that catalyze the formation of molecular hydrogen.  This
latter effect could partially offset the negative feedback on star
formation from photodissociation (e.g., \citealp{Haiman:2000};
\citealp{Ricotti:2002}; \citealp{Ahn:2007}).
\par  
The assembly and structure of the disks in our simulations 
would probably have been rather different had SN explosions been taken into account. 
Material ejected by SNe could sweep up and shock-heat the surrounding gas and
entrain strong outflows, even in those relatively massive halos whose evolution is hardly 
affected by photoionization (e.g., \citealp{MacLow:1999}; \citealp{Ferrara:2000};
\citealp{Mori:2002}; \citealp{Tassis:2003};
\citealp{DallaVecchia:2008}). SNe may drive 
turbulence and establish a multi-phase interstellar medium with hot
shock-heated chimneys and cold molecular spots inside a warm
photoionized gas (e.g., \citealp{McKee:1977}; \citealp{Wada:2001};
\citealp{Ricotti:2008}; \citealp{Wise:2008a}; \citealp{Greif:2010}).
The dynamics, morphology and thickness of gas disks 
may then critically depend on the distribution of mass over these
three phases (e.g., \citealp{Kaufmann:2007}). Note that 
feedback from SNe could be significantly amplified by previous
episodes of photoionization (e.g., \citealp{Kitayama:2005};
\citealp{Pawlik:2009b}; \citealp{Hambrick:2010}). 
\par
SN explosions affect the subsequent star formation process also by
enriching the interstellar and intergalactic gas with the metals
synthesized in stars (e.g., \citealp{Aguirre:2001};
\citealp{Madau:2001}; \citealp{Scannapieco:2002}; \citealp{Cen:2010}; \citealp{Wiersma:2010}). The increased
metallicity enables additional cooling which may help the gas 
to fragment (e.g., \citealp{Bromm:2001}; \citealp{Schneider:2006}; 
\citealp{Safranek:2010}). \cite{Jappsen:2009} have compared high-redshift low-mass halo 
simulations that include low-temperature cooling by both metals and molecular hydrogen with 
identical simulations that include only cooling by molecular hydrogen and found 
roughly equivalent levels of fragmentation in both types of simulations. 
Our molecular cooling simulation hence may have already captured some 
of the most important effects of low-temperature metal cooling.
\par
\par 
\section{Summary} 
\label{Sec:Summary}
Motivated by the exciting prospect of the direct detection of stellar
light from redshifts $z \gtrsim 10$ with the upcoming {\it JWST}, 
we investigated the assembly of the
first dwarf galaxies using high-resolution cosmological zoomed
smoothed particle hydrodynamics simulations of individual
halos. Previous works suggest that galaxies inside halos with masses
$M_{\rm vir} \lesssim 10^8 \Msun$ at $z \gtrsim 10$ are likely too
faint, by at least a factor of 10, to be observed in the proposed exposures with {\it
JWST}. Hence, the light collected in future observations with {\it
JWST} may come mostly from galaxies inside halos with
masses $M_{\rm vir} \sim 10^9 \Msun$. 
\par
We performed two simulations of
such galaxies that were identical except for differences in the
employed non-equilibrium primordial gas chemistry and cooling network. In the
first of these simulations, gas cooled by emission of 
radiation from both atomic hydrogen and helium and molecular hydrogen. 
We compared this simulation to one in which the formation of molecular hydrogen
was suppressed and, hence, the gas cooled only via atomic processes. We have post-processed the simulated galaxies using idealized models
for star formation and for the strength of the associated
recombination and UV continuum
radiation. We have extrapolated the results to galaxies inside halos
with lower and higher masses to estimate the observability of the
first galaxies with {\it JWST}. 
\par
Our main results are:
\begin{itemize}
\item
At the final simulation redshift $z = 10$, both simulated halos host two nested, extended, 
rotationally supported gas disks. The disks have radii of about 
$0.07$ and $0.3 \kpc$ and total masses of about $8 \times 10^7$ and
$1.2 \times 10^8\Msun$ and surround a central compact gas core with radius
of about $10 \pc$ and total mass of about $5 \times 10^7 \Msun$. 
\item
If $z > 10$ star-bursting galaxies are found in {\it JWST} exposures of less than $\sim 10^5 \s$, then these galaxies 
likely host stellar populations characterized by a top-heavy IMF, or 
reside in halos more massive than $\sim 10^9 \Msun$, or are magnified by gravitational lensing.
\item
Deep {\it JWST} exposures of $10^6 \s$ will find $\sim 10-100$ star-bursting galaxies with
redshifts $z > 10$, assuming a normal IMF. The same exposures will find up
to $\sim$ a thousand $z > 10$ star-bursting galaxies if stellar
populations are characterized by zero metallicity and a top-heavy IMF.
\end{itemize}
\par
Our simulations did not include star formation and associated feedback. They provide a useful
reference for comparison with simulations that include star formation and 
feedback that we will present in future works. 
\par
\acknowledgments
We are grateful to Volker Springel, Joop Schaye, and Claudio Dalla
Vecchia for letting us use their versions of {\sc gadget} as well as
their implementations of FOF and {\sc subfind}. We thank the referee
for comments which improved the discussion and presentation of the
present work. AHP thanks Joop Schaye, Claudio Dalla Vecchia, Marcel
Haas, Freeke van de Voort, and Athena Stacy for helpful discussions,
and Marcel Haas for a thorough reading of an early draft.  The
simulations presented here were carried out at the Texas Advanced
Computing Center (TACC).  This research is supported by NASA through
Astrophysics Theory and Fundamental Physics Program grants NNX08AL43G
and NNX09AJ33G and through NSF grants AST-0708795 and AST-1009928.

\clearpage

\end{document}